\documentclass[sn-basic]{sn-jnl}
\usepackage{mathptmx}
\usepackage{amsmath,amssymb,amsfonts}%
\usepackage{mathtools}

\usepackage{subcaption}
\usepackage{algorithm} 
\usepackage[title]{appendix}

\newcommand{\tamdosfig}{.49\textwidth}
\newcommand{\splegbef}{-2.5mm}
\newcommand{\splegaft}{.8mm}
\newcommand{\algfontsize}{\small}
\newcommand{\algvertspace}{.14\baselineskip} 

\DeclareCaptionLabelFormat{continued}{\textbf{#1} \textbf{#2} (cont.)} 
\DeclareMathOperator{\E}{E}
\DeclareMathOperator{\Var}{Var}
\DeclareMathOperator{\Cov}{Cov}
\newcommand{\diff}{\mathrm d} 
\newcommand{\RR}{\theta}
\newcommand{\LRR}{\Theta}
\newcommand{\OR}{\psi}
\newcommand{\LOR}{\Psi}
\newcommand{\parg}{\zeta} 
\newcommand{\vahRR}{\hat \RR} 
\newcommand{\vahLRR}{\hat\LRR}
\newcommand{\vahOR}{\hat \OR}
\newcommand{\vahLOR}{\hat \LOR}
\newcommand{\vahparg}{\hat \zeta} 
\newcommand{\sug}{r} 
\newcommand{\suf}{r} 
\newcommand{\sus}{s} 
\newcommand{\effic}{\eta}
\newcommand{\efficsep}{\effic_\mathrm{el}}
\newcommand{\efficgr}{\effic_\mathrm{gr}}
\newcommand{\csusco}{a} 
\newcommand{\csusva}{b} 
\newcommand{\tarvar}{A} 
\newcommand{\varaf}{X} 
\newcommand{\raf}{x} 
\newcommand{\varafbor}{\varaf_0} 
\newcommand{\varafn}{Y} 
\newcommand{\varafnbor}{\varafn_0} 
\newcommand{\rafn}{y} 
\newcommand{\cdeadd}{\delta} 
\newcommand{\cdemul}{\gamma} 
\newcommand{\cdesm}{\epsilon} 
\newcommand{\vasag}{N} 
\newcommand{\sag}{n} 
\newcommand{\vgi}{k} 
\newcommand{\vasaf}{M} 
\newcommand{\vasas}{N} 
\newcommand{\fisa}{n} 
\newcommand{\vasu}{S} 
\newcommand{\pco}[1]{\bar #1} 
\newcommand{\tarsara}{\lambda} 
\newcommand{\tarsagr}{l} 
\newcommand{\tarsapr}{\tarsagr_1 \tarsagr_2} 
\newcommand{\ferr}{e}
\newcommand{\fcurv}{c}
\newcommand{\disc}{D}
\newcommand{\roprpr}{\phi}
\newcommand{\rapr}{\RR}
\newcommand{\susrou}{\xi} 
\newcommand{\discu}{d_2}
\newcommand{\discv}{d_1}
\newcommand{\discw}{d_0}
\newcommand{\matIentry}{F}
\newcommand{\matI}{\mathbf \matIentry}
\newcommand{\matJ}{\mathbf J}
\newcommand{\transp}[1]{#1^\intercal}
\newcommand{\liho}{L}
\newcommand{\vagr}{G} 
\newcommand{\vg}{t} 
\newcommand{\vgx}{x} 
\newcommand{\vgy}{y} 
\newcommand{\matg}{\mathbf Q}
\newcommand{\nug}{x} 
\newcommand{\diffgr}{\Delta}
\newcommand{\diffgra}{\tilde \diffgr}
\newcommand{\betaf}{\mathrm B} 
\newcommand{\ribetaf}{I} 
\newcommand{\betappdf}{f}
\newcommand{\parbetau}{u}
\newcommand{\parbetav}{v}
\newcommand{\harm}[1]{H_{#1}}
\newcommand{\cerr}{\mu}
\newcommand{\adjsus}{\alpha}
\newcommand{\bfa}{(a)} 
\newcommand{\bfb}{(b)} 
\newcommand{\CR}{Cram{\'e}r--Rao}
\newcommand{\bfio}{\beta} 
\newcommand{\bfios}{\bfio_\mathrm s} 
\newcommand{\bfiof}{\bfio_\mathrm f} 
\newcommand{\pbfe}{\pi} 
\newcommand{\dgvi}{(i)} 
\newcommand{\dgvii}{(ii)} 
\newcommand{\dgviii}{(iii)} 

\makeatletter
\newcommand\cond{
 \@ifstar
  {\mathrel{}\middle|\mathrel{}}
  {\mid}%
}
\makeatother

\newcommand{\st}{\cond}

\graphicspath{{figures/}}

\begin{document}

\title{
Estimation of relative risk, odds ratio and their\\ logarithms with guaranteed accuracy and controlled sample size ratio
}


\author[*]{\fnm{Luis} \sur{Mendo}}\email{luis.mendo@upm.es}

\affil[*]{\orgdiv{Information Processing and Telecommunications Center}, \orgname{Universidad Polit\'ecnica de Madrid}, \orgaddress{\street{Avenida Complutense, 30}, \city{Madrid}, \postcode{28040}, \country{Spain}}}


\abstract{
Given two populations from which independent binary observations are taken with parameters $p_1$ and $p_2$ respectively, estimators are proposed for the relative risk $p_1/p_2$, the odds ratio $p_1(1-p_2)/(p_2(1-p_1))$ and their logarithms. The sampling strategy used by the estimators is based on two-stage sequential sampling applied to each population, where the sample sizes of the second stage depend on the results observed in the first stage. The estimators guarantee that the relative mean-square error, or the mean-square error for the logarithmic versions, is less than a target value for any $p_1, p_2 \in (0,1)$, and the ratio of average sample sizes from the two populations is close to a prescribed value. The estimators can also be used with group sampling, whereby samples are taken in batches of fixed size from the two populations simultaneously, each batch containing samples from the two populations. The efficiency of the estimators with respect to the \CR{} bound is good, and in particular it is close to $1$ for small values of the target error.
}



\maketitle

% ...




\section{Introduction}
\label{part: intro}

Let $p_1, p_2 \in (0,1)$ denote the probabilities of occurrence of a given dichotomous attribute in two different populations. The problem of estimating the \emph{relative risk} (RR) or \emph{risk ratio},
\begin{equation}
\label{eq: RR}
\RR = \frac{p_1}{p_2},
\end{equation}
from binary observations of the two populations arises frequently in medical and social sciences, as well as in other fields. For example, in a phase-III clinical trial of a vaccine \citep[chapter~18]{Armitage02} the relevant attribute is the presence of a disease, and the two populations are vaccinated and non-vaccinated people. The \emph{odds ratio} (OR),
\begin{equation}
\label{eq: OR}
\OR = \frac{p_1(1-p_2)}{p_2(1-p_1)},
\end{equation}
is sometimes used instead of RR, as are their logarithmic versions: \emph{log relative risk} (LRR), $\LRR= \log \RR$, and \emph{log odds ratio} (LOR), $\LOR = \log \OR$. The latter is especially important in connection with logistic regression \citep{Agresti02},
which is a prevalent tool in machine learning \citep{Bishop06}.

When estimating any of these parameters, it is crucial to have knowledge about the \emph{accuracy of the estimation}. Ideally, the estimation should guarantee a target accuracy, regardless of the unknown $p_1$ and $p_2$. A common measure of accuracy is the \emph{mean-square error} (MSE), or its square root, known as root-mean-square error (RMSE), which for unbiased estimators reduce to variance or standard deviation respectively. For non-logarithmic parameters such as RR and OR, these error measurements are meaningful in a \emph{relative} sense \citep{Mendo25b}, because the significance of a given estimation error can only be assessed by comparing it with the true value of the parameter. Thus, it is natural to require that the RMSE be proportional to the true value of the parameter. On the other hand, for LRR and LOR the estimation error is meaningful by itself, without comparing with the true value, as the logarithm transforms ratios into differences.

A second desirable feature, along with guaranteed accuracy, is to have control on the \emph{proportion of sample sizes} of the two populations. Consider first the case that the two populations are sampled individually, i.e.~with \emph{element sampling}, meaning that samples from any population can be taken one by one as needed. In many use cases, it may be required that the two sample sizes be similar, or that they approximately satisfy a given ratio. Another possible sampling procedure is \emph{group sampling}, whereby samples are collected in groups or batches, each containing  $\tarsagr_1$ samples from population $1$ and $\tarsagr_2$ from population $2$. This imposes a strict sample size ratio of $\tarsagr_1/\tarsagr_2$.  Either with element sampling or with group sampling, these conditions only refer to the sample size \emph{ratio}; the actual sample sizes should be chosen to fulfill the target accuracy.

Note that the difference of the group sampling scheme with respect to element sampling is not only that samples are taken in \emph{groups} (as opposed to one by one), but also that each group simultaneously contains samples from \emph{both} populations (instead of each population being sampled separately).

It will be assumed that each population is infinite, and observations are statistically independent. This implies that the observations can be modeled as Bernoulli trials. The requirement that the target accuracy, as defined earlier, be satisfied for all $p_1, p_2 \in (0,1)$ makes it necessary to use \emph{sequential sampling}, because any fixed sample size will fail to satisfy that requirement for low enough $p_1, p_2$. More specifically, this work will make extensive use of \emph{inverse binomial sampling} (IBS) \citetext{\citealp{Haldane45}; \citealp[chapter~2]{Lehmann98}}. Denoting the presence or absence of the attribute of interest in a sample as ``success'' or ``failure'', IBS consists in observing samples until a predefined number $\sug$ of successes is obtained. The number $\sug$ will be referred to as the parameter of the IBS procedure.

This paper presents unbiased estimators of the four parameters RR, LRR, OR and LOR, that guarantee a target accuracy and provide control on the sample size ratio, irrespective of $p_1$ and $p_2$. The estimators are based on the \emph{two-stage sampling} procedure suggested in a previous work \citep[section~4]{Mendo25b}. As argued above, the target accuracy, $\tarvar$, is defined as relative MSE for RR and OR, or as MSE for LRR and LOR. The control on the sample size ratio means that, for element sampling of each population, \emph{average} sample sizes will approximately satisfy a specified proportion. The estimators can also be applied with group sequential sampling, using groups of $\tarsagr_1$ and $\tarsagr_2$ samples from each population. This incurs, as will be seen, a small increase in the average sample size compared to element sampling, but ensures an \emph{exact} ratio of the (random) sample sizes. In both cases the estimation \emph{efficiency}, defined in terms of the \CR{} bound, is considerably large, and is close to $1$ for small $\tarvar$.

Estimation of the RR, OR or their logarithmic versions is an important problem in statistics. For a review of results in this area, see \citet[section~1]{Mendo25b}. Variants of the problem considering a desired ratio of sample sizes have been addressed in existing works (often assuming the ratio equal to $1$);
see for example \citet{Siegmund82}, \citet{Agresti99}, \citet{Cho13}, \citet{Cho19}.
However, to the author's knowledge, no previously proposed estimators for these parameters guarantee a target accuracy as defined above, i.e.~relative error for RR and OR or absolute error for LRR and LOR, while offering control on the proportion of sample sizes.

The following notation and basic identities will be used. The function $\log \vgx$ represents the natural logarithm of $\vgx$. The $\sag$-th harmonic number is denoted as
\begin{equation}
\label{eq: harm}
\harm{\sag} = \sum_{\vgi=1}^{\sag} \frac 1 {\vgi}.
\end{equation}
Matrices are written in boldface letters, and $\transp\matg$ represents the transpose of a matrix $\matg$. The regularized incomplete beta function is defined as 
\begin{equation}
\ribetaf(\raf; \parbetau, \parbetav) = \frac 1 {\betaf (\parbetau, \parbetav)} \int_0^\raf \vg^{\parbetau-1}(1-\vg)^{\parbetav-1}\,\diff\mkern .8mu\vg, \quad 0 < \raf < 1; \ \parbetau,\parbetav> 0,
\end{equation}
where $\betaf (\parbetau, \parbetav)$ is the beta function; and from \citet[equations (6.1.15), (6.2.2), (26.5.15)]{Abramowitz70} it follows that
\begin{align}
\label{eq: betaf ident 1}
(\parbetav-1)\betaf(\parbetau+1, \parbetav-1) &= \parbetau\, \betaf(\parbetau, \parbetav), \\
\label{eq: betaf ident 2}
(\parbetau-1)\betaf(\parbetau-1, \parbetav+1) & = \parbetav\, \betaf(\parbetau, \parbetav), \\
\label{eq: ribetaf ident 1}
\ribetaf(\raf; \parbetau, \parbetav) - \ribetaf(\raf; \parbetau+1, \parbetav-1) &= \frac{\raf^\parbetau (1-\raf)^{\parbetav-1}}{\parbetau\, \betaf(\parbetau, \parbetav)}, \\
\label{eq: ribetaf ident 2}
\ribetaf(\raf; \parbetau, \parbetav) - \ribetaf(\raf; \parbetau-1, \parbetav+1) &= -\frac{\raf^{\parbetau-1} (1-\raf)^\parbetav}{\parbetav\, \betaf(\parbetau, \parbetav)}.
\end{align}
The probability density function of a beta prime distribution with parameters $\parbetau$, $\parbetav$ is denoted as $\betappdf(\rafn; \parbetau, \parbetav)$:
\begin{equation}
\label{eq: betappdf}
\betappdf(\rafn; \parbetau, \parbetav) = \frac{\rafn^{\parbetau-1}}{\betaf(\parbetau, \parbetav) (1+\rafn)^{\parbetau+\parbetav}}, \quad \rafn>0; \ \parbetau,\parbetav> 0.
\end{equation}
For a random variable $\varafn$ with this distribution \citep[section~4.4]{Chattamvelli21a}, 
\begin{equation}
\label{eq: Pr betap}
\Pr[\varafn \leq \rafn]
= \int_0^\rafn \betappdf(\vg; \parbetau, \parbetav)\,\diff\mkern .8mu\vg
= \ribetaf\left(\frac{\rafn}{\rafn+1}; \parbetau, \parbetav \right).
\end{equation}
In IBS with success probability $p$, the number $\vasag$ of samples needed to obtain $\sug$ successes has a negative binomial distribution with parameters $\sug$ and $p$.
Then, for $p \in (0,1)$ \citetext{\citealp[equation~(3.1)]{Pathak84}; \citealp[section~4.8.2]{Ross10}},
\begin{align}
\label{eq: neg bin: E}
\E[\vasag] &= \frac{\sug}{p}, \\
\label{eq: neg bin: E inv minus 1}
\E\left[\frac 1{\vasag-1}\right] &= \frac{p}{\sug-1} \quad \text{for } \sug \geq 2, \\
\label{eq: neg bin: Var}
\Var[\vasag] &= \frac{\sug (1-p)}{p^2}, \\
\label{eq: neg bin: Var inv minus 1}
\Var\left[ \frac{1}{\vasag-1} \right] &\leq \frac{p^2 (1-p)}{(\sug-1)^2 (\sug-2+2p)} < \frac{p^2 (1-p)}{(\sug-1)^2 (\sug-2)}
\quad \text{for } \sug \geq 3.
\end{align}

The rest of the paper is organized as follows. Section~\ref{part: RR} describes the estimation procedure for RR, derives approximate expressions and bounds for the average sample sizes and estimation efficiency, and compares these with values obtained from Monte Carlo simulations. The estimation procedure and the results for LRR, OR and LOR are to some extent analogous, and are presented in Sections~\ref{part: LRR}, \ref{part: OR} and \ref{part: LOR}. Concluding remarks are given in Section~\ref{part: concl}.

\section{Estimation of relative risk}
\label{part: RR}

The estimation procedure for RR with element sampling of the two populations is considered in Section~\ref{part: RR sep}. First, the general estimation approach is motivated and described (Section~\ref{part: RR sep proc}). The precise definition of the estimator is then completed (Section~\ref{part: RR sep param ave}). Lastly, theoretical bounds are obtained for the average sample sizes and estimation efficiency, and these are compared with simulation results (Sections~\ref{part: RR sep bounds} and \ref{part: RR sep effic}).

Group sampling is addressed in Section~\ref{part: RR gr}. The estimation procedure used in this case is described (Section~\ref{part: RR gr proc}), and then approximations for the average number of groups and efficiency are obtained and compared with simulation results (Sections~\ref{part: RR gr ave} and \ref{part: RR gr effic}).

\subsection{Element sampling}
\label{part: RR sep}

\subsubsection{Estimation procedure}
\label{part: RR sep proc}

The estimator to be presented is unbiased, and uses two-stage sampling. Each of the two sampling stages is comprised of two independent IBS procedures, one for each population. Before explaining the purpose of each stage, it is necessary to define several parameters and variables. The IBS parameters of the first stage are denoted as $\suf_1$ and $\suf_2$ for the two populations respectively. The resulting numbers of samples are negative binomial random variables, $\vasaf_1$ and $\vasaf_2$. Similarly, the second-stage IBS procedures have parameters $\sus_1$ and $\sus_2$,
and the numbers of samples are $\vasas_1$ and $\vasas_2$. The parameters $\sus_1$ and $\sus_2$ are obtained from $\vasaf_1$ and $\vasaf_2$, as will be seen, and are thus random variables. The total number of samples used from each population $i = 1,2$ is then $\vasaf_i + \vasas_i$. From \eqref{eq: neg bin: E} it stems that $\E[\vasaf_i] = \suf_i/p_1$, $\E[\vasas_i \cond \sus_i] = \sus_i/p_i$, and
\begin{equation}
\label{eq: E vasa RR LRR}
\E[\vasaf_i + \vasas_i] = \frac{\suf_i+\E[\sus_i]}{p_i}.
\end{equation}
Consider a target relative MSE equal to $\tarvar$, and a desired ratio $\tarsara$ of average sample sizes. Let $\vahRR$ denote the estimation of $\RR$. Since MSE reduces to variance for an unbiased estimator, the conditions that $\vahRR$ must satisfy are
\begin{align}
\label{eq: cond tarvar RR}
\frac{\Var[\vahRR]}{\RR^2} &\leq \tarvar, \\
\label{eq: cond tarsara}
\frac{\E[\vasaf_1+\vasas_1]}{\E[\vasaf_2+\vasas_2]} &\approx \tarsara
\end{align}
for any $p_1, p_2 \in (0,1)$.

The purpose of the first sampling stage is to obtain two \emph{pilot} sets of samples,
one from each population, using predefined values for the IBS parameters $\suf_1$ and $\suf_2$; and from those acquire some knowledge about $\rapr$. With this knowledge, suitable values for the second-stage IBS parameters $\sus_1$ and $\sus_2$ are computed such that \eqref{eq: cond tarvar RR} and \eqref{eq: cond tarsara} are satisfied. The results of the second stage, i.e.~$\vasas_1$ and $\vasas_2$, are then used to produce the final estimate $\vahRR$. The rationale is as follows. According to \eqref{eq: E vasa RR LRR},
\begin{equation}
\label{eq: E1 E2 eq RR}
\frac{\E[\vasaf_1+\vasas_1]}{\E[\vasaf_2+\vasas_2]} = \frac{\suf_1+\E[\sus_1]}{(\suf_2+\E[\sus_2]) \rapr}.
\end{equation}
Each of the IBS procedures in the second stage provides information about one of the two probabilities $p_1$ and $p_2$. Given a target accuracy for the second-stage estimate $\vahRR$, there is a \emph{trade-off} between the parameters $\sus_1$ and $\sus_2$: decreasing one of them causes the information on the corresponding probability to be less accurate, which can be compensated for by increasing the other parameter. In view of \eqref{eq: E1 E2 eq RR}, this can be exploited for balancing $\E[\vasaf_1+\vasas_1]$ and $\E[\vasaf_2+\vasas_2]$; doing so requires knowledge about $\rapr$, which is provided by the first stage.

An initial idea to obtain information about $\rapr$ from the first-stage variables $\vasaf_1$ and $\vasaf_2$ is to compute an estimate of it using a generalization of the method described in \citet[section~2]{Mendo25b} for estimating $p/(1-p)$. Namely, from \eqref{eq: neg bin: E} and \eqref{eq: neg bin: E inv minus 1} it follows that $(\suf_1-1)/(\vasaf_1-1)$ is an unbiased estimator of $p_1$ and $\vasaf_2/\suf_2$ is an unbiased estimator of $1/p_2$. Therefore, since the observations used by those estimators are independent, 
\[
\frac{(\suf_1-1) \, \vasaf_2}{\suf_2 \, (\vasaf_1-1)}
\]
is an unbiased estimator of $\rapr$. Replacing $\rapr$ by this estimate and $\E[\sus_i]$ by $\sus_i$, $i=1,2$ in \eqref{eq: E1 E2 eq RR} suggests that the condition \eqref{eq: cond tarsara} will be roughly satisfied if $\sus_1$ and $\sus_2$ are chosen so that
\begin{equation}
\label{eq: cond nsus ratio roughly}
\frac{\suf_1+\sus_1}{\suf_2+\sus_2} \approx \tarsara \, \frac{(\suf_1-1) \, \vasaf_2}{\suf_2 \, (\vasaf_1-1)} .
\end{equation}
It is more convenient, however, to substitute the requirement \eqref{eq: cond nsus ratio roughly} by a generalized version:
\begin{equation}
\label{eq: cond nsus ratio}
\frac{\sus_1+\cdeadd_1}{\sus_2+\cdeadd_2} = \cdemul \, \varaf
\end{equation}
with $\varaf$ defined as
\begin{equation}
\label{eq: varaf RR LRR}
\varaf = \frac{\vasaf_2-\cdesm_2}{\vasaf_1-\cdesm_1},
\end{equation}
where $\cdemul > 0$, $\cdeadd_i \mathbb \in \mathbb R$, $\cdesm_i \in (0,1)$ for $i=1,2$ are design parameters, whose values can be selected to facilitate meeting \eqref{eq: cond tarsara} with good approximation.
(Observe that \eqref{eq: cond nsus ratio} indeed reduces to \eqref{eq: cond nsus ratio roughly} for $\cdemul = \tarsara(\suf_1-1)/\suf_2$, $\cdeadd_i = \suf_i$, $\cdesm_1 = 1$, $\cdesm_2 = 0$.)
In addition,  $\sus_i$ and $\cdeadd_i$ must satisfy
\begin{equation}
\label{eq: cond cdeadd >}
\sus_i + \cdeadd_i > 0, \quad i=1,2,
\end{equation}
to ensure that the left-hand side of \eqref{eq: cond nsus ratio} does not involve division by zero and remains positive.
It should be noted that \eqref{eq: cond nsus ratio} may give non-integer values for $\sus_1$ and $\sus_2$, which thus have to be rounded. This will introduce a small additional error in \eqref{eq: cond tarsara}.

For $\suf_1 \geq 3$, a simple analysis based on \eqref{eq: neg bin: Var} and \eqref{eq: neg bin: Var inv minus 1} shows that the relative variance of $\varaf$ is bounded uniformly on $p_1$, $p_2$. On the other hand, for $\suf_1 = 2$ it is easy to see that the relative variance takes arbitrarily large values as $p_1 \rightarrow 0$.
Thus, to ensure that the variability of $\varaf$ is not too large, the following additional requirement is imposed:
\begin{equation}
\label{eq: suf geq 3}
\suf_i \geq 3, \quad i=1,2.
\end{equation}

The second sampling stage uses the parameters $\sus_1$ and $\sus_2$, determined in the first stage, to obtain $\vasas_1$ and $\vasas_2$. By the same reasoning applied earlier,
\begin{equation}
\label{eq: vahRR}
\vahRR = \frac{(\sus_1-1) \, \vasas_2}{\sus_2 \, (\vasas_1-1)}
\end{equation}
is a conditionally unbiased estimation of $\RR$ given $\sus_1, \sus_2$; and thus it is also unconditionally unbiased. For $\sus_1 \geq 3$ the conditional variance of $(\sus_1-1)/(\vasas_1-1)$ can be bounded, using \eqref{eq: neg bin: Var inv minus 1}, as
\begin{equation}
\Var\left[\frac{\sus_1-1}{\vasas_1-1} \cond* \sus_1 \right] \leq \frac{p_1^2(1-p_1)}{\sus_1-2+2p_1};
\end{equation}
and for $\sus_2 \geq 1$ the conditional variance of $\vasas_2/\sus_2$ is, according to \eqref{eq: neg bin: Var},
\begin{equation}
\Var\left[\frac{\vasas_2}{\sus_2} \cond* \sus_2 \right] = \frac{1-p_2}{\sus_2 p_2^2}.
\end{equation}
Therefore,
\begin{equation}
\label{eq: Var varhRR cond no unif}
\begin{split}
\frac{\E\left[ \vahRR^2 \cond \sus_1, \sus_2 \right]}{\RR^2} 
&= \frac{p_2^2}{p_1^2} \left(\Var\left[\frac{\sus_1-1}{\vasas_1-1} \cond* \sus_1 \right] + p_1^2\right) \left(\Var\left[\frac{\vasas_2}{\sus_2} \cond* \sus_2 \right] + \frac 1 {p_2^2} \right) \\
& \leq \left(\frac{1-p_1}{\sus_1-2+2p_1} + 1 \right) \left( \frac{1-p_2}{\sus_2} + 1 \right).
\end{split}
\end{equation}
This implies that, for all $p_1, p_2 \in (0,1)$,
\begin{equation}
\label{eq: Var varhRR cond}
\frac{\E\left[ \vahRR^2 \cond \sus_1, \sus_2 \right]}{\RR^2} < \frac{1}{\sus_1-2} + \frac{1}{\sus_2} + \frac{1}{(\sus_1-2)\sus_2} + 1.
\end{equation}
In view of \eqref{eq: Var varhRR cond}, let the function $\ferr(\sus_1, \sus_2)$ be defined as
\begin{equation}
\label{eq: ferr}
\ferr(\sus_1,\sus_2) = \frac{1}{\sus_1-\cerr_1} + \frac{1}{\sus_2-\cerr_2} + \frac{\cerr_{12}}{(\sus_1-\cerr_1)(\sus_2-\cerr_2)},
\quad\sus_1>\cerr_1,\ \sus_2>\cerr_2,
\end{equation}
with
\begin{equation}
\label{eq: cerr RR}
\cerr_1 = 2, \qquad \cerr_2 = 0, \qquad \cerr_{12}= 1.
\end{equation}
This will be referred to as \emph{error function}. The parameters $\cerr_1$, $\cerr_2$ and $\cerr_{12}$ are introduced for convenience; this way the expression of $\ferr(\sus_1,\sus_2)$ for other estimators will be the same as \eqref{eq: ferr}, only with different values of these parameters. Then, requiring
\begin{equation}
\label{eq: ferr leq tarvar}
\ferr(\sus_1,\sus_2) \leq \tarvar
\end{equation}
guarantees that condition \eqref{eq: cond tarvar RR} holds; in fact with strict inequality. Namely, from \eqref{eq: Var varhRR cond}--\eqref{eq: ferr leq tarvar}, 
\begin{equation}
\label{eq: Var vahrr}
\frac{\Var\left[ \vahRR \right]}{\RR^2} = \frac{\E\left[ \vahRR^2 \right]}{\RR^2} - \frac{\left(\E\left[ \vahRR \right]\right)^2}{\RR^2} = \frac{\E\left[\E\left[ \vahRR^2  \cond* \sus_1, \sus_2 \right]\right]}{\RR^2} - 1
< \E\left[\ferr(\sus_1, \sus_2) \right] \leq \tarvar.
\end{equation}

\subsubsection{Estimator parameters and approximate average sample sizes}
\label{part: RR sep param ave}

According to \eqref{eq: E vasa RR LRR}, to achieve small average sample sizes, $\suf_i$ and $\sus_i$, $i=1,2$ should be as small as possible. In this section, approximate expressions are first derived for the average sample sizes, from which the choice of $\sus_1$, $\sus_2$ is addressed and the values of the estimator parameters $\cdemul$, $\cdeadd_1$, $\cdeadd_2$, $\cdesm_1$, $\cdesm_2$ are determined. (Section~\ref{part: RR sep bounds} will discuss how to select the values of $\suf_1$, $\suf_2$.)

\begin{figure}
\centering
\includegraphics[width=.56\textwidth]{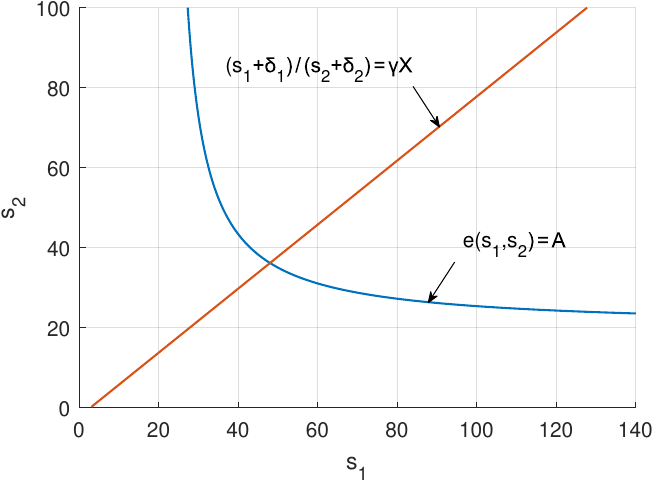}
\caption{Solutions $\sus_1, \sus_2$ to \eqref{eq: cond nsus ratio} and \eqref{eq: ferr tarvar} (example with $\tarvar=0.05$, $\cdemul=0.5$, $\cdeadd_1=1$, $\cdeadd_2=3$, $\varaf=2.5$, $\cerr_1 = 2$, $\cerr_2 = 0$, $\cerr_{12}= 1$)}
\label{fig: sus_sol}
\end{figure}

Since $\ferr(\sus_1, \sus_2)$ is a decreasing function of $\sus_1$ and $\sus_2$, to minimize these parameters \eqref{eq: ferr leq tarvar} should be treated as an equality. Thus, $\sus_1$ and $\sus_2$ are determined by
\begin{equation}
\label{eq: ferr tarvar}
\ferr(\sus_1,\sus_2) = \tarvar 
\end{equation}
together with \eqref{eq: cond nsus ratio} (where the values of the design parameters $\cdemul$, $\cdeadd_1$ and $\cdeadd_2$ are yet to be defined).
Solving this quadratic equation system yields
\begin{align}
\label{eq: nsus 1}
\sus_1 &=
\frac{\cdemul\varaf(\tarvar(\cdeadd_2+\cerr_2)+1) - \tarvar(\cdeadd_1-\cerr_1) + 1 +
\sqrt{\disc}}{2\tarvar}, \\
\label{eq: nsus 2}
\sus_2 &= \frac{\sus_1 + \cdeadd_1}{\cdemul\varaf} - \cdeadd_2,
\end{align}
where the discriminant $\disc$ is
\begin{equation}
\label{eq: disc}
\begin{split}
\disc &= (\cdemul\varaf(\tarvar(\cdeadd_2+\cerr_2)+1) - \tarvar(\cdeadd_1-\cerr_1)+1)^2 \\
&\quad - 4\tarvar\left(\cdemul\varaf((\tarvar\cerr_1+1)(\cdeadd_2+\cerr_2) + \cerr_1-\cerr_{12}) - (\tarvar\cerr_1+1)\cdeadd_1\right).
\end{split}
\end{equation}
There would be another pair of solutions where the square root in \eqref{eq: nsus 1} has a negative sign, but that pair is not valid because it does not satisfy $\sus_1 > \cerr_1$, $\sus_2 > \cerr_2$ as required by \eqref{eq: ferr}. This is illustrated in Figure~\ref{fig: sus_sol}, which makes it clear that there is only one solution pair in the valid range; and therefore it corresponds to the positive sign.

As indicated in Section~\ref{part: RR sep proc}, the solutions \eqref{eq: nsus 1} and \eqref{eq: nsus 2} have to be rounded, because only integer numbers can be used as IBS parameters. Depending on their specific values it may be necessary to round both of them up, or it may be sufficient to round one up and the other down, if that satisfies \eqref{eq: ferr leq tarvar}. A simple criterion, which will be assumed in the rest of the paper, is as follows. First, randomly choose $(i,j)=(1,2)$ or $(2,1)$ with equal probability; then round $\sus_i$ up and $\sus_j$ down, and check if those values satisfy \eqref{eq: ferr leq tarvar}. If not, try rounding $\sus_i$ down and $\sus_j$ up. If not valid either, round both values up, which necessarily satisfies \eqref{eq: ferr leq tarvar}.

\begin{figure}%
\centering%
\begin{subfigure}{\tamdosfig}%
\centering%
\includegraphics[scale=.68]{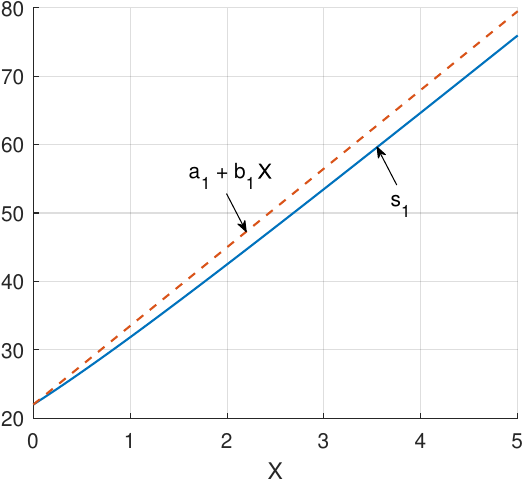}%
\caption{$\sus_1$}%
\label{fig: sus1_aprox}%
\end{subfigure}%
\hfill%
\begin{subfigure}{\tamdosfig}%
\centering%
\includegraphics[scale=.68]{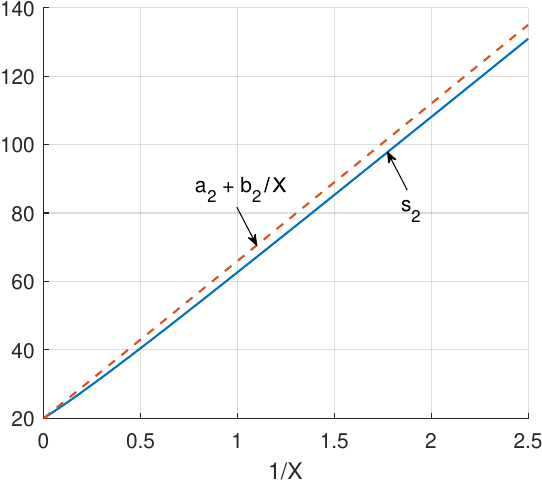}%
\caption{$\sus_2$}%
\label{fig: sus2_aprox}%
\end{subfigure}%
\caption{First-order approximations of $\sus_1$ and $\sus_2$ (example with $\tarvar=0.05$, $\cdemul=0.5$, $\cdeadd_1=1$, $\cdeadd_2=3$, $\cerr_1 = 2$, $\cerr_2 = 0$, $\cerr_{12}= 1$)}%
\label{fig: sus_approx}%
\end{figure}

Expressions \eqref{eq: nsus 1} and \eqref{eq: nsus 2} give $\sus_1$ and $\sus_2$ as functions of $\varaf$, up to the required rounding, but are difficult to deal with. A natural simplification, which will be helpful in fulfilling \eqref{eq: cond tarsara}, is to replace them by the first-order approximations
\begin{align}
\label{eq: nsus 1 approx}
\sus_1 & \approx \csusco_1 + \csusva_1 \varaf, \\
\label{eq: nsus 2 approx}
\sus_2 & \approx \csusco_2 + \frac{\csusva_2}{\varaf},
\end{align}
where the coefficients $\csusco_1$, $\csusva_1$, $\csusco_2$, $\csusva_2$ are obtained from \eqref{eq: nsus 1}--\eqref{eq: disc} as 
\begin{alignat}{2}
\label{eq: csufco 1}
\csusco_1 &= \lim_{\varaf \rightarrow 0} \sus_1 &&= 1/\tarvar + \cerr_1, \\
\label{eq: csufva 1}
\csusva_1 &= \lim_{\varaf \rightarrow \infty} \frac{\sus_1}{\varaf} &&= \cdemul(1/\tarvar + \cdeadd_2 + \cerr_2), \\
\label{eq: csufco 2}
\csusco_2 &= \lim_{\varaf \rightarrow \infty} \sus_2 &&= 1/\tarvar + \cerr_2, \\
\label{eq: csufva 2}
\csusva_2 &= \lim_{\varaf \rightarrow 0} \sus_2 \varaf &&= \frac{1 / \tarvar + \cdeadd_1 + \cerr_1}{\cdemul}. 
\end{alignat}
Figure~\ref{fig: sus_approx} represents these approximations using the example values (same as in Figure~\ref{fig: sus_sol}) $\tarvar= 0.05$, $\cdemul=0.5$, $\cdeadd_1=1$, $\cdeadd_2=3$, and with $\cerr_1$, $\cerr_2$, $\cerr_{12}$ given by \eqref{eq: cerr RR} as specified for RR. The figure illustrates that the accuracy of the approximations depends on the curvature of $\sus_1$ and $\sus_2$ as functions of $\varaf$ and $1/\varaf$. In fact, for certain values of $\tarvar$, $\cdeadd_1$ and $\cdeadd_2$ the approximations \eqref{eq: nsus 1 approx} and \eqref{eq: nsus 2 approx} are exact. This will be analyzed in Section~\ref{part: RR sep bounds}.

Substituting \eqref{eq: nsus 1 approx} and \eqref{eq: nsus 2 approx} into \eqref{eq: E vasa RR LRR}, and introducing an additional term $\susrou$ to account for the effect of rounding $\sus_1$ and $\sus_2$, the average sample sizes are approximated as
\begin{align}
\label{eq: E vasa 1 approx}
\E[\vasaf_1 + \vasas_1] &\approx \frac{\csusco_1 + \suf_1 + \susrou}{p_1} + \frac{\csusva_1}{p_1} \E[\varaf], \\
\label{eq: E vasa 2 approx}
\E[\vasaf_2 + \vasas_2] &\approx \frac{\csusco_2 + \suf_2 + \susrou}{p_2} + \frac{\csusva_2}{p_2} \E\left[\frac 1 {\varaf}\right].
\end{align}
The term $\susrou$ models the average increase in $\sus_1$ and $\sus_2$ due to rounding. The impact of this on $\E[\vasaf_i + \vasas_1]$, $i=1,2$ will usually be negligible, and thus $\susrou$ could be taken as $0$ with good approximation. However, choosing $\susrou=1$ (together with appropriate values of $\suf_1$ and $\suf_2$) will be useful in Section~\ref{part: RR sep bounds} to obtain upper bounds on $\E[\vasaf_i + \vasas_i]$.

The terms $\E[\varaf]$ and $\E[1/\varaf]$ in \eqref{eq: E vasa 1 approx} and \eqref{eq: E vasa 2 approx} can be obtained as follows. Since $\vasaf_1$ and $\vasaf_2$ in \eqref{eq: varaf RR LRR} are independent,
\begin{equation}
\label{eq: E varaf}
\E[\varaf] = \E[ \vasaf_2-\cdesm_2 ] \E\left[ \frac 1 {\vasaf_1-\cdesm_1} \right].
\end{equation}
From \eqref{eq: neg bin: E},
\begin{equation}
\label{eq: vasaf 2 cnsaadd 2}
\E[ \vasaf_2-\cdesm_2 ] = \frac{\suf_2}{p_2} - \cdesm_2 = \frac{\suf_2}{p_2} \left(1 - \frac{\cdesm_2 p_2}{\suf_2} \right).
\end{equation}
As regards $\E[ 1/(\vasaf_1-\cdesm_1) ]$, it is easy to see that for $\cdesm_1 \in (0,1)$ 
\begin{equation}
\label{eq: 1 vasaf cnsaadd}
\frac 1 {\vasaf_1} < \frac 1 {\vasaf_1-\cdesm_1} < \frac {\cdesm_1} {\vasaf_1-1} + \frac {1-\cdesm_1} {\vasaf_1},
\end{equation}
where $\E[1/(\vasaf_1-1)]$ is given by \eqref{eq: neg bin: E inv minus 1} and $\E[1/\vasaf_1]$ is bounded as \citep[section~II]{Mendo06}
\begin{equation}
\label{eq: E 1 vasaf 1}
\frac {p_1} {\suf_1-1} \left(1 - \frac{p_1}{\suf_1-2}\right) < \E\left[ \frac 1 {\vasaf_1} \right] < \frac {p_1} {\suf_1-1} \left(1 - \frac{p_1}{\suf_1-1+p_1}\right).
\end{equation}
Therefore,
\begin{equation}
\label{eq: E 1 vasaf cnsaadd bis}
\frac {p_1} {\suf_1-1} \left(1 - \frac{p_1}{\suf_1-2}\right) < \E\left[ \frac 1 {\vasaf_1-\cdesm_1} \right] < \frac {p_1} {\suf_1-1} \left(1 - \frac{(1-\cdesm_1)p_1}{\suf_1-1+p_1}\right).
\end{equation}
It follows from \eqref{eq: E varaf}, \eqref{eq: vasaf 2 cnsaadd 2} and \eqref{eq: E 1 vasaf cnsaadd bis} that $\E[\varaf]$ is bounded for $\cdesm_1, \cdesm_2 \in (0,1)$ as
\begin{equation}
\label{eq: E varaf bound}
\left(1 - \frac{\cdesm_2 p_2}{\suf_2} \right) \left(1 - \frac{p_1}{\suf_1-2}\right) < \frac{\suf_1-1}{\suf_2\rapr} \E[\varaf] < \left(1 - \frac{\cdesm_2 p_2}{\suf_2} \right) \left(1 - \frac{(1-\cdesm_1)p_1}{\suf_1-1+p_1}\right).
\end{equation}
By analogous arguments, $\E[1/\varaf]$ satisfies the following bound for $\cdesm_1, \cdesm_2 \in (0,1)$:
\begin{equation}
\label{eq: E 1/varaf bound}
\left(1 - \frac{\cdesm_1 p_1}{\suf_1} \right) \left(1 - \frac{p_2}{\suf_2-2}\right) < \frac{(\suf_2-1)\rapr}{\suf_1} \E\left[\frac 1{\varaf}\right]  < \left(1 - \frac{\cdesm_1 p_1}{\suf_1} \right) \left(1 - \frac{(1-\cdesm_2)p_2}{\suf_2-1+p_2}\right).
\end{equation}
A convenient choice for $\cdesm_1$ and $\cdesm_2$, which will be assumed in the sequel, is $\cdesm_1 = \cdesm_2 = 1/2$. Then, in view of \eqref{eq: E varaf bound} and \eqref{eq: E 1/varaf bound}, $\E[\varaf]$ and $\E[1/\varaf]$ are well approximated, especially for small $p_1$, $p_2$, as
\begin{align}
\label{eq: E varaf approx RR LRR}
\E[\varaf] &\approx \frac{\suf_2 \rapr}{\suf_1-1}, \\
\label{eq: E 1/varaf approx RR LRR}
\E\left[\frac 1 {\varaf}\right] &\approx \frac{\suf_1}{(\suf_2-1) \rapr}.
\end{align}
Substituting \eqref{eq: E varaf approx RR LRR} and \eqref{eq: E 1/varaf approx RR LRR} into \eqref{eq: E vasa 1 approx} and \eqref{eq: E vasa 2 approx} yields
\begin{align}
\label{eq: E vasa 1 approx RR bis}
\E[\vasaf_1 + \vasas_1] &\approx \frac{\csusco_1 + \suf_1 + \susrou}{p_1} + \frac{\csusva_1 \suf_2}{(\suf_1-1)p_2}, \\
\label{eq: E vasa 2 approx RR bis}
\E[\vasaf_2 + \vasas_2] &\approx \frac{\csusco_2 + \suf_2 + \susrou}{p_2} + \frac{\csusva_2 \suf_1}{(\suf_2-1)p_1}.
\end{align}

By means of \eqref{eq: E vasa 1 approx RR bis} and \eqref{eq: E vasa 2 approx RR bis}, a desired ratio of average sample sizes can be approximately achieved. Specifically, \eqref{eq: cond tarsara} will hold for all $p_1$, $p_2$ if
\begin{align}
\label{eq: igualdad cruzada 1 RR}
\csusco_1 + \suf_1 + \susrou &= \frac{\tarsara \csusva_2 \suf_1}{\suf_2-1}, \\
\label{eq: igualdad cruzada 2 RR}
\tarsara (\csusco_2 + \suf_2 + \susrou) &= \frac{\csusva_1 \suf_2}{\suf_1-1}.
\end{align}
This equation system can be expressed, making use of \eqref{eq: csufco 1}--\eqref{eq: csufva 2}, 
as
\begin{align}
\label{eq: igualdad cruzada 1 RR bis}
\frac 1 {\tarvar} + \suf_1 + \cerr_1 + \susrou &= \frac{\tarsara \suf_1}{\cdemul(\suf_2-1)} \left(\frac 1 {\tarvar} + \cdeadd_1 + \cerr_1\right), \\
\label{eq: igualdad cruzada 2 RR bis}
\frac{\tarsara}{\cdemul} \left(\frac 1 {\tarvar} + \suf_2 + \cerr_2 + \susrou\right) &= \frac{\suf_2}{\suf_1-1} \left(\frac 1 {\tarvar} + \cdeadd_2 + \cerr_2\right).
\end{align}
Given $\tarvar$, $\tarsara$, $\susrou$, and with $\cerr_1$, $\cerr_2$, $\cerr_{12}$ known, the design parameters $\suf_1$, $\suf_2$, $\cdeadd_1$, $\cdeadd_2$, $\cdemul$ provide degrees of freedom that help meet \eqref{eq: igualdad cruzada 1 RR bis} and \eqref{eq: igualdad cruzada 2 RR bis}. Namely, imposing the relationships
\begin{equation}
\label{eq: suf 1 suf 2, cdeadd 1 cdeadd 2}
\suf_2 = \suf_1 + \cerr_1 - \cerr_2, \qquad
\cdeadd_2 = \cdeadd_1 + \cerr_1 - \cerr_2,
\end{equation}
it is straightforward to solve for $\cdemul$ and $\cdeadd_1$ in \eqref{eq: igualdad cruzada 1 RR bis} and \eqref{eq: igualdad cruzada 2 RR bis}:
\begin{align}
\label{eq: cdemul}
\cdemul &= \tarsara \sqrt{\frac{\suf_1 (\suf_1-1)}{\suf_2 (\suf_2-1)}} = \tarsara \sqrt{\frac{\suf_1 (\suf_1-1)}{(\suf_1+\cerr_1-\cerr_2) (\suf_1+\cerr_1-\cerr_2-1)}}, \\
\begin{split}
\label{eq: cdeadd RR LRR}
\cdeadd_1 &= \left( \frac 1 {\tarvar} + \suf_1 + \cerr_1 + \susrou \right) \frac{\tarsara(\suf_1-1)}{\cdemul\, (\suf_1+\cerr_1-\cerr_2)} - \frac 1 {\tarvar} - \cerr_1 \\
& = \left( \frac 1 {\tarvar} +  \suf_1 + \cerr_1 + \susrou \right) \sqrt{\frac {(\suf_1-1)(\suf_1+\cerr_1-\cerr_2-1)} {\suf_1(\suf_1+\cerr_1-\cerr_2)}} - \frac 1 {\tarvar} - \cerr_1.
\end{split}
\end{align}

The parameters $\cerr_1$, $\cerr_2$ and $\cerr_{12}$ for the RR estimator, given by \eqref{eq: cerr RR}, satisfy
\begin{equation}
\label{eq: cond on cerr}
0 \leq \cerr_1, \cerr_2 \leq 2, \quad 0 \leq \cerr_{12},
\end{equation}
and this will also be the case for the other estimators to be described later. Combining \eqref{eq: cond on cerr} with \eqref{eq: suf geq 3}, \eqref{eq: suf 1 suf 2, cdeadd 1 cdeadd 2} and \eqref{eq: cdeadd RR LRR}, it follows that $\cdeadd_i > -1/\tarvar-\cerr_i$, $i=1,2$. On the other hand, \eqref{eq: ferr} and \eqref{eq: ferr leq tarvar} imply that $\sus_i-\cerr_i > 1/\tarvar$. From these two inequalities it stems that $\cdeadd_1$ and $\cdeadd_2$ as computed from \eqref{eq: suf 1 suf 2, cdeadd 1 cdeadd 2} and \eqref{eq: cdeadd RR LRR} fulfill the requirement \eqref{eq: cond cdeadd >}, and are thus valid.
Making use of \eqref{eq: suf 1 suf 2, cdeadd 1 cdeadd 2}, the expressions \eqref{eq: nsus 1}--\eqref{eq: disc} for $\sus_1$, $\sus_2$ 
are more conveniently written as
\begin{align}
\label{eq: nsus 1 bis}
\sus_1 &=
\frac{\cdemul\varaf(\tarvar(\cdeadd_1+\cerr_1)+1) - \tarvar(\cdeadd_1-\cerr_1) + 1 +
\sqrt{\disc}}{2\tarvar}, \\
\label{eq: nsus 2 bis}
\sus_2 &= \frac{\sus_1 + \cdeadd_1}{\cdemul\varaf} - \cdeadd_1-\cerr_1+\cerr_2, \\
\label{eq: disc bis}
\begin{split}
\disc &= (\cdemul\varaf(\tarvar(\cdeadd_1+\cerr_1)+1) - \tarvar(\cdeadd_1-\cerr_1)+1)^2 \\
&\quad - 4\tarvar\left(\cdemul\varaf((\tarvar\cerr_1+1)(\cdeadd_1+\cerr_1) + \cerr_1-\cerr_{12}) - (\tarvar\cerr_1+1)\cdeadd_1\right).
\end{split}
\end{align}

Substituting \eqref{eq: csufco 1}--\eqref{eq: csufva 2} 
and \eqref{eq: suf 1 suf 2, cdeadd 1 cdeadd 2}--\eqref{eq: cdeadd RR LRR} 
into \eqref{eq: E vasa 1 approx RR bis} and \eqref{eq: E vasa 2 approx RR bis} yields
\begin{align}
\label{eq: E vasa 1 res RR LRR}
\E[\vasaf_1 + \vasas_1] &\approx \left( \frac 1 {\tarvar} + \suf_1 + \cerr_1 + \susrou \right) \left( \frac 1 {p_1} + \frac {\tarsara} {p_2} \right), \\
\label{eq: E vasa 2 res RR LRR}
\E[\vasaf_2 + \vasas_2] &\approx \left( \frac 1 {\tarvar} + \suf_1 + \cerr_1 + \susrou \right) \left( \frac 1 {\tarsara p_1} + \frac 1 {p_2} \right).
\end{align}
It will be beneficial to use \emph{normalized} versions of the average sample sizes, $\E[\vasaf_i + \vasas_i] \sqrt{p_1 p_2}$, so that the expressions depend on $p_1$ and $p_2$ only through their ratio $\rapr$. Defining
\begin{equation}
\roprpr = \sqrt{p_1 p_2},
\end{equation}
the approximations \eqref{eq: E vasa 1 res RR LRR} and \eqref{eq: E vasa 2 res RR LRR} can be written as
\begin{align}
\label{eq: E vasa 1 norm res RR LRR}
\E[\vasaf_1 + \vasas_1] \roprpr &\approx
\left( \frac 1 {\tarvar} + \suf_1 + \cerr_1 + \susrou \right) \left( \frac 1 {\sqrt{\tarsara \rapr}} + \sqrt{\tarsara \rapr} \right) \sqrt{\tarsara}, \\
\label{eq: E vasa 2 norm res RR LRR}
\E[\vasaf_2 + \vasas_2] \roprpr &\approx
\left( \frac 1 {\tarvar} + \suf_1 + \cerr_1 + \susrou \right) \left( \frac 1 {\sqrt{\tarsara \rapr}} + \sqrt{\tarsara \rapr} \right) \frac 1 {\sqrt{\tarsara}}.
\end{align}

The average number of samples $\E[\vasaf_i+\vasas_i]$, $i=1,2$ is the sum of two terms inversely proportional to $p_1$ and $p_2$, according to \eqref{eq: E vasa 1 res RR LRR} and \eqref{eq: E vasa 2 res RR LRR}; and it is $\tarsara$ times more sensitive to $p_2$ than to $p_1$. That is, the parameter of the population for which a \emph{smaller} average sample size is desired
(as specified by \eqref{eq: cond tarsara}) has a \emph{stronger} influence on both average sample sizes. It is also noteworthy that, for $\tarsara$ fixed, \eqref{eq: E vasa 1 norm res RR LRR} and \eqref{eq: E vasa 2 norm res RR LRR} are minimized when $\tarsara\rapr = 1$, which according to \eqref{eq: cond tarsara} and \eqref{eq: E1 E2 eq RR} corresponds to $\suf_1+\E[\sus_1] \approx \suf_2+\E[\sus_2]$.

\subsubsection{Analysis of the approximation and bounds on average sample sizes}
\label{part: RR sep bounds}

Based on an analysis of the approximation error in \eqref{eq: nsus 1 approx} and \eqref{eq: nsus 2 approx}, this section obtains bounds on the average sample sizes and selects appropriate values for $\suf_1$ and $\suf_2$.

The error in approximating \eqref{eq: nsus 1} and \eqref{eq: nsus 2} by \eqref{eq: nsus 1 approx} and \eqref{eq: nsus 2 approx} vanishes when $\sus_1$ is an affine function of $\varaf$ and $\sus_2$ is an affine function of $1/\varaf$ (see Figure~\ref{fig: sus_approx}). This happens when $\tarvar$, $\suf_1$ and $\susrou$ satisfy a certain relationship, as discussed next. Expanding \eqref{eq: disc bis} and collecting terms, $\disc$ can be written as
$
\discu \varaf^2 + \discv \varaf + \discw
$
with
\begin{align}
\discu &= \cdemul^2(\tarvar(\cdeadd_1+\cerr_1)+1)^2, \\
\discv &= 2\cdemul(-\tarvar^2(\cdeadd_1+\cerr_1)^2 - 2\tarvar(\cdeadd_1+\cerr_1-\cerr_{12}) + 1), \\
\discw &= (\tarvar(\cdeadd_1+\cerr_1)+1)^2.
\end{align}
The condition that $\sus_1$ and $\sus_2$ are affine functions of $\varaf$ and $1/\varaf$ is equivalent to $\disc$ being a perfect square with respect to $\varaf$, that is,
\begin{equation}
\label{eq: disc: cond}
\discv = \pm 2\sqrt{\discu\discw}.
\end{equation}
The equality \eqref{eq: disc: cond} with negative sign has the solution $\tarvar=-1/\cerr_{12}$ if $\cerr_{12} \neq 0$. With the estimators considered in this paper, $\cerr_{12}$ is either $0$ or positive, and this results in no solution or a negative solution for $\tarvar$, which is not valid. On the other hand, \eqref{eq: disc: cond} with positive sign is satisfied if and only if
$
\tarvar(\cdeadd_1+\cerr_1)^2 + 2(\cdeadd_1+\cerr_1) - \cerr_{12} = 0.
$
Taking into account \eqref{eq: cdeadd RR LRR}, this can be expressed as $\fcurv(\tarvar, \suf_1, \susrou) = 0$ with
\begin{equation}
\label{eq: fcurv RR LRR}
\begin{split}
\fcurv(\tarvar, \suf_1, \susrou) &= \tarvar\left(\left( \frac 1 {\tarvar} +  \suf_1 + \cerr_1 + \susrou \right) \sqrt{\frac {(\suf_1-1)(\suf_1+\cerr_1-\cerr_2-1)} {\suf_1(\suf_1+\cerr_1-\cerr_2)}} - \frac 1 {\tarvar}\right)^2 \\
& \quad + 2\left(\left( \frac 1 {\tarvar} +  \suf_1 + \cerr_1 + \susrou \right) \sqrt{\frac {(\suf_1-1)(\suf_1+\cerr_1-\cerr_2-1)} {\suf_1(\suf_1+\cerr_1-\cerr_2)}} - \frac 1 {\tarvar}\right)
- \cerr_{12}.
\end{split}
\end{equation}
Furthermore, it is easy to see that $\fcurv(\tarvar, \suf_1, \susrou)$ being positive (negative) implies that $\sus_1$ and $\sus_2$ given by \eqref{eq: nsus 1 bis}--\eqref{eq: disc bis} 
are both convex (concave) functions of $\varaf$ and $1/\varaf$ respectively, which ensures that they are smaller (greater) than their approximations. Thus $\fcurv(\tarvar, \suf_1, \susrou)$ will be referred to as \emph{curvature function}.

This characterization of the error allows transforming \eqref{eq: E vasa 1 norm res RR LRR} and \eqref{eq: E vasa 2 norm res RR LRR} into upper bounds. Since rounding $\sus_1$ and $\sus_2$ never increases their values by $1$ or more, taking $\susrou = 1$ and choosing $\suf_1$ such that $\fcurv(\tarvar, \suf_1, \susrou) \geq 0$ implies that approximations \eqref{eq: E vasa 1 approx} and \eqref{eq: E vasa 2 approx} become inequalities:
\begin{align}
\label{eq: E vasa 1 <}
\E[\vasaf_1 + \vasas_1] &< \frac{\csusco_1 + \suf_1 + 1}{p_1} + \frac{\csusva_1}{p_1} \E[\varaf], \\
\label{eq: E vasa 2 <}
\E[\vasaf_2 + \vasas_2] &< \frac{\csusco_2 + \suf_2 + 1}{p_2} + \frac{\csusva_2}{p_2} \E\left[\frac 1 {\varaf}\right].
\end{align}
From \eqref{eq: E varaf bound} and \eqref{eq: E 1/varaf bound} it is clear that 
$\E[\varaf] < \suf_2 \rapr/(\suf_1-1)$ and $\E[1/\varaf] < \suf_1/((\suf_2-1) \rapr)$.
Substituting into \eqref{eq: E vasa 1 <} and \eqref{eq: E vasa 2 <} and using \eqref{eq: csufco 1}--\eqref{eq: csufva 2} 
and \eqref{eq: suf 1 suf 2, cdeadd 1 cdeadd 2}--\eqref{eq: cdeadd RR LRR} 
yields
\begin{align}
\label{eq: E vasa 1 norm res RR LRR <}
\E[\vasaf_1 + \vasas_1] \roprpr &< \left( \frac 1 {\tarvar} + \suf_1 + \cerr_1 + 1 \right) \left( \frac 1 {\sqrt{\tarsara \rapr}} + \sqrt{\tarsara \rapr} \right) \sqrt{\tarsara}, \\
\label{eq: E vasa 2 norm res RR LRR <}
\E[\vasaf_2 + \vasas_2] \roprpr &< \left( \frac 1 {\tarvar} + \suf_1 + \cerr_1 + 1 \right) \left( \frac 1 {\sqrt{\tarsara \rapr}} + \sqrt{\tarsara \rapr} \right) \frac 1 {\sqrt{\tarsara}}.
\end{align}
These bounds hold for $\susrou=1$ and for any $\suf_1$ such that $\fcurv(\tarvar,\suf_1, 1) \geq 0$. Considering the restriction \eqref{eq: suf geq 3}, the best (i.e.~lowest) upper bounds are obtained by choosing $\suf_1$ as
\begin{equation}
\label{eq: cho suf}
\suf_1 = \min\{\suf = 3,4,5,\ldots \st \fcurv(\tarvar, \suf, 1) \geq 0\}.
\end{equation}
Thus the values $\susrou=1$ and $\suf_1$ as in \eqref{eq: cho suf} will be used for the estimator.

\begin{figure}%
\centering%
\includegraphics[width=.68\textwidth]{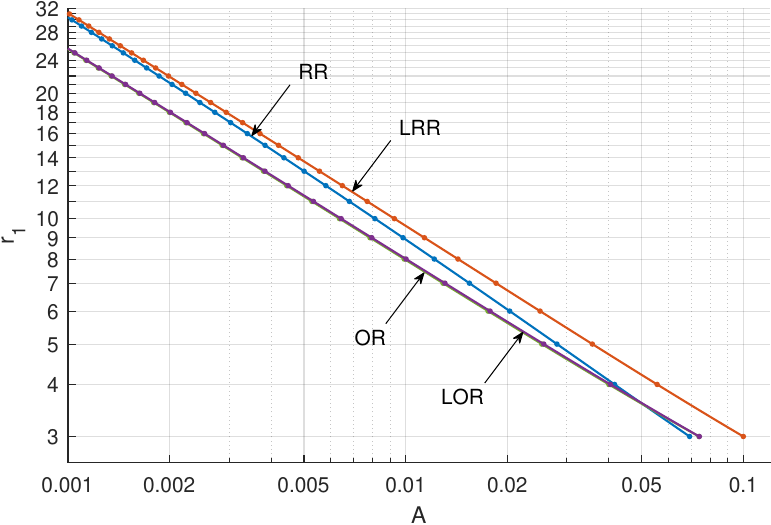}%
\caption{Pairs $(\tarvar, \suf_1)$ for which $\fcurv(\tarvar, \suf_1, 1) = 0$}
\label{fig: curv_zero_1}%
\end{figure}%

The condition $\fcurv(\tarvar, \suf_1, \susrou) = 0$ can be written, from \eqref{eq: fcurv RR LRR}, as
\begin{equation}
\label{eq: curv zero bis RR LRR}
\begin{split}
& (\suf_1+\cerr_1+\susrou)^2(\suf_1-1)(\suf_1+\cerr_1-\cerr_2-1) \tarvar^2 \\
&\quad + \left(2(\suf_1+\cerr_1+\susrou)(\suf_1-1)(\suf_1+\cerr_1-\cerr_2-1) - \cerr_{12}\suf_1(\suf_1+\cerr_1-\cerr_2)\right) \tarvar \\
&\quad + 1-2\suf_1-\cerr_1+\cerr_2 = 0.
\end{split}
\end{equation}
Considering $\suf_1$ as given, this is a quadratic equation in $\tarvar$ with a single positive solution. Figure~\ref{fig: curv_zero_1} shows the resulting curve for RR, i.e.~for $\cerr_1$, $\cerr_2$, $\cerr_{12}$ as in \eqref{eq: cerr RR}, with $\susrou=1$. (The figure also contains curves for other estimators, to be presented in following sections.) For the pairs $(\tarvar, \suf_1)$ in this curve the approximations \eqref{eq: nsus 1 approx} and \eqref{eq: nsus 2 approx} are exact. Note that $\suf_1$ is considered as a continuous variable for clarity of the representation, but only the points with integer $\suf_1$ (marked with dots in the graph) are feasible. Furthermore, it is seen from \eqref{eq: fcurv RR LRR} that $\fcurv(\tarvar, \suf_1, 1)$ increases with $\suf_1$, and therefore the region above (below) the curve corresponds to $\fcurv(\tarvar, \suf_1, 1)$ positive (negative). Thus, for a given $\tarvar$, once the value of $\suf_1$ for which $\fcurv(\tarvar, \suf_1, 1) = 0$ is known, rounding this up gives the minimum integer $\suf_1$ such that $\fcurv(\tarvar, \suf_1, 1) \geq 0$. That is, \eqref{eq: cho suf} is the integer on or immediately above the curve in Figure~\ref{fig: curv_zero_1}, limited from below by $3$.

With the values selected for $\cdemul$, $\cdeadd_1$, $\cdeadd_2$, $\cdesm_1$, $\cdesm_2$, $\susrou$, $\suf_1$, $\suf_2$, the RR estimator is completely specified. The estimation procedure is summarized in Algorithm~\ref{algo: RR LRR} (see Appendix~\ref{part: app algo}), together with a list of its properties, some of which will be derived in the remainder of this section. (The algorithm also includes the LRR case, to be presented in Section~\ref{part: LRR}.)

The performance of the RR estimator is evaluated by means of Monte Carlo simulations in the following. For each combination of parameters
a simulation is run consisting of $10^6$ realizations of the estimator. The \emph{empirical} MSE and average numbers of samples are computed from the simulation (i.e.~expectation is replaced by sample mean), and then they are compared with the theoretical expressions.

Figure~\ref{fig: Ei_RR_1_allwdh} shows the bounds \eqref{eq: E vasa 1 norm res RR LRR <} and \eqref{eq: E vasa 2 norm res RR LRR <} for the normalized average numbers of samples with $\suf_1$ given by \eqref{eq: cho suf} (lines), as well as simulation results for the same $\suf_1$ (bold, red dots) and for all $\suf_1 = 3,\ldots,50$ (light, grey dots). The graphs consider several combinations of $\tarsara$, $\rapr$ and $\roprpr$, with variable $\tarvar$. As a reference, \eqref{eq: cho suf} gives values $3,\ldots,31$ for the displayed range of $\tarvar$.

\begin{figure}%
\begin{subfigure}{\textwidth}%
\centering%
\includegraphics[width=\tamdosfig]{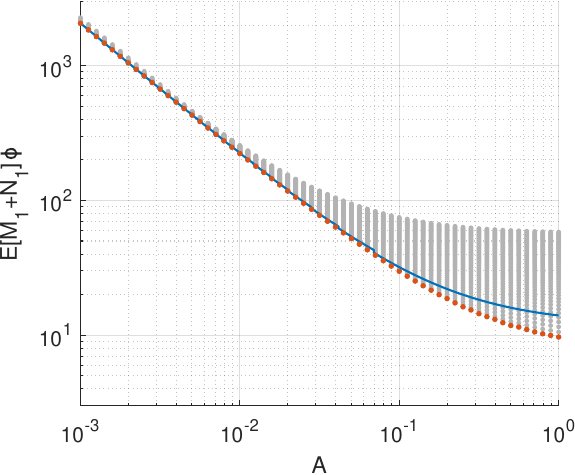}%
\hfill%
\includegraphics[width=\tamdosfig]{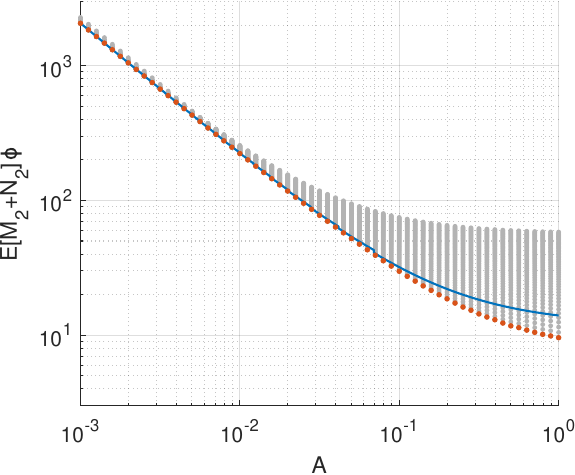}%
\caption{$\tarsara=1$, $\rapr=1$, $\roprpr=0.01$ ($p_1=p_2=0.01$)}%
\label{fig: Ei_RR_1_11_1_0p01_allwdh}%
\end{subfigure}%
\\[\splegaft]
\begin{subfigure}{\textwidth}%
\centering%
\includegraphics[width=\tamdosfig]{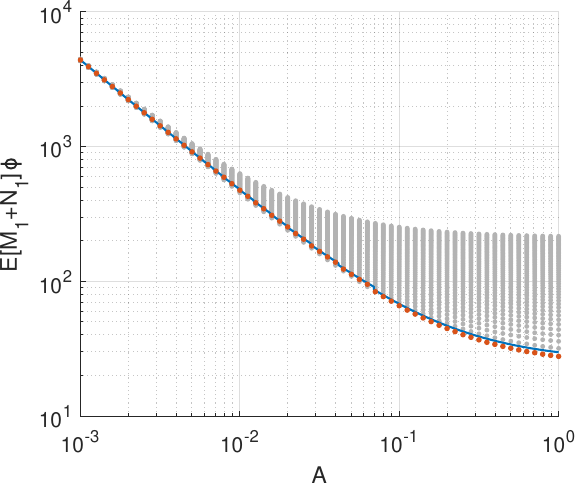}%
\hfill%
\includegraphics[width=\tamdosfig]{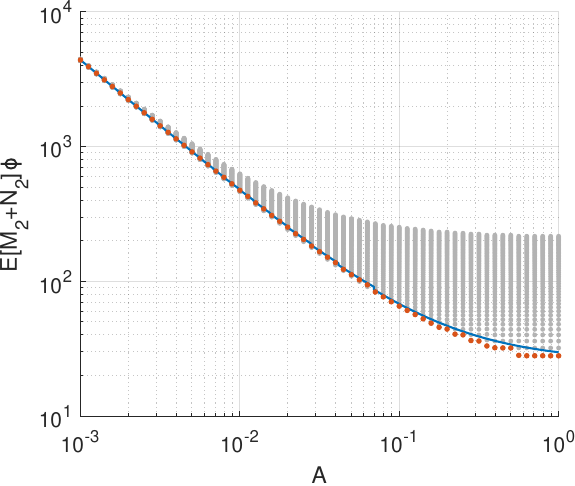}%
\caption{$\tarsara=1$, $\rapr=16$, $\roprpr=0.01$ ($p_1 = 0.04$, $p_2 = 0.0025$)}%
\label{fig: Ei_RR_1_11_16_0p01_allwdh}%
\end{subfigure}%
\\[\splegaft]
\begin{subfigure}{\textwidth}%
\centering%
\includegraphics[width=\tamdosfig]{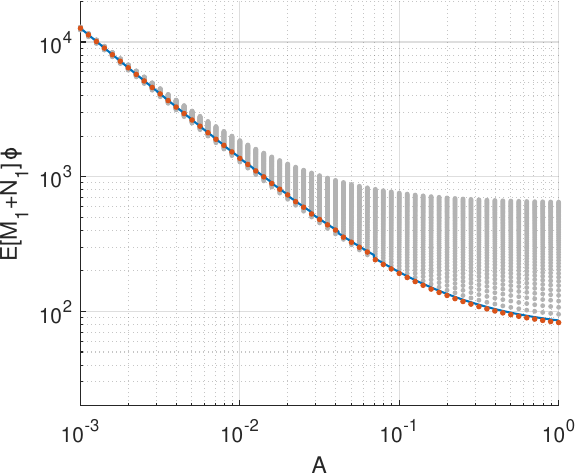}%
\hfill%
\includegraphics[width=\tamdosfig]{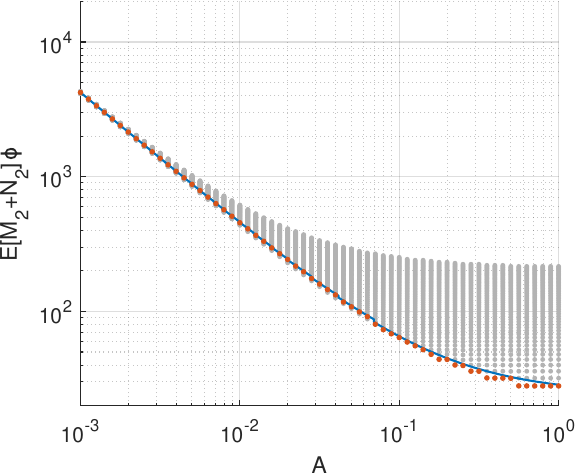}%
\caption{$\tarsara=3$, $\rapr=16$, $\roprpr=0.01$ ($p_1 = 0.04$, $p_2 = 0.0025$)}%
\label{fig: Ei_RR_1_31_16_0p01_allwdh}%
\end{subfigure}%
\\[\splegbef]%
\centering%
\begin{small}%
Line: bound. Bold dots: simulation. Light dots: simulation, other values of $\suf_1$.
\end{small}%
\\[\splegaft]%
\caption{Normalized average sample sizes for RR with element sampling, varying $\suf_1$}
\label{fig: Ei_RR_1_allwdh}%
\end{figure}%
\begin{figure}%
\ContinuedFloat
\begin{subfigure}{\textwidth}%
\centering%
\includegraphics[width=\tamdosfig]{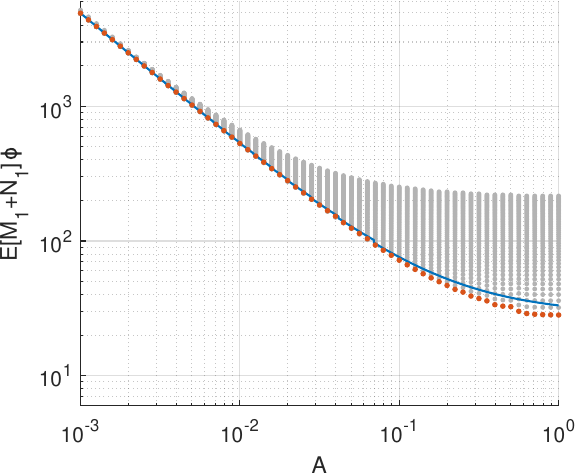}%
\hfill%
\includegraphics[width=\tamdosfig]{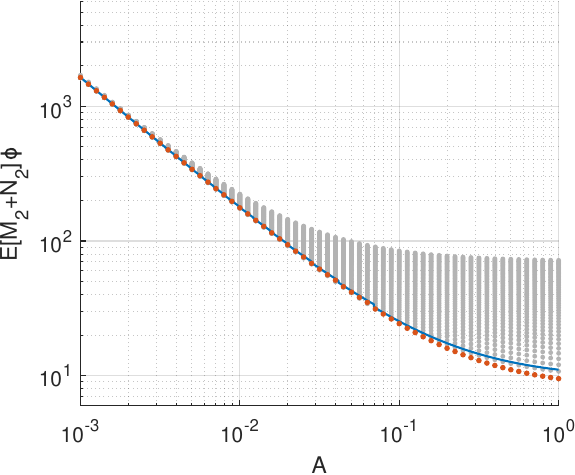}%
\caption{$\tarsara=3$, $\rapr=1/16$, $\roprpr=0.01$ ($p_1 = 0.0025$, $p_2 = 0.04$)}%
\label{fig: Ei_RR_1_31_1d16_0p01_allwdh}%
\end{subfigure}%
\\[\splegaft]
\begin{subfigure}{\textwidth}%
\centering%
\includegraphics[width=\tamdosfig]{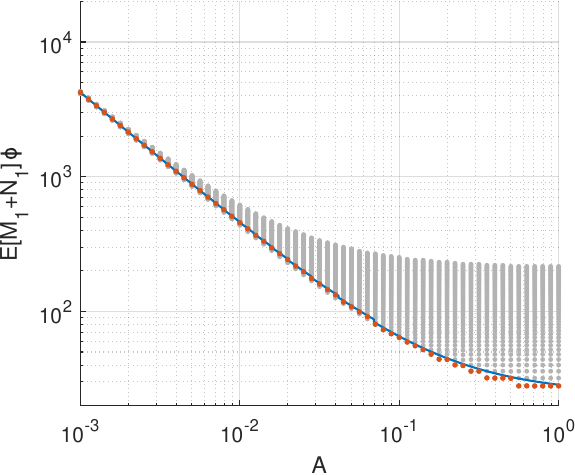}%
\hfill%
\includegraphics[width=\tamdosfig]{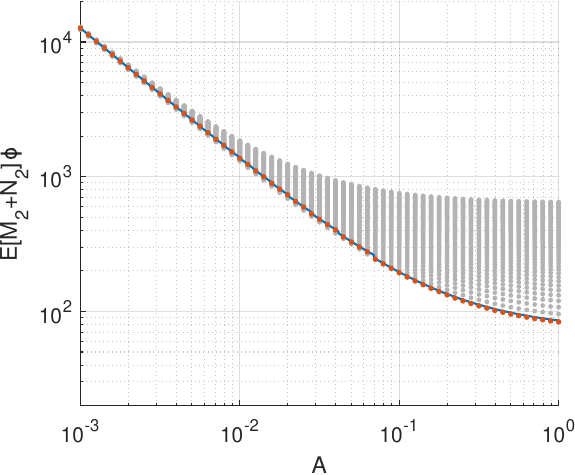}%
\caption{$\tarsara=1/3$, $\rapr=1/16$, $\roprpr=0.01$ ($p_1 = 0.0025$, $p_2 = 0.04$)}%
\label{fig: Ei_RR_1_13_1d16_0p01_allwdh}%
\end{subfigure}%
\\[\splegaft]
\begin{subfigure}{\textwidth}%
\centering%
\includegraphics[width=\tamdosfig]{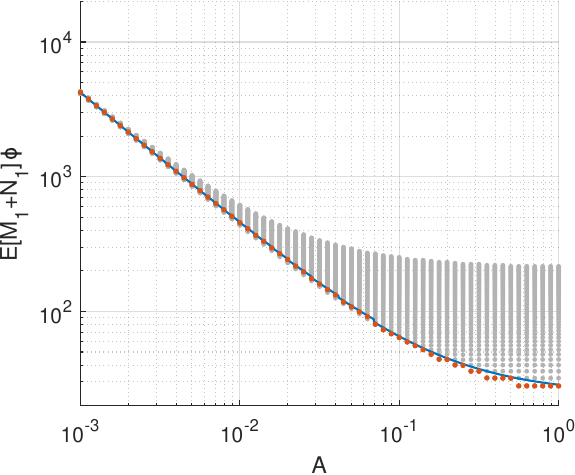}%
\hfill%
\includegraphics[width=\tamdosfig]{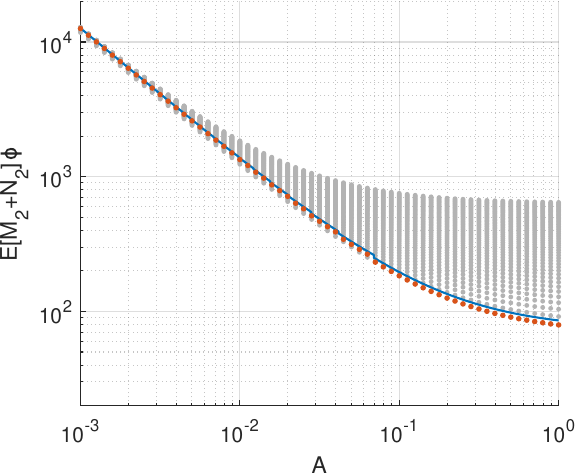}%
\caption{$\tarsara=1/3$, $\rapr=1/16$, $\roprpr=0.1$ ($p_1 = 0.025$, $p_2 = 0.4$)}%
\label{fig: Ei_RR_1_13_1d16_0p1_allwdh}%
\end{subfigure}%
\\[\splegbef]%
\centering%
\begin{small}%
Line: bound. Bold dots: simulation. Light dots: simulation, other values of $\suf_1$.
\end{small}%
\\[\splegaft]%
\caption{Normalized average sample sizes for RR with element sampling, varying $\suf_1$}
\end{figure}%

The following observations can be made from Figure~\ref{fig: Ei_RR_1_allwdh}. For $\suf_1$ as in \eqref{eq: cho suf}, simulated values are close to the theoretical bounds (and always below). When $\tarvar$ is not too large they are in fact very close, whereas for large $\tarvar$ the difference increases because $\suf_1$ is limited below by  $3$. In addition, comparing with simulations for other values of $\suf_1$, \eqref{eq: cho suf} achieves average sample sizes equal or very close to their minima with respect to $\suf_1$. For $\suf_1$ given by \eqref{eq: cho suf},
the ratio $\E[\vasaf_1+\vasas_1]/\E[\vasaf_2+\vasas_2]$ in the simulations is very close to $\tarsara$ in all cases (for reference, note that the theoretical curves  \eqref{eq: E vasa 1 norm res RR LRR <} and \eqref{eq: E vasa 2 norm res RR LRR <} are in the exact ratio $\tarsara$); this will be analyzed with more detail later. The effect of $\roprpr$ on the normalized average sample sizes $\E[\vasaf_i+\vasas_i] \roprpr$, $i=1,2$ is imperceptible (compare Figures~\ref{fig: Ei_RR_1_13_1d16_0p01_allwdh} and \ref{fig: Ei_RR_1_13_1d16_0p1_allwdh}). The parameter $\rapr$ has an impact on the average sample sizes, but not on their ratio, which remains close to $\tarsara$ (compare Figures~\ref{fig: Ei_RR_1_11_1_0p01_allwdh} and \ref{fig: Ei_RR_1_11_16_0p01_allwdh}, or \ref{fig: Ei_RR_1_31_16_0p01_allwdh} and \ref{fig: Ei_RR_1_31_1d16_0p01_allwdh}).
If both $\rapr$ and $\tarsara$ are replaced by their reciprocal values, $\E[\vasaf_1+\vasas_1]$ and $\E[\vasaf_2+\vasas_2]$ are simply swapped (see Figures~\ref{fig: Ei_RR_1_31_16_0p01_allwdh} and \ref{fig: Ei_RR_1_13_1d16_0p01_allwdh}). This is in agreement with \eqref{eq: E vasa 1 norm res RR LRR} and \eqref{eq: E vasa 2 norm res RR LRR}. Lastly, for $\tarvar$ small the average sample sizes are approximately inversely proportional to this parameter. Again, this can be observed in \eqref{eq: E vasa 1 norm res RR LRR} and \eqref{eq: E vasa 2 norm res RR LRR}, where for $\tarvar$ small the term $1/\tarvar$ dominates the other summands in the first factor.

The theoretical curves in Figure~\ref{fig: Ei_RR_1_allwdh} have small \emph{jump discontinuities} (see for example Figure~\ref{fig: Ei_RR_1_31_16_0p01_allwdh} near $\tarvar=0.07$), caused by the discrete character of $\suf_1$ and $\suf_2$. The jumps occur when the result of \eqref{eq: cho suf} changes by $1$. This effect is also present in the simulation results (bold dots), although less evident. Apart from this, for fixed $\suf_1$ the simulated results exhibit \emph{steep changes} in certain locations, produced by the rounding applied to $\sus_1$ and $\sus_2$ (see for example the rightmost region for $\E[\vasaf_2+\vasas_2]$ in Figure~\ref{fig: Ei_RR_1_31_16_0p01_allwdh} with $\suf_1=3$, which in that region corresponds to the bold dots). These are only observable in the simulation results, because the theoretical curves do not explicitly model the rounding of $\sus_1$ and $\sus_2$; and they are not discontinuities, but short sections where the slope is large in absolute value, as will be discussed later.

\begin{figure}%
\begin{subfigure}{\tamdosfig}%
\includegraphics[width=\textwidth]{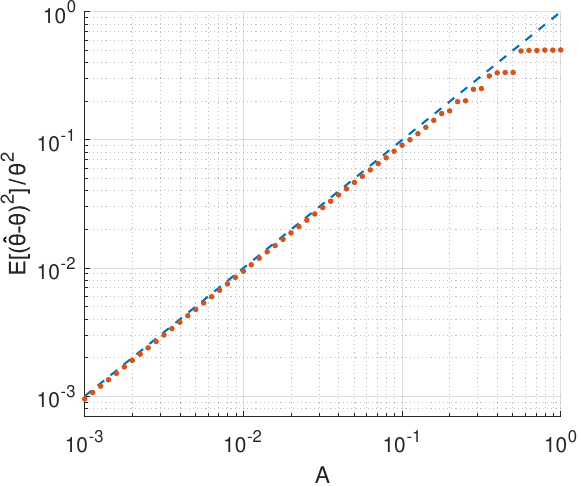}%
\captionsetup{justification=centering}
\caption{$\tarsara=1$, $\rapr=16$, $\roprpr=0.1$\\ ($p_1 = 0.4$, $p_2 = 0.025$)}%
\label{fig: MSE_RR_1_11_16_0p1_cho}%
\end{subfigure}%
\hfill
\begin{subfigure}{\tamdosfig}%
\includegraphics[width=\textwidth]{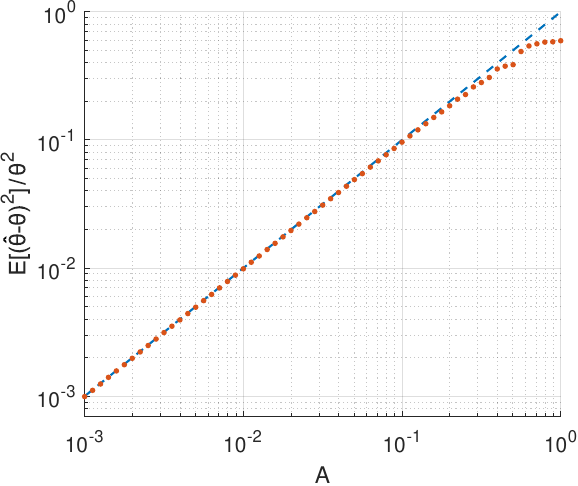}%
\captionsetup{justification=centering}%
\caption{$\tarsara=3$, $\rapr=1/16$, $\roprpr=0.01$\\ ($p_1 = 0.0025$, $p_2 = 0.04$)}%
\label{fig: MSE_RR_1_31_1d16_0p01_cho}%
\end{subfigure}%
\\[\splegbef]%
\centering%
\begin{small}%
Dashed line: target. Dots: simulation.
\end{small}%
\\[\splegaft]%
\caption{Relative MSE for RR with element sampling}
\label{fig: MSE_RR_1_cho}%
\end{figure}

Figure~\ref{fig: MSE_RR_1_cho} compares the target $\tarvar$ with the relative MSE, $\E[(\vahRR-\RR)^2]/\RR^2$, obtained from simulations
in two specific cases, for $\suf_1$ given by \eqref{eq: cho suf}.
As seen, the relative MSE is always less than $\tarvar$, in accordance with \eqref{eq: cond tarvar RR}. The difference between simulation and target is considerable when the latter is large. Again, this is a consequence of the limitation of $\suf_1$ in \eqref{eq: cho suf} to values not smaller than $3$. In addition, the difference increases slightly with $\roprpr$
(compare Figures~\ref{fig: MSE_RR_1_11_16_0p1_cho} and \ref{fig: MSE_RR_1_31_1d16_0p01_cho}). This is, at least in part, explained by the fact that the uniform bound \eqref{eq: Var varhRR cond} becomes less tight as $p_1$ or $p_2$ approach $1$, as can be seen by comparing it with \eqref{eq: Var varhRR cond no unif}.

The relative MSE, like the average sample sizes, has steep changes near certain values of $\tarvar$, caused by rounding $\sus_1$ and $\sus_2$. The effect is noticeable in the rightmost region of Figure~\ref{fig: MSE_RR_1_cho}, and is explained as follows. For large $\tarvar$, at least one of the two parameters $\sus_1$ and $\sus_2$ is small, according to \eqref{eq: ferr} and \eqref{eq: ferr tarvar}, so as to make the relative MSE similar to (but smaller than) the target. Consequently that parameter, $\sus_j$, has a very narrow distribution before rounding, which implies that rounding it almost always produces the same integer value. This will typically be the next greater integer, because for small $\sus_j$ rounding causes a large variation in the error function, and thus rounding down is not likely to satisfy \eqref{eq: ferr leq tarvar}. In these conditions, if $\tarvar$ is increased by a small amount, the change in the distribution of $\sus_j$ before rounding, even if correspondingly small, can be sufficient to cause the rounded value to decrease by $1$, producing a significant change in $\E[(\vahRR-\RR)^2]/\RR^2$, as well as in $\E[\vasaf_j+\vasas_j]$.

As a specific example, for $\tarsara=1$, $\rapr=16$ (Figure~\ref{fig: MSE_RR_1_11_16_0p1_cho}), consider $\tarvar = 0.562$ (leftmost dot in the highest ``plateau'' of simulated values), for which \eqref{eq: cho suf} gives $\suf_1=3$. The simulation yields an average value of $1.856$ for $\sus_2$ before rounding, and a standard deviation of $0.082$. The value after rounding is $2$ with probability $0.962$ (and greater with probability $0.038$). The parameter $\sus_1$ before rounding takes larger values, with average $109.2$ and standard deviation $76.6$. Reducing $\tarvar$ to $0.501$ (next simulated value towards the left), which still corresponds to $\suf_1=3$, the average of $\sus_2$ before rounding becomes $2.087$, and the standard deviation is $0.094$. After rounding, $\sus_2$ now equals $3$ with probability $0.999$ (and is greater with probability $0.001$). Again, $\sus_1$ before rounding takes larger values, with mean and standard deviation $112.5$ and $77.9$ respectively. The change of the rounded value of $\sus_2$, from being $2$ to being $3$ with probabilities near $1$, causes the rightmost vertical gap in Figure~\ref{fig: MSE_RR_1_11_16_0p1_cho}, from a relative MSE approximately equal to $0.50$ down to $0.34$. In contrast, for small $\tarvar$ both $\sus_1$ and $\sus_2$ are large and the effect of rounding is less marked, because, on one hand, the distribution of each parameter is wider, with many possible rounded values; and, on the other hand, rounding a large value only causes a small variation in the relative MSE.

The same behavior can be seen in the average sample sizes. Thus, in Figure~\ref{fig: Ei_RR_1_11_16_0p01_allwdh}, which also corresponds to $\tarsara=1$, $\rapr=16$, the simulation results for $\E[\vasaf_2+\vasas_2]$ with $\suf_1$ as in \eqref{eq: cho suf} (bold dots) show a vertical gap at the same horizontal position discussed in the preceding paragraph for the relative MSE. The effect is not discernible in the graph of $\E[\vasaf_1+\vasas_1]$ because $\sus_1$ is large.

\begin{figure}%
\begin{subfigure}{\tamdosfig}%
\includegraphics[width=\textwidth]{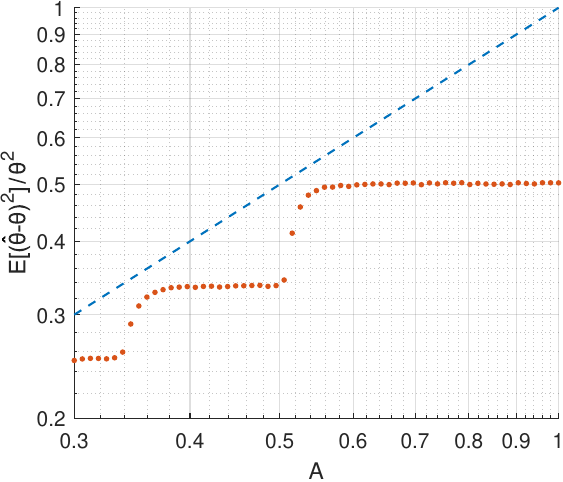}%
\captionsetup{justification=centering}%
\caption{Relative MSE}%
\label{fig: MSE_RR_1_11_16_0p1_si3}%
\end{subfigure}%
\hfill
\begin{subfigure}{\tamdosfig}%
\includegraphics[width=\textwidth]{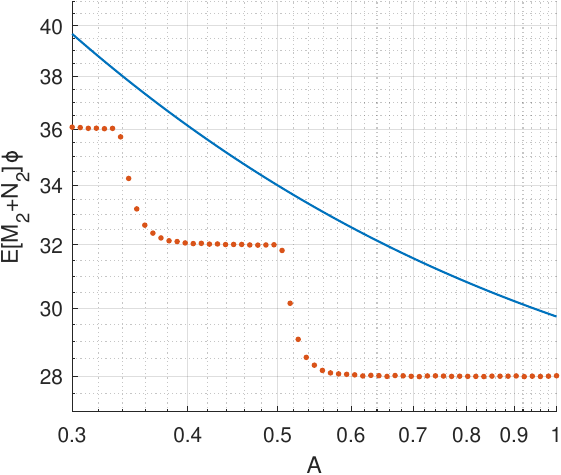}%
\captionsetup{justification=centering}%
\caption{$\E[\vasaf_2+\vasas_2] \roprpr$}%
\label{fig: E2_RR_1_11_16_0p1_si3}%
\end{subfigure}%
\\[\splegbef]%
\centering%
\begin{small}%
Dashed line: target. Solid line: approximation. Dots: simulation.
\end{small}%
\\[\splegaft]%
\caption{Detail of relative MSE and normalized average sample size for RR with element sampling, for large $\tarvar$; $\tarsara=1$, $\rapr=16$, $\roprpr=0.1$ ($p_1=0.4$, $p_2=0.025$)}
\label{fig: RR_1_si}%
\end{figure}

It follows from the above that the effect of rounding $\sus_1$ and $\sus_2$ is not a discontinuity, but rather a short section with large slope in $\E[(\vahRR-\RR)^2]/\RR^2$, or in $\E[\vasaf_i+\vasas_i]$, as a function of $\tarvar$. Indeed, since \eqref{eq: nsus 1 bis} and \eqref{eq: nsus 2 bis} are continuous functions of $\tarvar$, the distributions of $\sus_1$ and $\sus_2$ also vary continuously with $\tarvar$, and so do the relative MSE and average sample sizes. This can be seen in Figure~\ref{fig: MSE_RR_1_11_16_0p1_si3}, which is a detailed view of the rightmost part of Figure~\ref{fig: MSE_RR_1_11_16_0p1_cho} with smaller spacing along the horizontal axis, and in Figure~\ref{fig: E2_RR_1_11_16_0p1_si3}, which shows the corresponding values of $\E[\vasaf_2+\vasas_2]$.

\begin{figure}%
\begin{subfigure}{\tamdosfig}%
\includegraphics[width=\textwidth]{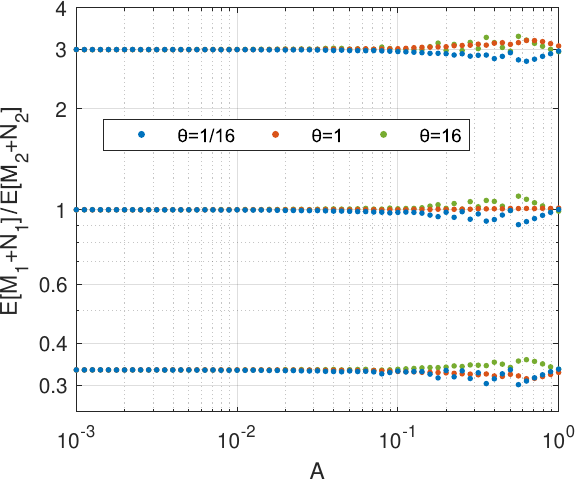}%
\caption{$\roprpr=0.01$}%
\label{fig: rss_RR_1_0p01}%
\end{subfigure}%
\hfill
\begin{subfigure}{\tamdosfig}%
\includegraphics[width=\textwidth]{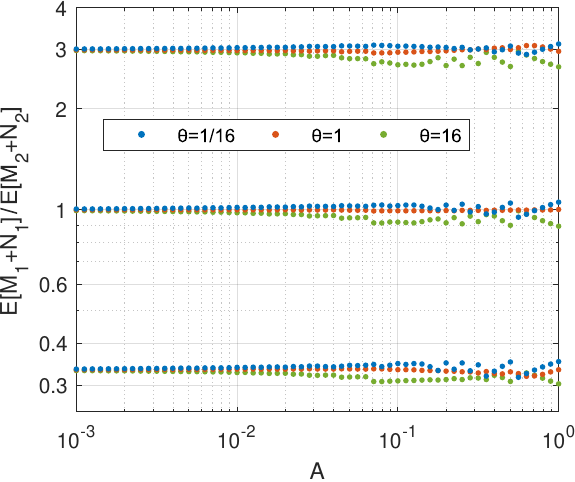}%
\caption{$\roprpr=0.1$}%
\label{fig: rss_RR_1_0p1}%
\end{subfigure}%
\\[\splegbef]%
\centering
\begin{small}%
$\tarsara=1/3$, $1$, $3$ (bottom to top). $\rapr=1/16$, $1$, $16$ (see legend).
\end{small}%
\\[\splegaft]%
\caption{Ratio of average sample sizes for RR with element sampling}
\label{fig: rss_RR_1}%
\end{figure}

The ratio $\E[\vasaf_1+\vasas_1] / \E[\vasaf_2+\vasas_2]$ obtained from simulation is represented in Figure~\ref{fig: rss_RR_1}. As can be seen, it is in general close to $\tarsara$, and very close for small or moderate values of $\tarvar$. The relatively larger deviations in the rightmost region of the graphs can again be attributed to the fact that for large $\tarvar$, since $\suf_1$ is limited from below by $3$ in \eqref{eq: cho suf}, the value of $\suf_1$ does not result in $\fcurv(\tarvar, \suf_1, \susrou)$ close to $0$. In any case, the deviation between the actual ratio and $\tarsara$ is less than $11\%$ for all combinations of parameters shown in the figure, and much lower for $\tarvar$ small.

\subsubsection{Estimation efficiency}
\label{part: RR sep effic}

The efficiency of an unbiased estimator can be defined by comparing its variance against the minimum variance that can be achieved by any unbiased estimator with the same sample size, or with the same average sample size. For a fixed-size estimator based on independent observations from a single population, the minimum possible variance is given by the \CR{} bound \citep{Ghosh97, Kay93}. This bound applies under certain regularity conditions on the distribution of the samples, which are satisfied when the observations are binary. For sequential estimators, there exists a generalization of the \CR{} bound, obtained by \citet{Wolfowitz47} \citetext{see also \citealp[section~4.3]{Ghosh97}}. The actual bound is the same, except that the \emph{average} sample size is used instead of a fixed sample size, and the regularity conditions are different.

For estimators based on independent observations from \emph{two} populations, as considered in this paper, a variation of the \CR{} bound can be applied in the fixed-size case \citep[chapter~3]{Kay93}. However, to the author's knowledge, there is no analogue of Wolfowitz's result for sequential estimators when the observations are obtained from more than one population. Therefore, the efficiency of the RR estimator (and of the estimators to be presented in subsequent sections) will be defined by comparing its variance with the lowest variance that can be attained by any \emph{fixed-size} estimator with the same average size for each population.

Consider fixed numbers $\fisa_1$ and $\fisa_2$ of independent binary samples taken from two populations with parameters $p_1$ and $p_2$. The number of successes observed from population $i=1,2$ follows a binomial distribution with parameters $\fisa_i$ and $p_i$. Thus
\begin{equation}
\label{eq: liho}
\liho(\vasu_1, \vasu_2; p_1, p_2) = \binom{\fisa_1}{\vasu_1} \binom{\fisa_2}{\vasu_2} p_1^{\vasu_1} (1-p_1)^{\fisa_1-\vasu_1} p_2^{\vasu_2} (1-p_2)^{\fisa_2-\vasu_2}
\end{equation}
is the probability of observing $\vasu_1$ and $\vasu_2$ successes from the two populations respectively. For a generic parameter $\parg$ that is a function of $p_1, p_2$, and an unbiased estimator $\vahparg$ of $\parg$, the \CR{} bound is \citep[section~3.8]{Kay93}
\begin{equation}
\label{eq: CR gen}
\Var[\vahparg] \geq \matJ \: \matI^{-1}\, \transp\matJ
\end{equation}
where $\matJ = \left[ \partial \parg/ \partial p_1 \;\; \partial \parg/ \partial p_2\right]$ is the $1 \times 2$ Jacobian vector and $\matI$ is the $2 \times 2$ Fisher information matrix, defined as
\begin{equation}
\label{eq: matI}
\matIentry_{i,j} = -\E\left[\frac{\partial^2 \log \liho(\vasu_1, \vasu_2; p_1, p_2)}{\partial p_i \partial p_j}\right], \quad i,j = 1,2. 
\end{equation}
This matrix is readily computed from \eqref{eq: liho} as $\matIentry_{i,i} = \fisa_i/(p_i(1-p_i))$, $\matIentry_{i,j} = 0$ for $i \neq j$ (which reflects the fact that observations from one population give no information about the other), and therefore \eqref{eq: CR gen} becomes 
\begin{equation}
\label{eq: CR gen bis}
\Var[\vahparg] \geq \left(\frac{\partial \parg}{\partial p_1}\right)^2 \frac{p_1(1-p_1)}{\fisa_1} + \left(\frac{\partial \parg}{\partial p_2}\right)^2 \frac{p_2(1-p_2)}{\fisa_2}.
\end{equation}

Equating $\fisa_i$ to the average number of observations of population $i$ used by the considered estimator, $\E[\vasaf_i+\vasas_i]$, the efficiency with element sampling $\efficsep$ is obtained as
\begin{equation}
\label{eq: effic sep}
\efficsep = \frac {\displaystyle \left(\frac{\partial \parg}{\partial p_1}\right)^2 \frac{p_1(1-p_1)}{\E[\vasaf_1+\vasas_1]} + \left(\frac{\partial \parg}{\partial p_2}\right)^2 \frac{p_2(1-p_2)}{\E[\vasaf_2+\vasas_2]} } {\Var[\vahparg]}.
\end{equation}
Particularizing \eqref{eq: effic sep} for RR, that is $\parg = \RR$, with $\partial\RR/\partial p_1 = \RR/p_1$, $\partial\RR/\partial p_2 = -\RR/p_2$,
\begin{equation}
\label{eq: effic RR sep}
\efficsep = \frac {\displaystyle \frac{1-p_1}{\E[\vasaf_1+\vasas_1] p_1} + \frac{1-p_2}{\E[\vasaf_2+\vasas_2] p_2} } {\Var[\vahRR]/\RR^2}.
\end{equation}
For $\susrou=1$ and $\suf_1$ as in \eqref{eq: cho suf}, substituting \eqref{eq: cond tarvar RR}, \eqref{eq: E vasa 1 norm res RR LRR <} and \eqref{eq: E vasa 2 norm res RR LRR <} into \eqref{eq: effic RR sep} yields the following bound for the efficiency of the RR estimator with element sampling:
\begin{equation}
\label{eq: effic RR LRR sep bound}
\begin{split}
\efficsep > \frac 1 {1 + \tarvar(\suf_1 + \cerr_1 + 1)} \left( 1 - \roprpr \frac{ 1/\sqrt{\tarsara} + \sqrt{\tarsara} } { 1/\sqrt{\tarsara\rapr} + \sqrt{\tarsara\rapr} } \right).
\end{split}
\end{equation}

\begin{figure}%
\begin{subfigure}{\tamdosfig}%
\includegraphics[width=\textwidth]{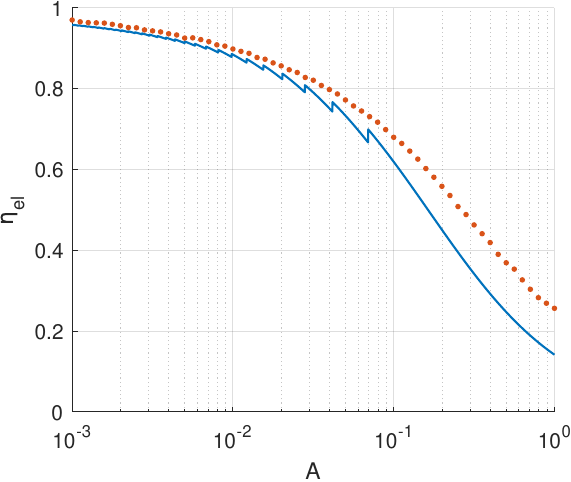}%
\captionsetup{justification=centering}%
\caption{$\tarsara=1$, $\rapr=1$, $\roprpr=0.01$\\ ($p_1 = p_2 = 0.01$)}%
\label{fig: efs_RR_1_11_1_0p01_cho}%
\end{subfigure}%
\hfill
\begin{subfigure}{\tamdosfig}%
\includegraphics[width=\textwidth]{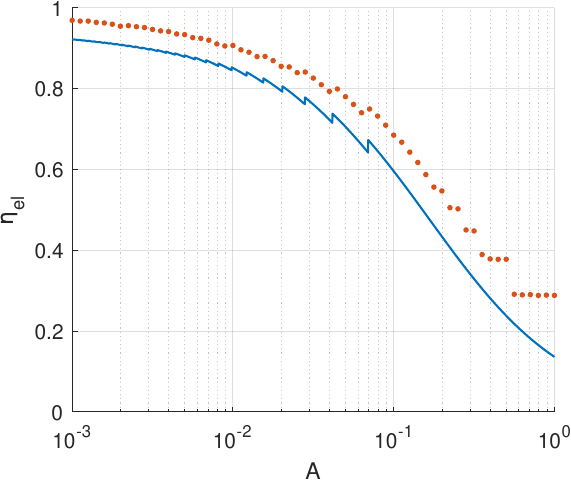}%
\captionsetup{justification=centering}%
\caption{$\tarsara=1$, $\rapr=16$, $\roprpr=0.1$\\ ($p_1 = 0.4$, $p_2 = 0.025$)}%
\label{fig: efs_RR_1_11_16_0p1_cho}%
\end{subfigure}%
\\[\splegaft]%
\begin{subfigure}{\tamdosfig}%
\includegraphics[width=\textwidth]{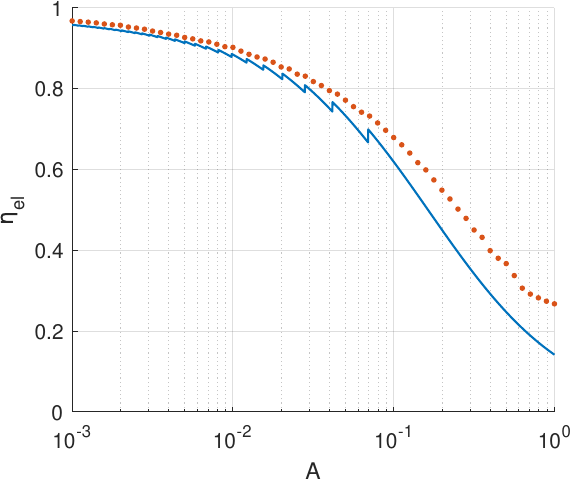}%
\captionsetup{justification=centering}%
\caption{$\tarsara=3$, $\rapr=1$, $\roprpr=0.01$\\ ($p_1 = p_2 = 0.01$)}%
\label{fig: efs_RR_1_31_1_0p01_cho}%
\end{subfigure}%
\hfill
\begin{subfigure}{\tamdosfig}%
\includegraphics[width=\textwidth]{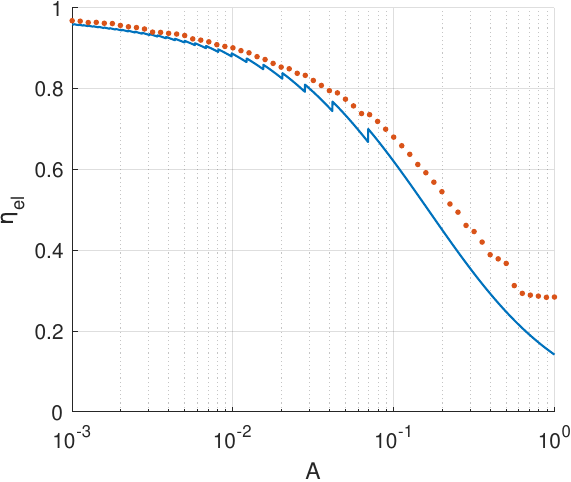}%
\captionsetup{justification=centering}%
\caption{$\tarsara=3$, $\rapr=1/16$, $\roprpr=0.01$\\ ($p_1 = 0.0025$, $p_2 = 0.04$)}%
\label{fig: efs_RR_1_31_1d16_0p01_cho}%
\end{subfigure}%
\\[\splegbef]%
\centering%
\begin{small}%
Line: bound. Dots: simulation.
\end{small}%
\\[\splegaft]%
\caption{Efficiency for RR with element sampling}
\label{fig: efs_RR_1_cho}%
\end{figure}

Figure~\ref{fig: efs_RR_1_cho} shows the efficiency $\efficsep$ obtained from Monte Carlo simulation. Its value is computed using \eqref{eq: effic RR sep} with $\E[\vasaf_i+\vasas_i]$, $i=1,2$ replaced by the corresponding sample means and $\Var[\vahRR]$ replaced by the sample MSE.
The bound \eqref{eq: effic RR LRR sep bound} is also plotted. As seen in the figure, the efficiency is high for the values of the target $\tarvar$ commonly used in practice, and in particular it is close to $1$ for small $\tarvar$. For example, $\tarvar = 0.04$, corresponding to a relative RMSE of $20\%$, gives values of $\efficsep$ around $80\%$. The efficiency as a function of $\tarvar$ exhibits the same discontinuities and steep changes that have been identified for the relative MSE and for $\E[\vasaf_i+\vasas_i]$. The simulation results deviate more from the theoretical bound for large $\tarvar$ (rightmost part of the graphs) or for large $\roprpr$ (Figure~\ref{fig: efs_RR_1_11_16_0p1_cho}). This is explained by the fact that in these conditions the relative MSE is considerably smaller than the target, and the average sample sizes are also considerably smaller than their bounds, as discussed previously. On the other hand, the bound is quite tight for small or moderate values of $\tarvar$ and $\roprpr$, which is precisely when it is most important to characterize $\efficsep$ accurately, as sample sizes are large in that case. 

It is easily seen that $\lim_{\tarvar \rightarrow 0} \tarvar \suf_1^2 = 1$ for $\tarvar$ and $\suf_1$ related by \eqref{eq: curv zero bis RR LRR}, and thus also for $\suf_1$ obtained from \eqref{eq: cho suf}. Combining this result with \eqref{eq: effic RR LRR sep bound}, it follows that $\efficsep$ tends to $1$ when $\tarvar$ and $\roprpr$ tend to $0$. This is in consonance with the values shown in Figure~\ref{fig: efs_RR_1_cho}.

\subsection{Group sampling}
\label{part: RR gr}

\subsubsection{Estimation procedure}
\label{part: RR gr proc}

In group sampling, samples are taken in groups or batches, each of which contains $\tarsagr_1$ and $\tarsagr_2$ samples from the two populations respectively. In consequence, the numbers of samples from both populations must be the same integer multiple of $\tarsagr_1$ and $\tarsagr_2$, i.e.~$\vagr \tarsagr_1$ and $\vagr \tarsagr_2$, where $\vagr$ is the number of groups.

Let $\tarsara$ be defined for group sampling as
\begin{equation}
\label{eq: tarsara tarsagr}
\tarsara = \frac{\tarsagr_1}{\tarsagr_2}.
\end{equation}
This definition is consistent with that used for element sampling, because $\tarsara$ still represents the ratio of average sample sizes. In addition, for group sampling $\vasaf_i$, $\vasas_i$, $i=1,2$ will continue to refer to the numbers of samples that \emph{would} result if the estimation procedure discussed thus far, using element sampling, were applied with $\tarsara$ given by \eqref{eq: tarsara tarsagr} (and therefore these variables do not correspond to the numbers of samples actually required with group sampling).

The estimation process with group sampling is as follows. Samples are \emph{used} individually, following the same procedure as with element sampling; but are in fact \emph{taken} in groups. As individual samples from either population become necessary, groups of samples are taken, each group providing $\tarsagr_1$ and $\tarsagr_2$ samples of the two populations. During this process, a group may provide more samples than needed for one or the two populations. In that case the ``surplus'' samples are stored for later use. When a sample from a given population is required, a new group is taken only if the surplus samples from that population have been exhausted. Once the procedure has finished, there may remain some stored samples, which will be discarded (and at least for one of the two populations, with index $j$, the number of discarded samples will be less than $\tarsagr_j$).

Estimation with group sampling thus proceeds as in Section~\ref{part: RR sep}, but with an added ``outer layer'' that translates group sampling into element sampling as described. With the above definition of $\vasaf_i$ and $\vasas_i$, it follows that the number of required groups is
\begin{equation}
\label{eq: vagr}
\vagr = \max \left\{ \left\lceil \frac{\vasaf_1+\vasas_1}{\tarsagr_1} \right\rceil, \left\lceil \frac{\vasaf_2+\vasas_2}{\tarsagr_2} \right\rceil \right\}.
\end{equation}

\subsubsection{Average number of groups}
\label{part: RR gr ave}

The relevant measure of sample size with group sampling is the average number of groups, $\E[\vagr]$. To make its characterization more tractable, it is helpful to approximate \eqref{eq: vagr} by assuming $\roprpr$ small, which implies that $\E[\vasaf_i+\vasas_i]$, $i=1,2$ are correspondingly large. In addition to simplifying computations, this is the most important case in practice, as argued previously. For large $\vasaf_i$ and $\vasas_i$, the rounding operations in \eqref{eq: vagr} can be removed with negligible error. Thus, defining
\begin{equation}
\label{eq: diffgr}
\diffgr = \frac{\vasaf_1+\vasas_1}{\tarsagr_1} - \frac{\vasaf_2+\vasas_2}{\tarsagr_2},
\end{equation}
and noting that $\max\{\vgx,\vgy\} = (\vgx+\vgy)/2+|\vgx-\vgy|/$2, it is possible to express $\E[\vagr]$ as
\begin{equation}
\label{eq: gr approx RR bis}
\E[\vagr] \approx \E\left[\max\left\{\frac{\vasaf_1+\vasas_1}{\tarsagr_1},  \frac{\vasaf_2+\vasas_2}{\tarsagr_2}\right\}\right] = \frac{\E[\vasaf_1+\vasas_1]}{2\tarsagr_1} + \frac{\E[\vasaf_2+\vasas_2]}{2\tarsagr_2} + \frac {\E\left[|\diffgr|\right]} 2.
\end{equation}
The terms $\E[\vasaf_1+\vasas_1]$ and $\E[\vasaf_2+\vasas_2]$ in \eqref{eq: gr approx RR bis} are approximately given by \eqref{eq: E vasa 1 norm res RR LRR} and \eqref{eq: E vasa 2 norm res RR LRR}. Regarding $\E[|\diffgr|]$, it is shown in Appendix~\ref{part: app RR LRR gr var} that, for $\roprpr$ small, 
\begin{equation}
\label{eq: pre diffgr RR}
\E\left[| \diffgr |\right] \approx
\E\left[\left|\frac{\suf_1+\sus_1}{\tarsagr_1 p_1} - \frac{\suf_2+\sus_2}{\tarsagr_2 p_2}\right|\right].
\end{equation}

Define
\begin{equation}
\label{eq: diffgra RR}
\diffgra = \frac{\suf_1+\sus_1}{\tarsagr_1 p_1} - \frac{\suf_2+\sus_2}{\tarsagr_2 p_2}.
\end{equation}
Using approximations \eqref{eq: nsus 1 approx} and \eqref{eq: nsus 2 approx} for $\sus_1, \sus_2$ and including the rounding term $\susrou$ as in Section~\ref{part: RR sep param ave} (this has very little effect on the approximations obtained here, but it is done for consistency), \eqref{eq: diffgra RR} becomes
\begin{equation}
\label{eq: diffgra RR approx}
\diffgra \roprpr \approx
\frac{\csusco_1 + \suf_1 + \susrou + \csusva_1 \varaf}{\tarsagr_1 \sqrt{\rapr}} - \frac{\csusco_2 + \suf_2 + \susrou + \csusva_2/\varaf}{\tarsagr_2} \sqrt{\rapr}.
\end{equation}
Let $\varafbor$ denote the value of $\varaf$ for which the right-hand side of \eqref{eq: diffgra RR approx} equals $0$, and let $\varafn = \varaf/\rapr$. Taking into account \eqref{eq: tarsara tarsagr}, $\varafbor$ is seen to be the only positive solution of
\begin{equation}
\label{eq: eq, varafbor}
\csusva_1 \varafbor^2 + (\csusco_1 + \suf_1 + \susrou- \tarsara\rapr(\csusco_2 + \suf_2 + \susrou)) \varafbor - \tarsara\rapr\csusva_2 = 0.
\end{equation}
Making use of \eqref{eq: csufco 1}--\eqref{eq: csufva 2} 
and \eqref{eq: suf 1 suf 2, cdeadd 1 cdeadd 2}--\eqref{eq: cdeadd RR LRR}, 
this solution is obtained as $\varafbor= \rapr \varafnbor$ with
\begin{equation}
\label{eq: varafnbor RR LRR bis}
\varafnbor
= \frac {\textstyle \suf_1+\cerr_1-\cerr_2} {\textstyle 2\tarsara\rapr(\suf_1-1)} \left( \tarsara\rapr-1 + \sqrt{(\tarsara\rapr-1)^2 + \frac{4\tarsara\rapr (\suf_1-1)(\suf_1+\cerr_1-\cerr_2-1)}{\suf_1(\suf_1+\cerr_1-\cerr_2)}} \right).
\rule[-1.6em]{0pt}{0pt}
\end{equation}
As established in Appendix~\ref{part: app RR LRR gr var}, for $\roprpr$ small the variable $\varafn$ approximately follows a beta prime distribution with parameters $\suf_2$, $\suf_1$. Thus, from \eqref{eq: pre diffgr RR}--\eqref{eq: diffgra RR approx}, 
\begin{multline}
\label{eq: abs diffgr RR}
\E[|\diffgr|] \roprpr \approx
\E[|\diffgra|] \roprpr = \\
\int_{\varafnbor}^\infty \left( \frac{\csusco_1 + \suf_1 + \susrou + \csusva_1 \rapr \rafn}{\tarsagr_1 \sqrt{\rapr}} - \frac{\csusco_2 + \suf_2 + \susrou + \csusva_2/(\rapr \rafn)}{\tarsagr_2} \sqrt{\rapr} \right) \betappdf(\rafn; \suf_2, \suf_1) \, \diff\rafn \\
- \int_0^{\varafnbor} \left( \frac{\csusco_1 + \suf_1 + \susrou + \csusva_1 \rapr \rafn}{\tarsagr_1 \sqrt{\rapr}} - \frac{\csusco_2 + \suf_2 + \susrou + \csusva_2/(\rapr \rafn)}{\tarsagr_2} \sqrt{\rapr} \right) \betappdf(\rafn; \suf_2, \suf_1) \, \diff\rafn,
\end{multline}
with $\betappdf(\rafn; \suf_2, \suf_1)$ defined by \eqref{eq: betappdf}. Applying \eqref{eq: Pr betap},
\begin{equation}
\label{eq: int 0}
\int_0^{\varafnbor} \betappdf(\rafn; \suf_2, \suf_1) \,\diff\rafn = \ribetaf\left(\frac{\varafnbor}{\varafnbor+1}; \suf_2, \suf_1 \right) = 1 - \int_{\varafnbor}^\infty \betappdf(\rafn; \suf_2, \suf_1) \,\diff\rafn.
\end{equation}
Similarly, combining \eqref{eq: betappdf} and \eqref{eq: Pr betap} with the identities \eqref{eq: betaf ident 1} and \eqref{eq: betaf ident 2} gives
\begin{align}
\label{eq: int 1}
\begin{multlined}
\int_0^{\varafnbor} \rafn \betappdf(\rafn; \suf_2, \suf_1) \,\diff\rafn
= \frac{\suf_2}{\suf_1-1} \int_0^{\varafnbor} \betappdf(\rafn; \suf_2+1, \suf_1-1) \,\diff\rafn \\
= \frac{\suf_2}{\suf_1-1} \ribetaf\left(\frac{\varafnbor}{\varafnbor+1}; \suf_2+1, \suf_1-1 \right)
= \frac{\suf_2}{\suf_1-1} - \int_{\varafnbor}^\infty \rafn \betappdf(\rafn; \suf_2, \suf_1) \,\diff\rafn,
\end{multlined} \\
\begin{multlined}
\label{eq: int -1}
\int_0^{\varafnbor} \frac 1 {\rafn} \betappdf(\rafn; \suf_2, \suf_1) \,\diff\rafn
= \frac{\suf_1}{\suf_2-1} \int_0^{\varafnbor} \betappdf(\rafn; \suf_2-1, \suf_1+1) \,\diff\rafn \\
= \frac{\suf_1}{\suf_2-1} \ribetaf\left(\frac{\varafnbor}{\varafnbor+1}; \suf_2-1, \suf_1+1 \right)
= \frac{\suf_1}{\suf_2-1} - \int_{\varafnbor}^\infty \frac 1 {\rafn} \betappdf(\rafn; \suf_2, \suf_1) \,\diff\rafn.
\end{multlined}
\end{align}
From \eqref{eq: abs diffgr RR}--\eqref{eq: int -1}, 
\begin{equation}
\label{eq: abs diffgr RR bis}
\begin{split}
\E[|\diffgr|] \roprpr &\approx \left(\frac{\csusco_1 + \suf_1 + \susrou}{\tarsagr_1\sqrt{\rapr}} - \frac{\csusco_2 + \suf_2 + \susrou}{\tarsagr_2}\sqrt{\rapr}\right) \left(1 - 2\ribetaf\left(\frac{\varafnbor}{\varafnbor+1}; \suf_2, \suf_1 \right)\right) \\
&\quad + \frac{\csusva_1 \suf_2 \sqrt{\rapr}}{(\suf_1-1)\tarsagr_1} \left(1 - 2\ribetaf\left(\frac{\varafnbor}{\varafnbor+1}; \suf_2+1, \suf_1-1 \right)\right) \\
&\quad - \frac{\csusva_2 \suf_1}{(\suf_2-1)\tarsagr_2 \sqrt{\rapr}} \left(1 - 2\ribetaf\left(\frac{\varafnbor}{\varafnbor+1}; \suf_2-1, \suf_1+1 \right)\right).
\end{split}
\end{equation}
Using \eqref{eq: csufco 1}, \eqref{eq: csufco 2}, \eqref{eq: igualdad cruzada 1 RR}, \eqref{eq: igualdad cruzada 2 RR}, \eqref{eq: suf 1 suf 2, cdeadd 1 cdeadd 2} and the identities \eqref{eq: ribetaf ident 1} and \eqref{eq: ribetaf ident 2} into \eqref{eq: abs diffgr RR bis} yields
\begin{equation}
\label{eq: abs diffgr RR ter}
\begin{split}
\E[|\diffgr|] \roprpr & \approx 2 \left( \frac 1 {\tarvar} + \suf_1 + \cerr_1 + \susrou \right)
\frac{\varafnbor^{\suf_1+\cerr_1-\cerr_2-1}}{(\varafnbor+1)^{2\suf_1+\cerr_1-\cerr_2-1} \betaf(\suf_1+\cerr_1-\cerr_2,\suf_1)} \\
& \quad \cdot \left( \frac {1}{\suf_1 \tarsagr_1 \sqrt{\rapr}} + \frac {\varafnbor \sqrt{\rapr}} {(\suf_1+\cerr_2-\cerr_2) \tarsagr_2}  \right).
\end{split}
\end{equation}
Substituting \eqref{eq: E vasa 1 norm res RR LRR}, \eqref{eq: E vasa 2 norm res RR LRR} and \eqref{eq: abs diffgr RR ter} into \eqref{eq: gr approx RR bis}, and using \eqref{eq: tarsara tarsagr}, the average number of required groups for the RR estimator is obtained as 
\begin{multline}
\label{eq: gr approx RR LRR ter}
\E[\vagr] \roprpr
\approx
\left( \frac 1 {\tarvar} + \suf_1 + \cerr_1 + \susrou \right) \left( \frac 1 {\tarsagr_1 \sqrt{\rapr}} + \frac{\sqrt{\rapr}}{\tarsagr_2} \right. \\
\left. + \frac{\varafnbor^{\suf_1+\cerr_1-\cerr_2-1}}
{(\varafnbor+1)^{2\suf_1+\cerr_1-\cerr_2-1} \betaf(\suf_1+\cerr_1-\cerr_2,\suf_1)} \left( \frac 1 {\suf_1 \tarsagr_1 \sqrt{\rapr}} + \frac {\varafnbor \sqrt{\rapr}} {(\suf_1+\cerr_1-\cerr_2) \tarsagr_2}  \right) \right),
\end{multline}
where $\varafnbor$ is given by \eqref{eq: varafnbor RR LRR bis}.

\begin{figure}%
\begin{subfigure}{\tamdosfig}%
\includegraphics[width=\textwidth]{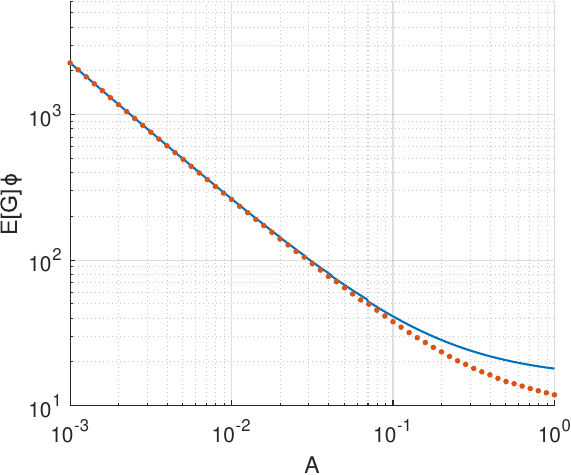}%
\captionsetup{justification=centering}%
\caption{$\tarsagr_1=1$, $\tarsagr_2=1$, $\rapr=1$, $\roprpr=0.01$\\ ($p_1=0.01$, $p_2=0.01$)}%
\label{fig: Eg_RR_1_11_1_0p01_cho}%
\end{subfigure}%
\hfill
\begin{subfigure}{\tamdosfig}%
\includegraphics[width=\textwidth]{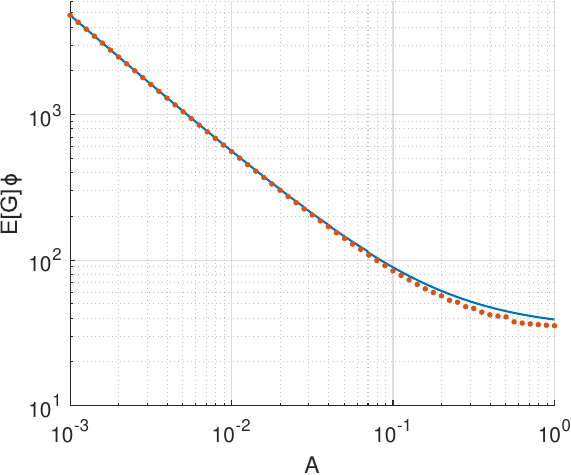}%
\captionsetup{justification=centering}%
\caption{$\tarsagr_1=1$, $\tarsagr_2=1$, $\rapr=16$, $\roprpr=0.01$\\ ($p_1 = 0.04$, $p_2 = 0.0025$)}%
\label{fig: Eg_RR_1_11_16_0p01_cho}%
\end{subfigure}%
\\[\splegaft]%
\begin{subfigure}{\tamdosfig}%
\includegraphics[width=\textwidth]{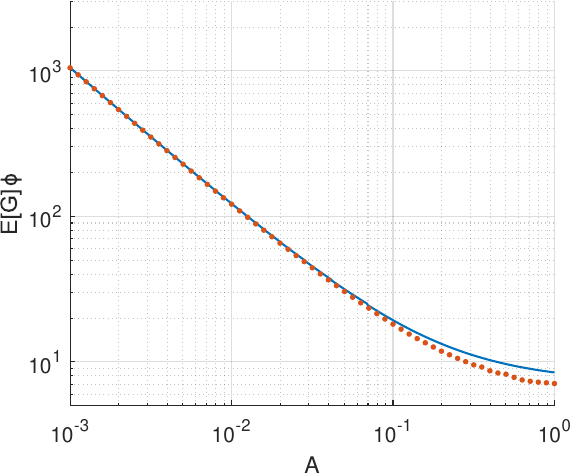}%
\captionsetup{justification=centering}%
\caption{$\tarsagr_1=2$, $\tarsagr_2=5$, $\rapr=16$, $\roprpr=0.01$\\ ($p_1 = 0.04$, $p_2 = 0.0025$)}%
\label{fig: Eg_RR_1_25_16_0p01_cho}%
\end{subfigure}%
\hfill
\begin{subfigure}{\tamdosfig}%
\includegraphics[width=\textwidth]{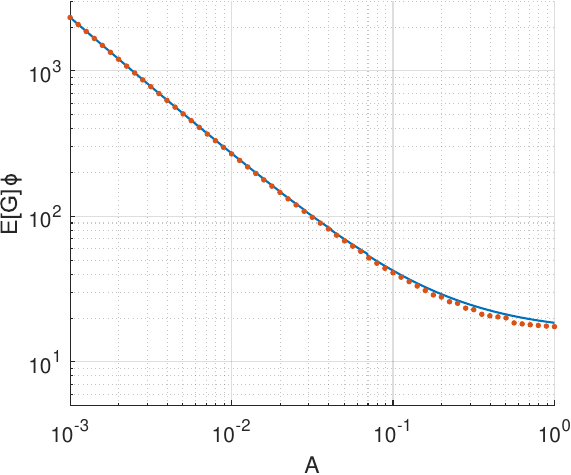}%
\captionsetup{justification=centering}%
\caption{$\tarsagr_1=2$, $\tarsagr_2=5$, $\rapr=1/16$, $\roprpr=0.01$\\ ($p_1 = 0.0025$, $p_2 = 0.04$)}%
\label{fig: Eg_RR_1_25_1d16_0p01_cho}%
\end{subfigure}%
\\[\splegbef]%
\centering%
\begin{small}%
Line: approximation. Dots: simulation.
\end{small}%
\\[\splegaft]%
\caption{Normalized average number of groups for RR with group sampling}
\label{fig: Eg_RR_1_cho}%
\end{figure}

Figure~\ref{fig: Eg_RR_1_cho} represents the normalized average number of groups obtained from simulation, as well as its theoretical approximation \eqref{eq: gr approx RR LRR ter}. Overall, the graphs follow similar patterns to those observed for $\E[\vasaf_i+\vasas_i] \roprpr$ with element sampling. The approximation is accurate for small or moderate $\tarvar$, and less so for large $\tarvar$. Discontinuities due to the choice of $\suf_1$ are minimally visible, and steep changes due to rounding $\sus_1$ and $\sus_2$ can be observed in the simulation results for large $\tarvar$. Comparing $\E[\vagr]$ for different values of $\rapr$ only makes sense when $\tarsagr_1$ and $\tarsagr_2$ are fixed; that is, between Figures~\ref{fig: Eg_RR_1_11_1_0p01_cho} and \ref{fig: Eg_RR_1_11_16_0p01_cho} or between Figures~\ref{fig: Eg_RR_1_25_16_0p01_cho} and \ref{fig: Eg_RR_1_25_1d16_0p01_cho}. For $\tarsagr_1$ and $\tarsagr_2$ given, it can be seen from \eqref{eq: gr approx RR LRR ter} that the minimum of $\E[\vagr]$ with respect to $\rapr$ does not necessarily occur when $\tarsara \rapr = 1$, as was the case for element sampling.

\subsubsection{Estimation efficiency}
\label{part: RR gr effic}

 The efficiency of the estimator is defined, as in Section~\ref{part: RR sep effic}, by comparing the estimation variance with the lowest variance that could be achieved by a fixed-size estimator with the same average numbers of samples, as given by the \CR{} bound. With group sampling, the average number of samples required from population $i$ is $\E[\vagr] \tarsagr_i$, $i=1,2$. Equating $\fisa_i$ to $\E[\vagr] \tarsagr_i$ in \eqref{eq: CR gen bis}, the efficiency $\efficgr$ for an unbiased estimator of a generic parameter $\parg$ with group sampling is obtained as
\begin{equation}
\label{eq: effic gr}
\efficgr = \frac {\displaystyle \left(\frac{\partial \parg}{\partial p_1}\right)^2 \frac{p_1(1-p_1)}{\tarsagr_1} + \left(\frac{\partial \parg}{\partial p_2}\right)^2 \frac{p_2(1-p_2)}{\tarsagr_2} } {\E[\vagr] \Var[\vahparg]}.
\end{equation}
Particularizing for RR, i.e.~for $\parg = \RR$,
\begin{equation}
\label{eq: effic RR LRR gr exact}
\efficgr = \frac { (1-p_1)/(\tarsagr_1 p_1) + (1-p_2)/(\tarsagr_2 p_2) } {\E[\vagr] \Var[\vahRR]/\RR^2}.
\end{equation}
Approximating $\Var[\RR]/\RR^2 \approx \tarvar$, the efficiency for RR with group sampling is expressed as
\begin{equation}
\label{eq: effic RR LRR gr}
\efficgr \approx
\frac{(1-p_1)/(\tarsagr_1 \sqrt{\rapr}) + (1-p_2)\sqrt{\rapr}/\tarsagr_2} { \tarvar \E[\vagr] \roprpr},
\end{equation}	
where $\E[\vagr] \roprpr$ is given by \eqref{eq: gr approx RR LRR ter}.

\begin{figure}%
\begin{subfigure}{\tamdosfig}%
\includegraphics[width=\textwidth]{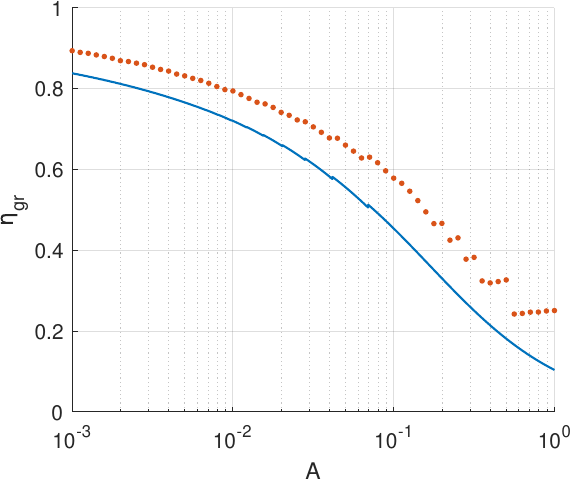}%
\captionsetup{justification=centering}%
\caption{$\tarsagr_1=1$, $\tarsagr_2=1$, $\rapr=16$, $\roprpr=0.1$\\ ($p_1 = 0.4$, $p_2 = 0.025$)}%
\label{fig: efg_RR_1_11_16_0p1_cho}%
\end{subfigure}%
\hfill
\begin{subfigure}{\tamdosfig}%
\includegraphics[width=\textwidth]{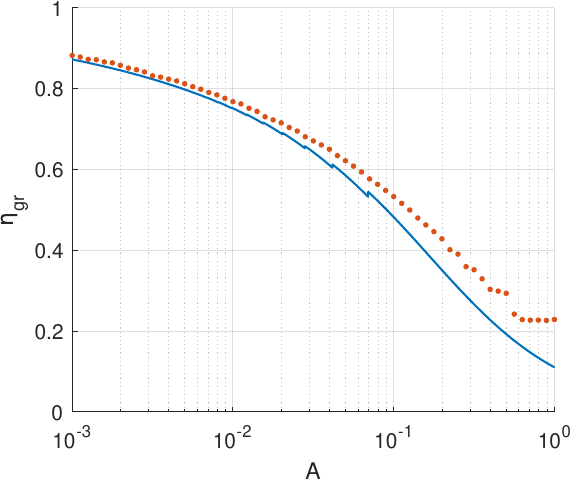}%
\captionsetup{justification=centering}%
\caption{$\tarsagr_1=3$, $\tarsagr_2=1$, $\rapr=1/16$, $\roprpr=0.01$\\ ($p_1 = 0.0025$, $p_2 = 0.04$)}%
\label{fig: efg_RR_1_31_1d16_0p01_cho}%
\end{subfigure}%
\\[\splegbef]%
\centering%
\begin{small}%
Line: approximation. Dots: simulation.
\end{small}%
\\[\splegaft]%
\caption{Efficiency for RR with group sampling}
\label{fig: efg_RR_1_cho}%
\end{figure}

Simulation results for the efficiency are shown in Figure~\ref{fig: efg_RR_1_cho}. The values are obtained using \eqref{eq: effic RR LRR gr exact} with $\E[\vagr]$ and $\Var[\vahRR]$ replaced by the sample mean and sample MSE. The approximation \eqref{eq: effic RR LRR gr} is also displayed. The figure contains only two specific cases for brevity. The results are in general little sensitive to $\tarsara$ and $\rapr$; but, similarly to what was observed with element sampling, the theoretical approximation becomes more conservative for large $\tarvar$ or $\roprpr$. Comparing with Figure~\ref{fig: efs_RR_1_cho}, group sampling is seen to be less efficient than element sampling, in accordance with the number of required groups being the maximum over the two populations, as given by \eqref{eq: vagr}. This causes an efficiency loss of approximately $0.15$ for values of $\tarvar$ in the range $0.01$--$0.1$, and a less substantial loss for $\tarvar$ small.

\section{Estimation of log relative risk}
\label{part: LRR}

This section describes the LRR estimator and analyzes its properties. The presentation, as will also be the case for subsequent estimators, follows the same logical course as in Section~\ref{part: RR}, but can be shorter, thanks to the similarities with RR. Section~\ref{part: LRR sep} addresses the estimation procedure with element sampling, and Section~\ref{part: LRR gr} considers group sampling.

\subsection{Element sampling}
\label{part: LRR sep}

The process for estimating the LRR, $\LRR= \log(p_1/p_2)$, is analogous to that for RR: it consists of two stages, each of which applies IBS to each population. The first stage uses fixed IBS parameters, $\suf_1$, $\suf_2$, and the second uses IBS parameters $\sus_1$, $\sus_2$ computed from the results $\vasaf_1$, $\vasaf_2$ of the first stage by means of the variable $\varaf$ defined in \eqref{eq: varaf RR LRR}, with $\cdesm_1 = \cdesm_2 = 1/2$. An unbiased estimation $\vahLRR$ is computed from the second-stage results $\vasas_1$, $\vasas_2$, according to the expression given next; and the error requirement is in this case defined in terms of MSE (rather than relative MSE), or equivalently variance:
\begin{equation}
\label{eq: cond tarvar LRR}
\Var[\vahLRR] \leq \tarvar.
\end{equation}

In the second stage, writing $\LRR = \log p_1 - \log p_2$, the estimator for the logarithm of a probability described in \citet[section~3]{Mendo25b} can be used for each of these two terms. Specifically, $-\harm{\vasas_i-1}+\harm{\sus_i-1}$, where $\harm{\sag}$ is the $\sag$-th harmonic number defined in \eqref{eq: harm}, is an unbiased estimator of $\log p_i$,
with variance less than $1/(\sus_i-1)$ for any $p_i \in (0,1)$
\citep{Mendo25b}.
Therefore,
\begin{equation}
\label{eq: vahLRR}
\vahLRR = -\harm{\vasas_1-1} + \harm{\vasas_2-1} + \harm{\sus_1-1} - \harm{\sus_2-1}
\end{equation}
is a conditionally unbiased estimator of $\LRR$ given $\sus_1, \sus_2$, which implies that it is also unconditionally unbiased; and, since the observations of the two populations are independent, 
\begin{equation}
\label{eq: cerr LRR pre}
\Var\left[\vahLRR \cond* \sus_1, \sus_2\right] < \frac 1 {\sus_1-1} + \frac 1 {\sus_2-1}.
\end{equation}
Defining the error function for LRR as the right-hand side of \eqref{eq: cerr LRR pre}, i.e.~as \eqref{eq: ferr} with 
\begin{equation}
\label{eq: cerr LRR}
\cerr_1 = 1, \qquad \cerr_2 = 1, \qquad \cerr_{12}= 0,
\end{equation}
and assuming \eqref{eq: ferr leq tarvar}, it follows that
\begin{equation}
\label{eq: Var varhLRR cond}
\Var\left[\vahLRR^2\right] = \E\left[\E\left[\vahLRR^2 \cond* \sus_1, \sus_2\right]\right] - \LRR^2 < \E\left[\ferr(\sus_1,\sus_2) + \LRR^2\right] - \LRR^2 \leq \tarvar.
\end{equation}
That is, condition \eqref{eq: ferr leq tarvar} ensures that \eqref{eq: cond tarvar LRR} holds for any $p_1, p_2 \in(0,1)$.

As in the RR case, $\sus_1$ and $\sus_2$ are obtained from the equation system formed by \eqref{eq: cond nsus ratio} and \eqref{eq: ferr tarvar}, except with $\cerr_1, \cerr_2, \cerr_{12}$ given by \eqref{eq: cerr LRR}. The solution is \eqref{eq: nsus 1 bis}--\eqref{eq: disc bis}. 
The values of $\sus_1$ and $\sus_2$ should then be rounded while satisfying \eqref{eq: ferr leq tarvar}. Using the first-order approximations \eqref{eq: nsus 1 approx}--\eqref{eq: csufva 2}, 
$\cdemul$ and $\cdeadd_1$ are obtained as in \eqref{eq: cdemul} and \eqref{eq: cdeadd RR LRR}; and \eqref{eq: cond cdeadd >} holds. The average sample sizes are approximately given by \eqref{eq: E vasa 1 norm res RR LRR} and \eqref{eq: E vasa 2 norm res RR LRR}.

The curvature function is defined in the same way as for RR, and the condition $\fcurv(\tarvar, \suf_1, \susrou) = 0$ is expressed by \eqref{eq: curv zero bis RR LRR}. The fact that $\cerr_1=\cerr_2$ and $\cerr_{12}=0$ for LRR implies that the positive solution of this equation has a simple expression,
\begin{equation}
\label{eq: cho suf LRR antes}
\tarvar= \frac 1 {(\suf_1+\cerr_1+\susrou)(\suf_1-1)}.
\end{equation}
In analogy with RR, the value $\susrou=1$ is used for the LRR estimator, with $\suf_1$ given by \eqref{eq: cho suf}, which in this case can be written explicitly as
\begin{equation}
\label{eq: cho suf LRR}
\suf_1 =
\max\left\{ 3, \left\lceil -1/2 + \sqrt{(3/2)^2 + 1/\tarvar} \right\rceil \right\}.
\end{equation}
The pairs $(\tarvar, \suf_1)$ determined by \eqref{eq: cho suf LRR antes} for $\susrou=1$ are shown in Figure~\ref{fig: curv_zero_1}.
With $\susrou=1$ and $\suf_1$ given by \eqref{eq: cho suf LRR}, the average sample sizes are bounded by \eqref{eq: E vasa 1 norm res RR LRR <} and \eqref{eq: E vasa 2 norm res RR LRR <}.

Algorithm~\ref{algo: RR LRR} (see Appendix~\ref{part: app algo}) describes the estimation procedure for LRR, as well as the properties of the estimator.

The simulation results for $\E[\vasaf_i+\vasas_i] \roprpr$, $i=1,2$ with $\susrou=1$ and varying $\suf_1$ are analogous to those shown for RR in Figure~\ref{fig: Ei_RR_1_allwdh} (light dots), and are omitted for brevity. As in that case, with $\suf_1$ chosen as in \eqref{eq: cho suf LRR} the bounds \eqref{eq: E vasa 1 norm res RR LRR <} and \eqref{eq: E vasa 2 norm res RR LRR <} are close to the actual average numbers of samples, and these approximately take their minimum values with respect to $\suf_1$. The simulated MSE, also omitted, has a similar behavior to that in Figure~\ref{fig: MSE_RR_1_cho}, with values always smaller than the target $\tarvar$, and very close to it unless $\tarvar$ is large.

\begin{figure}%
\begin{subfigure}{\textwidth}%
\includegraphics[width=\tamdosfig]{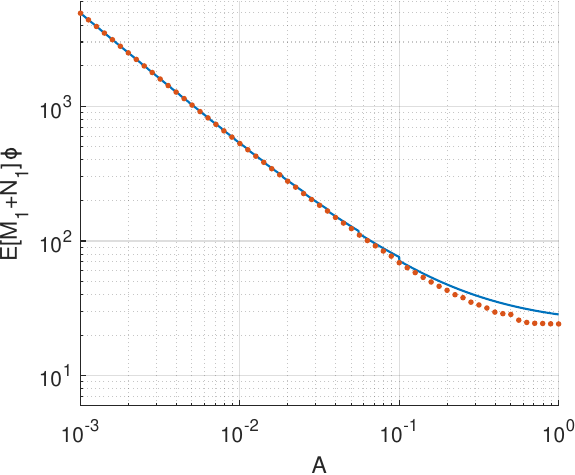}%
\hfill%
\includegraphics[width=\tamdosfig]{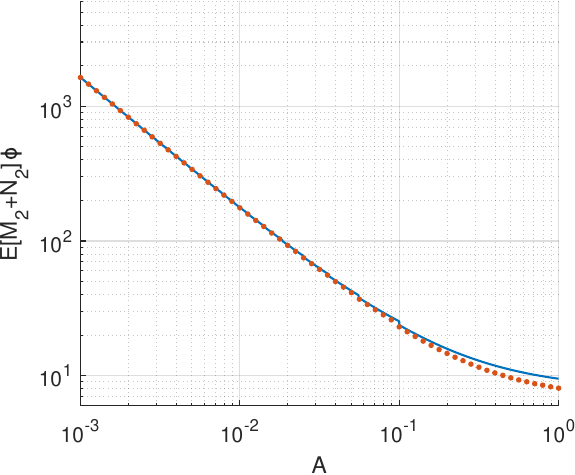}%
\caption{$\tarsara=3$, $\rapr=1/16$, $\roprpr=0.01$ ($p_1 = 0.0025$, $p_2 = 0.04$)}%
\label{fig: Ei_LRR_1_31_1d16_0p01_cho}%
\end{subfigure}%
\\[\splegaft]%
\begin{subfigure}{\textwidth}%
\includegraphics[width=\tamdosfig]{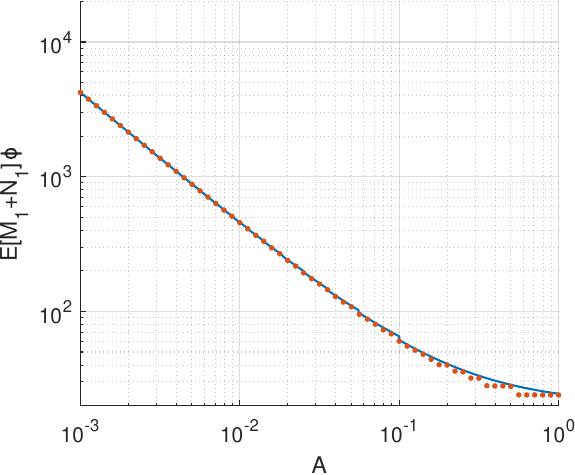}%
\hfill%
\includegraphics[width=\tamdosfig]{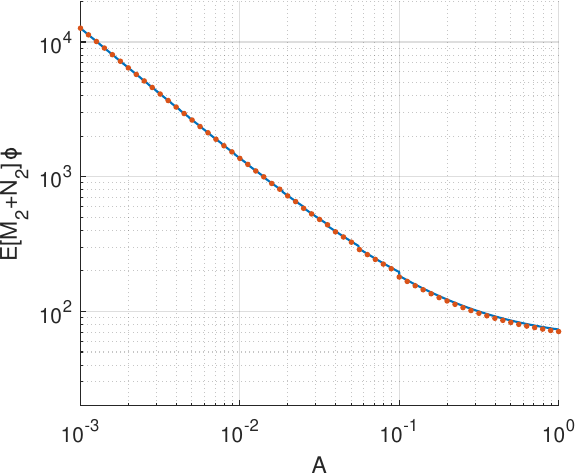}%
\caption{$\tarsara=1/3$, $\rapr=1/16$, $\roprpr=0.01$ ($p_1 = 0.0025$, $p_2 = 0.04$)}%
\label{fig: Ei_LRR_1_13_1d16_0p01_cho}%
\end{subfigure}%
\\[\splegbef]%
\centering%
\begin{small}%
Line: bound. Dots: simulation.
\end{small}%
\\[\splegaft]%
\caption{Normalized average sample sizes for LRR with element sampling}
\label{fig: Ei_LRR_1_cho}%
\end{figure}

Figure~\ref{fig: Ei_LRR_1_cho} shows, in two specific cases, the theoretical bounds \eqref{eq: E vasa 1 norm res RR LRR <} and \eqref{eq: E vasa 2 norm res RR LRR <} and simulated $\E[\vasaf_i+\vasas_i] \roprpr$ for $\susrou = 1$ and with  $\suf_1$ given by \eqref{eq: cho suf LRR}. The plotted values have a variation pattern similar to that observed for RR (Section~\ref{part: RR sep bounds}), with jump discontinuities and steep changes; again, the latter are only visible in the simulation results, and most apparent for large $\tarvar$. The values are slightly lower than those for RR in the rightmost region of the graphs (compare Figures~\ref{fig: Ei_LRR_1_31_1d16_0p01_cho} and \ref{fig: Ei_LRR_1_13_1d16_0p01_cho} with \ref{fig: Ei_RR_1_31_1d16_0p01_allwdh} and \ref{fig: Ei_RR_1_13_1d16_0p01_allwdh} respectively), due to the fact that $\cerr_1$ is $1$ for LRR and $2$ for RR. The effect of $\cerr_1$ is only appreciable for large $\tarvar$, as it stems from \eqref{eq: E vasa 1 norm res RR LRR <} and \eqref{eq: E vasa 2 norm res RR LRR <}. The ratio between average sample sizes obtained from simulation is close to $\tarsara$, as can be seen in Figure~\ref{fig: Ei_LRR_1_cho} (note that the ratio between the bounds is exactly $\tarsara$). More specifically, the achieved ratio deviates from $\tarsara$ by small or very small percentages, very similar to those in RR (results not shown).

\begin{figure}%
\begin{subfigure}{\tamdosfig}%
\includegraphics[width=\textwidth]{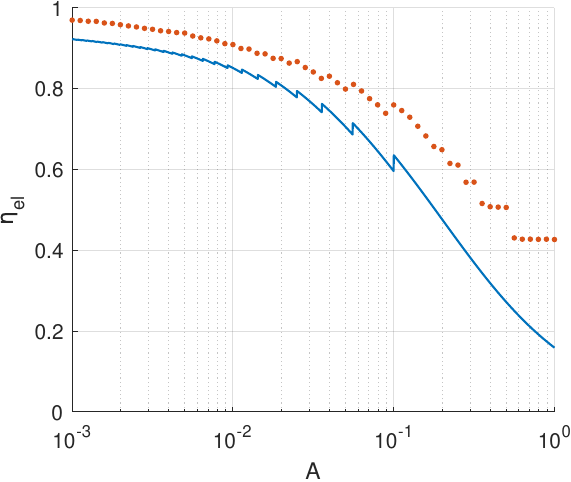}%
\captionsetup{justification=centering}%
\caption{$\tarsara=1$, $\rapr=16$, $\roprpr=0.1$\\ ($p_1 = 0.4$, $p_2 = 0.025$)}%
\label{fig: efs_LRR_1_11_16_0p1_cho}%
\end{subfigure}%
\hfill
\begin{subfigure}{\tamdosfig}%
\includegraphics[width=\textwidth]{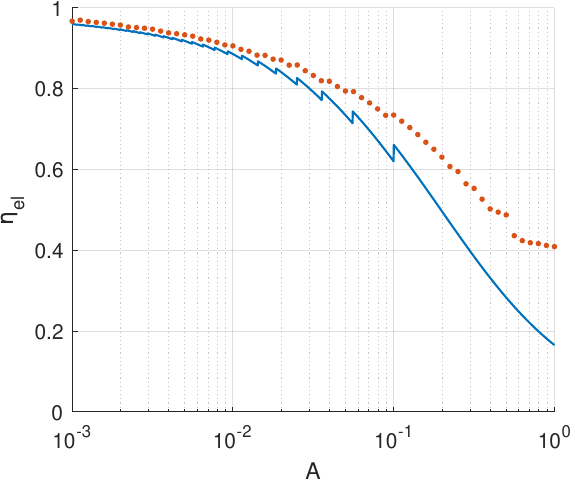}%
\captionsetup{justification=centering}%
\caption{$\tarsara=3$, $\rapr=1/16$, $\roprpr=0.01$\\ ($p_1 = 0.0025$, $p_2 = 0.04$)}%
\label{fig: efs_LRR_1_31_1d16_0p01_cho}%
\end{subfigure}%
\\[\splegbef]%
\centering%
\begin{small}%
Line: bound. Dots: simulation.
\end{small}%
\\[\splegaft]%
\caption{Efficiency for LRR with element sampling}
\label{fig: efs_LRR_1_cho}%
\end{figure}

As for the efficiency with element sampling, particularizing \eqref{eq: effic sep} for $\parg = \LRR$, with $\partial\LRR/\partial p_1 = 1/p_1$, $\partial\LRR/\partial p_2 = -1/p_2$ and substituting \eqref{eq: E vasa 1 norm res RR LRR <}, \eqref{eq: E vasa 2 norm res RR LRR <} and \eqref{eq: cond tarvar LRR} results in the same lower bound \eqref{eq: effic RR LRR sep bound} as for RR (although the obtained values will be slightly different because $\cerr_1$ and $\suf_1$ are), and the efficiency approaches $1$ when $\tarvar$ and $\roprpr$ tend to $0$. Figure~\ref{fig: efs_LRR_1_cho} compares this bound with simulation results, considering only two specific cases for brevity. The efficiency values from simulation are computed as explained in Section~\ref{part: RR sep effic}. Results are similar to those for RR (Figures~\ref{fig: efs_RR_1_11_16_0p1_cho} and \ref{fig: efs_RR_1_31_1d16_0p01_cho}), except that for large $\tarvar$ the efficiency is larger in LRR, as is the difference between simulation and bound.

\subsection{Group sampling}
\label{part: LRR gr}

The estimation of LRR with group sampling uses, as in the RR case, a number of groups $\vagr$ given by \eqref{eq: vagr} in order to provide the necessary amounts of individual samples of the two populations. The procedure is the same as in Section~\ref{part: RR gr proc}: a new group is taken whenever a sample of either population is required and no surplus samples of that population are available from previous groups. At the end of the process, any leftover samples are discarded.

The analysis in Appendix~\ref{part: app RR LRR gr var} shows that the approximation \eqref{eq: pre diffgr RR} is also valid for LRR; and $\E[\vagr]$ is then given by \eqref{eq: gr approx RR LRR ter}, with $\varafnbor$ as in \eqref{eq: varafnbor RR LRR bis} (and with the values of $\cerr_1$, $\cerr_2$ corresponding to LRR). Simulation results for the average number of groups, omitted for brevity, are very similar to those for RR (Figure~\ref{fig: Eg_RR_1_cho}), except that for large $\tarvar$ they are slightly lower than in RR, for the same reason as with element sampling.

\begin{figure}%
\begin{subfigure}{\tamdosfig}%
\includegraphics[width=\textwidth]{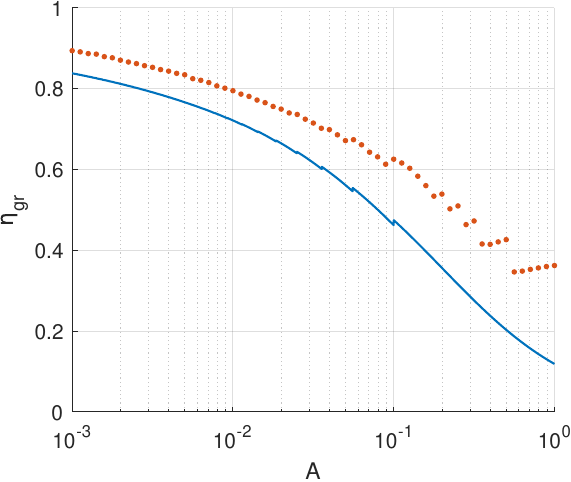}%
\captionsetup{justification=centering}%
\caption{$\tarsagr_1=1$, $\tarsagr_2=1$, $\rapr=16$, $\roprpr=0.1$\\ ($p_1 = 0.4$, $p_2 = 0.025$)}%
\label{fig: efg_LRR_1_11_16_0p1_cho}%
\end{subfigure}%
\hfill
\begin{subfigure}{\tamdosfig}%
\includegraphics[width=\textwidth]{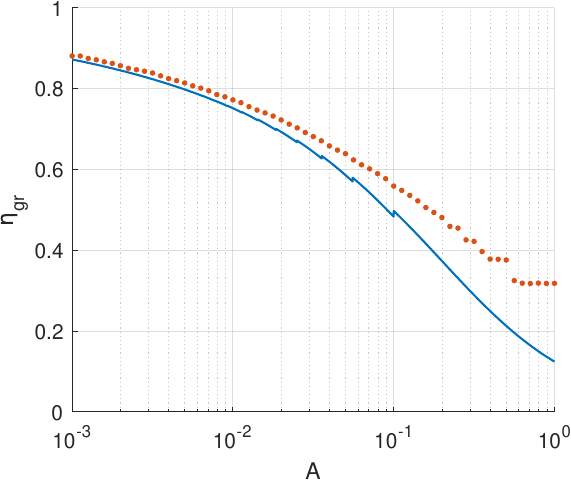}%
\captionsetup{justification=centering}%
\caption{$\tarsagr_1=3$, $\tarsagr_2=1$, $\rapr=1/16$, $\roprpr=0.01$\\ ($p_1 = 0.0025$, $p_2 = 0.04$)}%
\label{fig: efg_LRR_1_31_1d16_0p01_cho}%
\end{subfigure}%
\\[\splegbef]%
\centering%
\begin{small}%
Line: approximation. Dots: simulation.
\end{small}%
\\[\splegaft]%
\caption{Efficiency for LRR with group sampling}
\label{fig: efg_LRR_1_cho}%
\end{figure}

The efficiency with group sampling has the same approximate expression \eqref{eq: effic RR LRR gr} as for RR, with $\E[\vagr]$ computed as indicated above. The simulation results, shown in Figure~\ref{fig: efg_LRR_1_cho}, follow the same pattern observed for element sampling: the theoretical approximation is accurate when $\tarvar$ and $\roprpr$ are small, and conservative otherwise; and for large $\tarvar$ the efficiency deviates more from the theoretical curve than in RR (compare with Figure~\ref{fig: efg_RR_1_cho}).

\section{Estimation of odds ratio}
\label{part: OR}

\subsection{Element sampling}
\label{part: OR sep}

Several approaches are conceivable to estimate the OR $\OR$ defined in \eqref{eq: OR}. One method that could be employed is to estimate $p_1/p_2$ and $(1-p_2)/(1-p_1)$ separately, treating each as a RR and using the two-stage procedure described in Section~\ref{part: RR}. With this approach, the MSE in the estimation of $\OR$ depends on the errors in the two RR estimations; and the problem is how to distribute the target MSE between these two components so as to approximately achieve a desired ratio of average sample sizes. To this end, another sampling stage could be introduced before the RR estimations, but that would complicate the process. A better approach, which only requires two stages and results in good estimation efficiency, is based on estimating the \emph{odds} $p_1/(1-p_1)$ and $(1-p_2)/p_2$ separately, using in each case the method in \citet[section~2]{Mendo25b}. This is detailed next.

The estimation method consists, as in previous sections, of two stages with IBS. The second stage estimates $p_1/(1-p_1)$ for the first population, and $(1-p_2)/p_2$ for the second population. However, there are two differences with respect to the RR and LRR estimators. The \emph{first} difference is that in the second stage \emph{two} IBS procedures are used for each population. Given $\sus_1$, $\sus_2$ (which will be computed from the results of the first stage), IBS is applied to population $1$ to obtain $\sus_1$ successes, which requires $\vasas'_1$ samples. Then IBS is applied again to population $1$ to obtain $\sus_1-\adjsus$ failures, with $\adjsus=2$ (a different value of $\adjsus$ will be used for LOR estimation). This requires $\vasas''_1$ samples, for a total of $\vasas_1 = \vasas'_1 + \vasas''_1$ samples. For population $2$, IBS is applied to obtain $\sus_2-\adjsus$ successes, which requires $\vasas'_2$ samples; and then to obtain $\sus_2$ failures, which requires $\vasas''_2$ samples, for a total of $\vasas_2 = \vasas'_2 + \vasas''_2$ samples. The average numbers of samples used by the second stage are computed as follows. From \eqref{eq: neg bin: E}, the conditional mean of $\vasas_i$, $i=1,2$ given $\sus_i$ is
\begin{align}
\label{eq: E vasa 1 cond OR}
\E[\vasas_1 \cond \sus_1] &= \frac{\sus_1}{p_1} + \frac{\sus_1 - \adjsus}{1-p_1} = \frac{\sus_1 - \adjsus p_1}{p_1(1-p_1)}, \\
\label{eq: E vasa 2 cond OR}
\E[\vasas_2 \cond \sus_2] &= \frac{\sus_2 - \adjsus}{p_2} + \frac{\sus_2}{1-p_2} = \frac{\sus_2 - \adjsus (1-p_2)}{p_2(1-p_2)},
\end{align}
where $\adjsus=2$ for OR. Then, defining
\begin{equation}
\label{eq: pco p}
\pco p_i = p_i(1-p_i),
\end{equation}
and considering a generic $\adjsus \geq 0$, it stems from \eqref{eq: E vasa 1 cond OR} and \eqref{eq: E vasa 2 cond OR} that $\E[\vasas_i]$ is approximately inversely proportional to $\pco p_i$:
\begin{align}
\label{eq: E vasa 1 OR}
\E[\vasas_1] &= \E\left[\E[\vasas_1 \cond \sus_1]\right] = \frac{\E[\sus_1] - \adjsus p_1}{\pco p_1} \leq \frac{\E[\sus_1]}{\pco p_1}, \\
\label{eq: E vasa 2 OR}
\E[\vasas_2] &= \E\left[\E[\vasas_2 \cond \sus_2]\right] = \frac{\E[\sus_2] - \adjsus (1-p_2)}{\pco p_2} \leq \frac{\E[\sus_2]}{\pco p_2}.
\end{align}
Using the same ideas as for the RR estimator (Section~\ref{part: RR sep proc}), it can be seen that to approximately achieve a given ratio of average sample sizes, considering only the samples used by the second stage for the moment, $\E[\sus_1] / \E[\sus_2]$ should be roughly proportional to
\begin{equation}
\label{eq: pco rapr}
\pco\rapr = \frac{\pco p_1}{\pco p_2}.
\end{equation}
In view of this, the \emph{second} difference from previous estimators is that the first stage in this case needs to use samples with parameters $\pco p_i$, $i=1,2$, rather than $p_i$, so as to acquire information about $\pco\rapr$. Specifically, for $i=1,2$, the first stage applies IBS with $\suf_i$ successes to a sequence of samples with parameter $\pco p_i$. Denoting the number of samples used from this sequence by $\pco \vasaf_i$, it is clear from \eqref{eq: neg bin: E} that $\E[\pco \vasaf_i] = \suf_i/\pco p_i$. The samples with parameter $\pco p_i$ must be generated from samples with parameter $p_i$. A simple, efficient procedure for this will be given later, and it will be shown that with this procedure the total number of samples with parameter $p_i$ required by the first stage, $\vasaf_i$, has an average equal to $3\suf_i/(2 \pco p_i)$. Therefore, considering both sampling stages, the average numbers of samples from the two populations satisfy
\begin{equation}
\label{eq: E1 E2 eq OR}
\frac{\E[\vasaf_1+\vasas_1]}{\E[\vasaf_2+\vasas_2]} = \frac{3\suf_1/2+\E[\sus_1]}{(3\suf_2/2+\E[\sus_2]) \pco \rapr}.
\end{equation}
This is analogous to \eqref{eq: E1 E2 eq RR}, and in consequence $\sus_1$ and $\sus_2$ can be chosen using a similar approach as for RR, described next.

The first stage produces the variables $\pco\vasaf_1$ and $\pco\vasaf_2$, from which $\varaf$ is defined as
\begin{equation}
\label{eq: varaf OR LOR}
\varaf = \frac{\pco\vasaf_2-\cdesm_2}{\pco\vasaf_1-\cdesm_1},
\end{equation}
where $\cdesm_1 = \cdesm_2 = 1/2$ as in Section~\ref{part: RR}. For a target relative MSE given by $\tarvar$, the second-stage IBS parameters $\sus_1$ and $\sus_2$ are obtained from $\varaf$ by solving the equation system formed by \eqref{eq: cond nsus ratio} and \eqref{eq: ferr tarvar}, where the values of the design parameters $\cdemul$, $\cdeadd_1$ and $\cdeadd_2$ are yet to be specified; and the solutions will then have to be rounded to integer values, as usual. Once $\sus_1$, $\sus_2$ are known, the second stage is carried out, from which $\vasas'_1$, $\vasas''_1$, $\vasas'_2$, $\vasas''_2$ are obtained. According to \citet[section~2]{Mendo25b}, and taking into account that $\adjsus=2$,
\[
\frac{(\sus_1-1) \, \vasas''_1}{(\sus_1-2)\, (\vasas'_1-1)}
\]
is a conditionally unbiased estimator of $p_1/(1-p_1)$ given $\sus_1$;
\[
\frac{(\sus_2-1)\, \vasas'_2}{(\sus_2-2)\, (\vasas''_2-1)}
\]
is a conditionally unbiased estimator of $(1-p_2)/p_2$ given $\sus_2$; and for $\sus_1, \sus_2 \geq 3$ the conditional variances of these estimators are respectively less than
\[
\frac{p_1^2}{(\sus_1-2) (1-p_1)^2} \left(1-\frac{\pco p_1}{\sus_1-2+2p_1}\right)
\]
and
\[
\frac{(1-p_2)^2}{(\sus_2-2) p_2^2} \left(1-\frac{\pco p_2}{\sus_2-2p_2}\right).
\]
Therefore, since the observations are independent,
\begin{equation}
\label{eq: vahOR}
\vahOR = \frac
{(\sus_1-1)(\sus_2-1) \, \vasas''_1 \vasas'_2}
{(\sus_1-2)(\sus_2-2) \, (\vasas'_1-1) (\vasas''_2-1)}
\end{equation}
is an unbiased estimator of $\OR = p_1(1-p_2)/(p_2(1-p_1))$, and for $\sus_1, \sus_2 \geq 3$
\begin{equation}
\label{eq: Var varhOR cond no unif}
\frac{\E\left[ \vahOR^2 \cond* \sus_1, \sus_2 \right]}{\OR^2}
\leq \left( \frac{1}{\sus_1-2} \left(1-\frac{\pco p_1}{\sus_1-2+2p_1}\right) + 1 \right) \left( \frac{1}{\sus_2-2} \left(1-\frac{\pco p_2}{\sus_2-2p_2}\right) + 1 \right),
\end{equation}
which implies that, for any $p_1,p_2 \in (0,1)$,
\begin{equation}
\label{eq: Var varhOR cond}
\frac{\E\left[ \vahOR^2 \cond \sus_1, \sus_2 \right]}{\OR^2} < \frac{1}{\sus_1-2} + \frac{1}{\sus_2-2} + \frac{1}{(\sus_1-2)(\sus_2-2)} + 1.
\end{equation}
Based on \eqref{eq: Var varhOR cond}, the error function $\ferr(\sus_1,\sus_2)$ for OR is defined as in \eqref{eq: ferr} with
\begin{equation}
\label{eq: cerr OR}
\cerr_1 = 2, \qquad \cerr_2 = 2, \qquad \cerr_{12}= 1;
\end{equation}
and then, by the same reasoning as in \eqref{eq: Var vahrr}, if the rounded values of $\sus_1$ and $\sus_2$ satisfy \eqref{eq: ferr leq tarvar} this guarantees that $\Var[\vahOR] / \OR^2 < \tarvar$ for any $p_1, p_2 \in (0,1)$.

The procedure to generate samples with parameter $\pco p_i$ is as follows. Taking two samples with parameter $p_i$ as inputs is clearly sufficient to produce a sample with parameter $\pco p_i$. Namely, one possible criterion is \bfa{} to output success if and only if the first input is a success and the second is a failure. However, if the first input happens to be a failure the second input need not be observed. This occurs with probability $1-p_i$. Alternatively, \bfb{} the output can be defined to be success if the first input is a failure and the second is a success.
In this case, the second input is not needed if the first is a success, which occurs with probability $p_i$. If criteria \bfa{} or \bfb{} are randomly chosen with equal probabilities, the average number of inputs required to produce an output is $1 + (1-p_i)/2 + p_i/2 = 3/2$. This is an instance of a \emph{Bernoulli factory} \citep{Keane94}.

The $\pco \vasaf_i$ samples with parameter $\pco p_i$, $i=1,2$ required by the first stage are generated with the above method, requiring a total of $\vasaf_i$ samples of population $i$ as inputs. For the subsequent analysis it is necessary to characterize the relationship between $\E[\vasaf_i]$ and $\E[\pco \vasaf_i]$. It cannot be directly concluded from the preceding paragraph that $\E[\vasaf_i]/\E[\pco\vasaf_i] = 3/2$, because the IBS stopping rule could conceivably introduce some deviation in this ratio. However, this equality turns out to be true. More generally, for an arbitrary Bernoulli factory, if the average number of inputs needed to produce an output is equal to some constant $\bfio$, it can be seen that
\begin{equation}
\label{eq: vasaf pco vasaf}
\E[\vasaf_i] = \bfio \E[\pco\vasaf_i] = \frac{\bfio \suf_i}{\pco p_i},
\end{equation}
where $\bfio = 3/2$ for the described method. To prove this, let $\bfios$ and $\bfiof$ be the average number of inputs required to produce an output, conditioned on the output being success or failure respectively. Then
\begin{equation}
\label{eq: bfio bfios bfiof}
\bfio = \bfios \pco p_i + \bfiof (1-\pco p_i).
\end{equation}
IBS with parameter $\suf_i$ is applied to the outputs of the Bernoulli factory, and consumes $\pco\vasaf_i$ of those outputs, of which $\suf_i$ are successes and $\pco\vasaf_i-\suf_i$ are failures. Therefore, using \eqref{eq: neg bin: E},
\begin{equation}
\E[\vasaf_i] = \bfios \suf_i + \bfiof \left(\E[\pco\vasaf_i]-\suf_i \right) = \bfios \suf_i + \frac{\bfiof \suf_i (1-\pco p_i)}{\pco p_i},
\end{equation}
which combined with \eqref{eq: bfio bfios bfiof} gives \eqref{eq: vasaf pco vasaf}.

The characterization of $\E[\vasaf_i+\vasas_i]$, $i=1,2$, as well as the ensuing selection  of $\cdemul$, $\cdeadd_1$ and $\cdeadd_2$, relies, as in the RR case, on using first-order approximations for $\sus_1$ and $\sus_2$ as functions of $\varaf$ and $1/\varaf$ respectively. Expressions \eqref{eq: nsus 1 approx}--\eqref{eq: csufva 2} 
remain valid for OR, and \eqref{eq: E varaf approx RR LRR} and \eqref{eq: E 1/varaf approx RR LRR} hold with $\rapr$ replaced by $\pco \rapr$:
\begin{align}
\label{eq: E varaf approx OR LOR}
\E[\varaf] &\approx \frac{\suf_2 \pco\rapr}{\suf_1-1}, \\
\label{eq: E 1/varaf approx OR LOR}
\E\left[\frac 1 {\varaf}\right] &\approx \frac{\suf_1}{(\suf_2-1) \pco\rapr}.
\end{align}
In addition, \eqref{eq: E vasa 1 approx RR bis} and \eqref{eq: E vasa 2 approx RR bis} have to be modified to take into account that the first stage uses samples with parameter $\pco p_i$, obtained by transforming samples with parameter $p_i$, $i=1,2$. Specifically, from \eqref{eq: nsus 1 approx}, \eqref{eq: nsus 2 approx}, \eqref{eq: E vasa 1 OR}, \eqref{eq: E vasa 2 OR}, \eqref{eq: vasaf pco vasaf}, \eqref{eq: E varaf approx OR LOR} and \eqref{eq: E 1/varaf approx OR LOR}, and introducing the term $\susrou$ to account for the effect of rounding  $\sus_1$ and $\sus_2$,
\begin{align}
\label{eq: E vasa 1 approx OR}
\E[\vasaf_1 + \vasas_1] &= \frac{3\suf_1/2 + \E[\sus_1] - 2p_1}{\pco p_1}
\approx \frac{3\suf_1/2 + \E[\sus_1]}{\pco p_1}
\approx \frac{\csusco_1 + 3\suf_1/2 + \susrou}{\pco p_1} + \frac{\csusva_1 \suf_2}{(\suf_1-1)\pco p_2}, \\
\label{eq: E vasa 2 approx OR}
\E[\vasaf_2 + \vasas_2] &= \frac{3\suf_2/2 + \E[\sus_2] - 2(1-2p_2)}{\pco p_2}\approx \frac{\csusco_2 + 3\suf_2/2 + \susrou}{\pco p_2} + \frac{\csusva_2 \suf_1}{(\suf_2-1)\pco p_1}.
\end{align}
According to  \eqref{eq: csufco 1}--\eqref{eq: csufva 2}, 
\eqref{eq: E vasa 1 approx OR} and \eqref{eq: E vasa 2 approx OR}, the condition \eqref{eq: cond tarsara} for the ratio of average sample sizes will be satisfied regardless of $p_1, p_2$ if
\begin{align}
\label{eq: igualdad cruzada 1 OR bis}
\frac 1 {\tarvar} + \frac{3\suf_1} 2 + \cerr_1 + \susrou &= \frac{\tarsara \suf_1}{\cdemul(\suf_2-1)} \left( \frac 1 {\tarvar} + \cdeadd_1 + \cerr_1 \right), \\
\label{eq: igualdad cruzada 2 OR bis}
\frac{\tarsara}{\cdemul} \left(\frac 1 {\tarvar} + \frac{3\suf_2} 2 + \cerr_2 + \susrou \right) &= \frac{\suf_2}{\suf_1-1} \left(\frac 1 {\tarvar} + \cdeadd_2 + \cerr_2 \right).
\end{align}
By analogy with \eqref{eq: igualdad cruzada 1 RR bis} and \eqref{eq: igualdad cruzada 2 RR bis}, a simple solution to \eqref{eq: igualdad cruzada 1 OR bis} and \eqref{eq: igualdad cruzada 2 OR bis} is obtained if $3\suf_1/2 + \cerr_1 = 3\suf_2/2 + \cerr_2$ and $\cdeadd_1 + \cerr_1 = \cdeadd_2 + \cerr_2$. The former condition is compatible with $\suf_1, \suf_2 \in \mathbb N$ only if $2(\cerr_1-\cerr_2)/3 \in \mathbb Z$. In particular, this holds if
\begin{equation}
\label{eq: cerr 1 eq cerr 2}
\cerr_1 = \cerr_2.
\end{equation}
This is the case for OR estimation, and will also be true for LOR estimation (Section~\ref{part: LOR}). Thus, in the sequel \eqref{eq: cerr 1 eq cerr 2} will be assumed to hold. The two indicated conditions then reduce to
\begin{equation}
\label{eq: suf 1 suf 2, cdeadd 1 cdeadd 2, OR LOR}
\suf_2 = \suf_1, \qquad \cdeadd_2 = \cdeadd_1,
\end{equation}
which gives the solution to \eqref{eq: igualdad cruzada 1 OR bis} and \eqref{eq: igualdad cruzada 2 OR bis} as
\begin{align}
\label{eq: cdemul OR LOR}
\cdemul &= \tarsara, \\
\label{eq: cdeadd OR LOR}
\cdeadd_1 &= \left( \frac 1 {\tarvar} + \frac{3\suf_1}{2} + \cerr_1 + \susrou \right) \frac{\suf_1-1}{\suf_1} - \frac 1 {\tarvar} - \cerr_1.
\end{align}
An argument analogous to that used in Section~\ref{part: RR sep param ave} shows that \eqref{eq: cond cdeadd >} is satisfied.

The values of $\sus_1$, $\sus_2$ before rounding, obtained by solving \eqref{eq: cond nsus ratio} and \eqref{eq: ferr tarvar} with $\cdeadd_1$, $\cdeadd_2$ and $\cdemul$ as in \eqref{eq: suf 1 suf 2, cdeadd 1 cdeadd 2, OR LOR}--\eqref{eq: cdeadd OR LOR}, 
are again given by \eqref{eq: nsus 1 bis}--\eqref{eq: disc bis}. 
The curvature function is defined as in \eqref{eq: fcurv RR LRR} with the terms $1/\tarvar + \suf_1 + \cerr_1 + \susrou$ replaced by $1/\tarvar + 3\suf_1/2 + \cerr_1 + \susrou$ and with $\cerr_1=\cerr_2$:
\begin{equation}
\label{eq: fcurv OR LOR}
\begin{split}
\fcurv(\tarvar, \suf_1, \susrou) &= \tarvar\left(\left( \frac 1 {\tarvar} + \frac{3 \suf_1} 2 + \cerr_1 + \susrou \right) \frac {\suf_1-1} {\suf_1} - \frac 1 {\tarvar}\right)^2 \\
& \quad + 2\left(\left( \frac 1 {\tarvar} + \frac{3 \suf_1} 2 + \cerr_1 + \susrou \right) \frac {\suf_1-1} {\suf_1} - \frac 1 {\tarvar}\right) - \cerr_{12},
\end{split}
\end{equation}
from which the condition $\fcurv(\tarvar, \suf_1, \susrou)=0$ is
\begin{equation}
\label{eq: curv zero bis OR LOR}
\begin{split}
& \left(\frac{3\suf_1} 2 + \cerr_1 + \susrou\right)^2(\suf_1-1)^2 \tarvar^2 \\
&\quad + \left(2\left(\frac{3\suf_1} 2 + \cerr_1 + \susrou\right)(\suf_1-1)^2 - \cerr_{12}\suf_1^2\right) \tarvar + 1-2\suf_1 = 0.
\end{split}
\end{equation}
Similarly to previous estimators, $\susrou$ is set to $1$ and $\suf_i$ is chosen as in \eqref{eq: cho suf}, where $\fcurv(\tarvar, \suf_1, \susrou)$ is given by \eqref{eq: fcurv OR LOR}. Equivalently, the curve formed by the pairs $(\tarvar, \suf_1)$ that solve \eqref{eq: curv zero bis OR LOR} can be plotted, as shown in Figure~\ref{fig: curv_zero_1}; and then, for a given $\tarvar$, \eqref{eq: cho suf} corresponds to rounding up the value obtained from the curve, with a minimum of $3$. Defining
\begin{equation}
\label{eq: pco roprpr}
\pco\roprpr = \sqrt{\pco p_1 \pco p_2},
\end{equation}
this choice of $\susrou$ and $\suf_1$ results, by analogy with Section~\ref{part: RR sep bounds}, in the upper bounds
\begin{align}
\label{eq: E vasa 1 norm res OR LOR <}
\E[\vasaf_1 + \vasas_1] \pco\roprpr &< \left( \frac 1 {\tarvar} + \frac{3\suf_1} 2 + \cerr_1 + 1 \right) \left( \frac 1 {\sqrt{\tarsara\pco\rapr}} + \sqrt{\tarsara\pco\rapr} \right) \sqrt{\tarsara}, \\
\label{eq: E vasa 2 norm res OR LOR <}
\E[\vasaf_2 + \vasas_2] \pco\roprpr &< \left( \frac 1 {\tarvar} + \frac{3\suf_1} 2 + \cerr_1 + 1 \right) \left( \frac 1 {\sqrt{\tarsara\pco\rapr}} + \sqrt{\tarsara\pco\rapr} \right) \frac 1 {\sqrt{\tarsara}}.
\end{align}

The estimation procedure for OR is specified in Algorithm~\ref{algo: OR LOR} (see Appendix~\ref{part: app algo}), where the properties of the estimator are also indicated. (The algorithm covers the LOR case as well, to be presented in Section~\ref{part: LOR}.)

\begin{figure}%
\begin{subfigure}{\tamdosfig}%
\includegraphics[width=\textwidth]{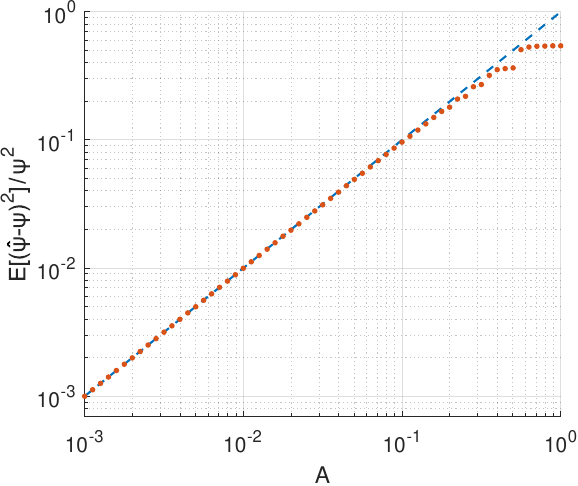}%
\captionsetup{justification=centering}%
\caption{$\tarsara=1$, $\rapr=16$, $\roprpr=0.1$\\ ($p_1 = 0.4$, $p_2 = 0.025$)}%
\label{fig: MSE_OR_1_11_16_0p1_cho}%
\end{subfigure}%
\hfill
\begin{subfigure}{\tamdosfig}%
\includegraphics[width=\textwidth]{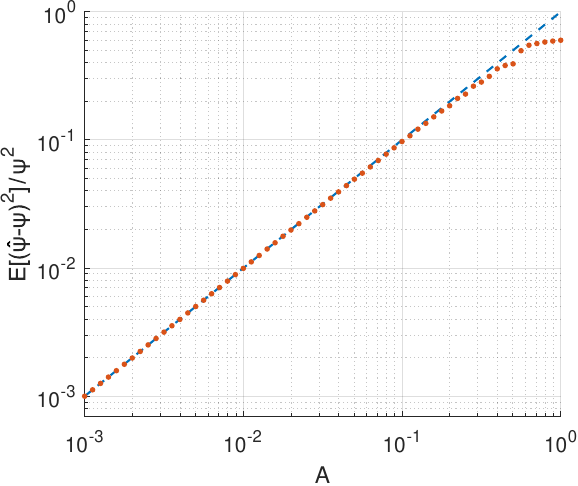}%
\captionsetup{justification=centering}%
\caption{$\tarsara=3$, $\rapr=1/16$, $\roprpr=0.01$\\ ($p_1 = 0.0025$, $p_2 = 0.04$)}%
\label{fig: MSE_OR_1_31_1d16_0p01_cho}%
\end{subfigure}%
\\[\splegbef]%
\centering%
\begin{small}%
Dashed line: target. Dots: simulation.
\end{small}%
\\[\splegaft]%
\caption{Relative MSE for OR with element sampling}
\label{fig: MSE_OR_1_cho}%
\end{figure}

The relative MSE
obtained from simulations is compared with the target $\tarvar$ in Figure~\ref{fig: MSE_OR_1_cho}. The simulation consists, as with previous estimators, of $10^6$ realizations for each combination of parameters. The difference between simulation results and target is seen to increase with $\tarvar$ and with $\roprpr$, as for RR. However, in OR the difference observed for large $\roprpr$ vanishes when $\tarvar$ is small, unlike in RR (compare the leftmost parts of Figures~\ref{fig: MSE_RR_1_11_16_0p1_cho} and \ref{fig: MSE_OR_1_11_16_0p1_cho}). This is related to the fact that the factors $1-\pco p_1/(\sus_1-2+2p_1)$ and $1-\pco p_2/(\sus_2-2p_2)$ in \eqref{eq: Var varhOR cond no unif}, which are replaced by $1$ in the uniform bound \eqref{eq: Var varhOR cond}, approach $1$ for large $\sus_1$, $\sus_2$. Thus when $\tarvar$ is small, which gives large values of $\sus_1$ and $\sus_2$, the uniform bound \eqref{eq: Var varhOR cond} is almost as good as \eqref{eq: Var varhOR cond no unif}. In RR, on the other hand, the corresponding factors are $(\sus_1-2)(1-p_1)/(\sus_1-2+2p_1) < 1-p_1$ and $1-p_2$, as is seen comparing \eqref{eq: Var varhRR cond no unif} and \eqref{eq: Var varhRR cond}, and these do not tend to $1$ for large $\sus_1$, $\sus_2$. (In LRR the factors are more cumbersome, see \citet[theorem~2]{Mendo25b}; but for large $\sus_1$, $\sus_2$ they tend to the same values $1-p_1$, $1-p_2$ as in RR.)

\begin{figure}%
\begin{subfigure}{\textwidth}%
\includegraphics[width=\tamdosfig]{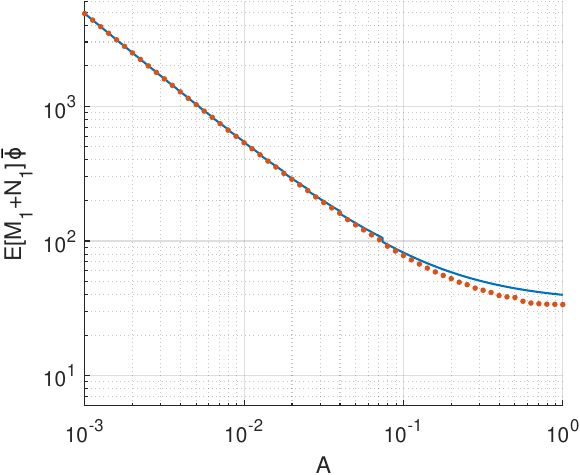}%
\hfill%
\includegraphics[width=\tamdosfig]{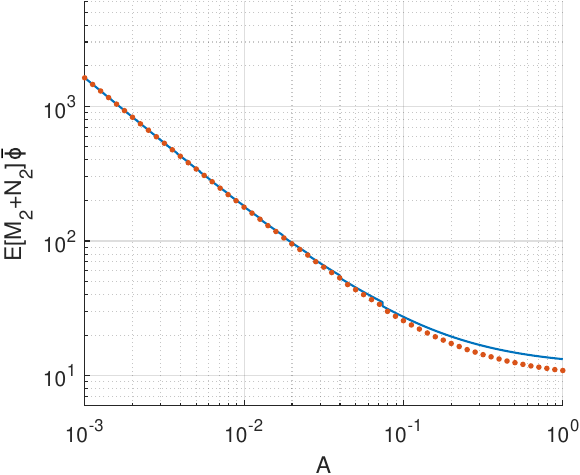}%
\caption{$\tarsara=3$, $\rapr=1/16$, $\roprpr=0.01$ ($p_1 = 0.0025$, $p_2 = 0.04$)}%
\label{fig: E1_OR_0_31_1d16_0p01_cho}%
\end{subfigure}%
\\[\splegaft]%
\begin{subfigure}{\textwidth}%
\includegraphics[width=\tamdosfig]{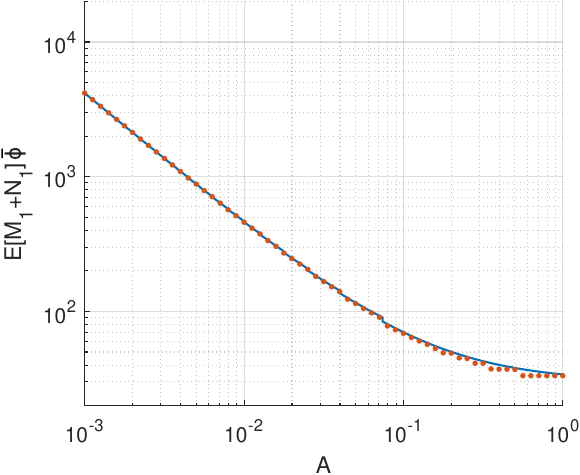}%
\hfill%
\includegraphics[width=\tamdosfig]{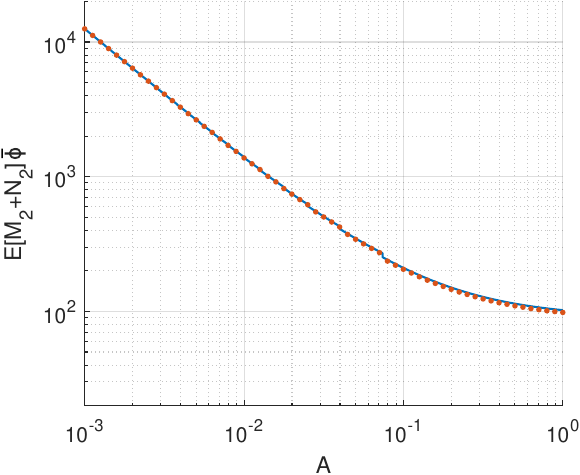}%
\caption{$\tarsara=1/3$, $\rapr=1/16$, $\roprpr=0.01$ ($p_1 = 0.0025$, $p_2 = 0.04$)}%
\label{fig: E1_OR_0_13_1d16_0p01_cho}%
\end{subfigure}%
\\[\splegbef]%
\centering%
\begin{small}%
Line: bound. Dots: simulation.
\end{small}%
\\[\splegaft]%
\caption{Normalized average sample sizes for OR with element sampling}
\label{fig: Ei_OR_0_cho}%
\end{figure}

Figure~\ref{fig: Ei_OR_0_cho} shows simulation results for $\E[\vasaf_i+\vasas_i] \pco \roprpr$, $i=1,2$. The values are very similar to those for $\E[\vasaf_i+\vasas_i] \roprpr$ in RR (Figures~\ref{fig: Ei_RR_1_31_1d16_0p01_allwdh} and \ref{fig: Ei_RR_1_13_1d16_0p01_allwdh}) except that for OR they are somewhat larger in the rightmost part of the graphs. This is due to the $3/2$ factor in OR, whose effect on the average sample sizes is only significant for large $\tarvar$.

The efficiency of the OR estimator with element sampling results from particularizing \eqref{eq: effic sep} to $\parg = \OR$, with $\partial\OR/\partial p_1 = \OR/\pco p_1$, $\partial\OR/\partial p_2 = -\OR/\pco p_2$:
\begin{equation}
\label{eq: effic OR gen}
\efficsep = \frac {\displaystyle \frac{1}{\E[\vasaf_1+\vasas_1] \pco p_1} + \frac{1}{\E[\vasaf_2+\vasas_2] \pco p_2} } {\Var[\OR]/\OR^2}.
\end{equation}
Then, using \eqref{eq: E vasa 1 norm res OR LOR <} and \eqref{eq: E vasa 2 norm res OR LOR <} and considering that $\Var[\vahOR] / \OR^2 < \tarvar$, the efficiency can be bounded as
\begin{equation}
\label{eq: effic OR LOR sep bound}
\efficsep > \frac{ 1 } { 1 + \tarvar(3\suf_1/2 + \cerr_1 + 1) }.
\end{equation}
This is similar to the bound \eqref{eq: effic RR LRR sep bound} for RR and LRR, the only differences being the $3/2$ factor in \eqref{eq: effic OR LOR sep bound}, arising from the Bernoulli factory, and the last factor in \eqref{eq: effic RR LRR sep bound} (but note that the latter is approximately $1$ for small $\roprpr$). By the same arguments used for RR and LRR, $\efficsep$ for OR approaches $1$ when $\tarvar$ tends to $0$.

\begin{figure}%
\begin{subfigure}{\tamdosfig}%
\includegraphics[width=\textwidth]{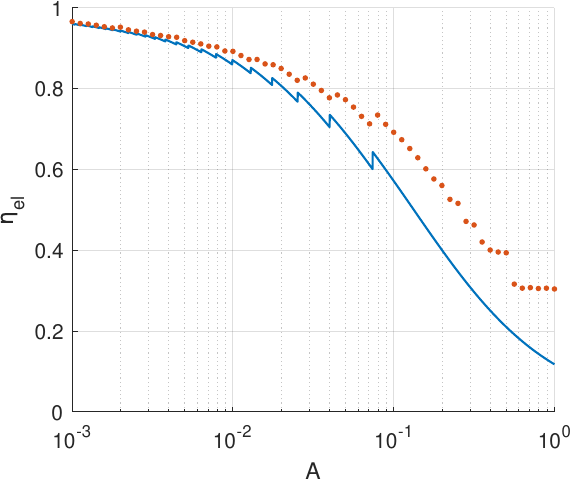}%
\captionsetup{justification=centering}%
\caption{$\tarsara=1$, $\rapr=16$, $\roprpr=0.1$\\ ($p_1 = 0.4$, $p_2 = 0.025$)}%
\label{fig: efs_OR_0_11_16_0p1_cho}%
\end{subfigure}%
\hfill
\begin{subfigure}{\tamdosfig}%
\includegraphics[width=\textwidth]{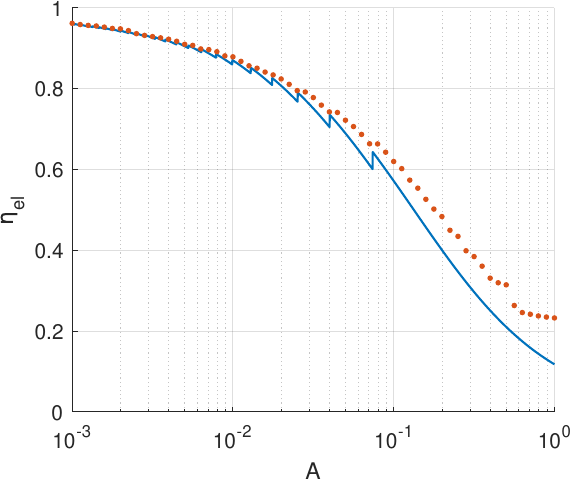}%
\captionsetup{justification=centering}%
\caption{$\tarsara=3$, $\rapr=1/16$, $\roprpr=0.01$\\ ($p_1 = 0.0025$, $p_2 = 0.04$)}%
\label{fig: efs_OR_0_31_1d16_0p01_cho}%
\end{subfigure}%
\\[\splegbef]%
\centering%
\begin{small}%
Line: bound. Dots: simulation.
\end{small}%
\\[\splegaft]%
\caption{Efficiency for OR with element sampling}
\label{fig: efs_OR_0_cho}%
\end{figure}

Figure~\ref{fig: efs_OR_0_cho} shows simulation results for $\efficsep$, and compares them with the bound \eqref{eq: effic OR LOR sep bound}. In the simulation, $\efficsep$ is computed as in previous sections, i.e.~using sample averages in \eqref{eq: effic OR gen}. The values are seen to be similar to those for RR and LRR, except that with large $\tarvar$ the efficiency is slightly lower for OR. Again, this is a consequence of the $3/2$ factor. As in RR and LRR, the difference between simulation and bound tends to be larger when $\roprpr$ is increased. However, in OR that difference vanishes for small $\tarvar$, unlike in the other cases (compare Figure~\ref{fig: efs_OR_0_11_16_0p1_cho} with Figure~\ref{fig: efs_RR_1_11_16_0p1_cho} or \ref{fig: efs_LRR_1_11_16_0p1_cho}). This agrees with the behavior of the MSE discussed earlier.

\subsection{Group sampling}
\label{part: OR gr}

Group sampling for OR estimation consumes a number of groups $\vagr$ given by \eqref{eq: vagr}, as with previous estimators, in order to provide the required amounts of samples $\vasaf_i+\vasas_i$ of each population $i=1,2$; and $\E[\vagr]$ can be approximately computed from \eqref{eq: gr approx RR bis}. It is shown in Appendix~\ref{part: app OR LOR gr var} that, for $\roprpr$ small, the term $\E[|\diffgr|]$ in \eqref{eq: gr approx RR bis} can be expressed as
\begin{equation}
\label{eq: pre diffgr OR}
\E\left[|\diffgr|\right] \approx
\E\left[\left|\frac{3\suf_1/2+\sus_1}{\tarsagr_1 \pco p_1} - \frac{3\suf_2/2+\sus_2}{\tarsagr_2 \pco p_2}\right|\right].
\end{equation}
Thus, proceeding as in Section~\ref{part: RR gr ave} and taking into account \eqref{eq: cerr 1 eq cerr 2},
\begin{equation}
\label{eq: gr approx OR LOR ter}
\begin{split}
\E[\vagr] \pco\roprpr
&\approx \left( \frac 1 {\tarvar} + \frac{3\suf_1} 2 + \cerr_1 + \susrou \right) \left( \frac 1 {\tarsagr_1 \sqrt{\pco\rapr}} + \frac{\sqrt{\pco\rapr}}{\tarsagr_2} \right. \\
&\quad \left. + \frac{\varafnbor^{\suf_1-1}}{(\varafnbor+1)^{2\suf_1-1} \betaf(\suf_1,\suf_1) \suf_1} \left( \frac {1}{\tarsagr_1 \sqrt{\pco\rapr}} + \frac {\varafnbor \sqrt{\pco\rapr}} {\tarsagr_2}  \right) \right),
\end{split}
\end{equation}
where
\begin{equation}
\label{eq: varafnbor OR LOR bis}
\varafnbor
= \frac {\suf_1} {2\tarsara\pco\rapr(\suf_1-1)} 
\left( \tarsara\pco\rapr-1 + \sqrt{(\tarsara\pco\rapr-1)^2 + \frac{4\tarsara\pco\rapr (\suf_1-1)^2}{\suf_1^2}} \right).
\end{equation}
The simulation results for $\E[\vagr] \pco \roprpr$, not shown, are similar to those for $\E[\vagr] \roprpr$ in RR and LRR, with the difference that for large $\tarvar$ the values are slightly greater in OR compared with those two cases, as it has been observed with element sampling.

\begin{figure}%
\begin{subfigure}{\tamdosfig}%
\includegraphics[width=\textwidth]{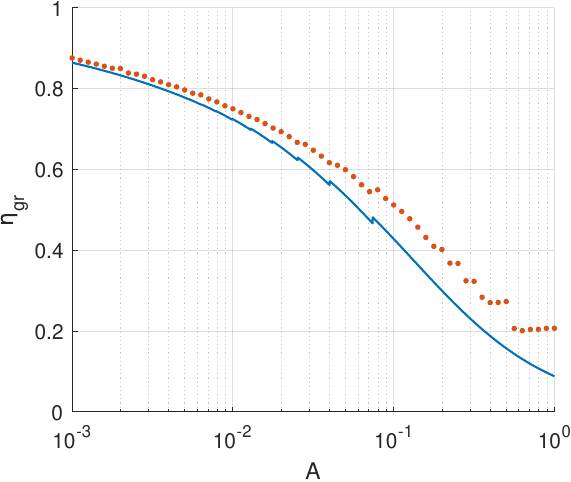}%
\captionsetup{justification=centering}%
\caption{$\tarsagr_1=1$, $\tarsagr_2=1$, $\rapr=16$, $\roprpr=0.1$\\ ($p_1 = 0.4$, $p_2 = 0.025$)}%
\label{fig: efg_OR_1_11_16_0p1_cho}%
\end{subfigure}%
\hfill
\begin{subfigure}{\tamdosfig}%
\includegraphics[width=\textwidth]{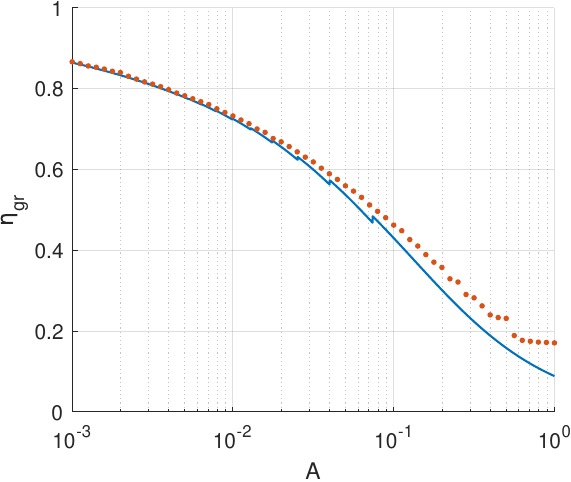}%
\captionsetup{justification=centering}%
\caption{$\tarsagr_1=3$, $\tarsagr_2=1$, $\rapr=1/16$, $\roprpr=0.01$\\ ($p_1 = 0.0025$, $p_2 = 0.04$)}%
\label{fig: efg_OR_1_31_1d16_0p01_cho}%
\end{subfigure}%
\\[\splegbef]%
\centering%
\begin{small}%
Line: approximation. Dots: simulation.
\end{small}%
\\[\splegaft]%
\caption{Efficiency for OR with group sampling}
\label{fig: efg_OR_1_cho}%
\end{figure}

The efficiency with group sampling is obtained particularizing \eqref{eq: effic gr} for $\parg = \OR$ and approximating $\Var[\vahOR] / \OR^2 \approx \tarvar$:
\begin{equation}
\label{eq: effic OR LOR gr}
\efficgr
\approx \frac
{1 / (\tarsagr_1 \sqrt{\pco\rapr}) + \sqrt{\pco\rapr} / \tarsagr_2 }
{ \tarvar \E[\vagr] \pco\roprpr},
\end{equation}
where $\E[\vagr] \pco\roprpr$ is given by \eqref{eq: gr approx OR LOR ter}. Figure~\ref{fig: efg_OR_1_cho} represents the theoretical approximation \eqref{eq: effic OR LOR gr}, as well as results from simulation. As with element sampling, the efficiency for large $\tarvar$ is slightly lower than in RR and LRR (Figures~\ref{fig: efg_RR_1_cho} and \ref{fig: efg_LRR_1_cho}), and for small $\tarvar$ there is almost no difference between simulation results and theoretical approximation even if $\roprpr$ is large.

\section{Estimation of log odds ratio}
\label{part: LOR}

\subsection{Element sampling}

The estimation method for the LOR, $\LOR= \log (p_1(1-p_2)/(p_2(1-p_1)))$, has few differences compared to that presented for OR in Section~\ref{part: OR}. The second stage estimates $\log(p_1/(1-p_1))$ and $\log(p_2/(1-p_2))$ separately, using for each the method described in \citet[section~3]{Mendo25b}, which consists of two IBS procedures with the same parameter.
Specifically, to estimate $\log(p_i/(1-p_i))$, $i=1,2$, IBS is applied to population $i$ until $\sus_i$ successes are obtained, which requires a random number $\vasas_i'$ of observations from this population, and then IBS is applied until $\sus_i$ failures are obtained, which requires $\vasas_i''$ additional observations. These are the same steps as for OR, but with $\adjsus=0$. Then, conditioned on $\sus_1$ and $\sus_2$, which are obtained in the first stage,
$-\harm{\vasas_i'-1}+\harm{\vasas_i''-1}$ is a conditionally unbiased estimator of  $\log(p_i/(1-p_i))$ with conditional variance less than $1/(\sus_i-5/4)$ for any $p_i \in (0,1)$ \citep[theorem~3]{Mendo25b}. Therefore,
\begin{equation}
\label{eq: vahLOR}
\vahLOR = -\harm{\vasas_1'-1}+\harm{\vasas_1''-1}+\harm{\vasas_2'-1}-\harm{\vasas_2''-1}
\end{equation}
is a conditionally unbiased estimator of $\LOR$ with
\begin{equation}
\Var[\vahLOR \cond \sus_1, \sus_2] < \frac 1 {\sus_1-5/4} + \frac 1 {\sus_2-5/4}.
\end{equation}
It follows that $\vahLOR$ is unconditionally unbiased, and, defining its error function as in \eqref{eq: ferr} with
\begin{equation}
\label{eq: cerr LOR}
\cerr_1 = 5/4, \qquad \cerr_2 = 5/4, \qquad \cerr_{12}= 0,
\end{equation}
inequality \eqref{eq: ferr leq tarvar} guarantees that $\Var[\vahLOR] < \tarvar$ for any $p_1, p_2 \in(0,1)$.

The number of samples from population $i=1,2$ used in the second stage, $\vasas_i = \vasas_i'+\vasas_i''$, has an average given by \eqref{eq: E vasa 1 OR} and \eqref{eq: E vasa 2 OR} with $\adjsus=0$, that is, $\E[\vasas_i] = \E[\sus_i]/\pco p_i$, where $\pco p_i$ is defined by \eqref{eq: pco p}. Thus, for the same reason as in OR estimation, the first stage for LOR estimation must be based on samples with parameters $\pco p_i$, $i=1,2$, generated from observations with parameters $p_i$. For each $i=1,2$, given $\suf_i$, a random number $\pco \vasaf_i$ of samples with parameter $\pco p_i$ are generated until $\suf_i$ successes are obtained, which in turn requires $\vasaf_i$ observations from population $i$. Then $\varaf$ is computed from $\pco \vasaf_1$ and $\pco \vasaf_2$ as in \eqref{eq: varaf OR LOR}. Since \eqref{eq: cerr 1 eq cerr 2} holds, imposing the additional condition \eqref{eq: suf 1 suf 2, cdeadd 1 cdeadd 2, OR LOR}, $\cdemul$ and $\cdeadd_1$ are obtained from \eqref{eq: cdemul OR LOR} and \eqref{eq: cdeadd OR LOR} as for OR, and \eqref{eq: cond cdeadd >} is satisfied. Following this, $\sus_1$ and $\sus_2$ are computed
using \eqref{eq: nsus 1 bis}--\eqref{eq: disc bis}, 
as for the other estimators, and then rounded with the restriction \eqref{eq: ferr leq tarvar}. The resulting average numbers of samples are bounded by \eqref{eq: E vasa 1 norm res OR LOR <} and \eqref{eq: E vasa 2 norm res OR LOR <}.

The condition $\fcurv(\tarvar,\suf_1,\susrou)=0$, from which $\suf_1$ and $\suf_2$ are obtained, is expressed by \eqref{eq: curv zero bis OR LOR}. Because $\cerr_1=\cerr_2$ and $\cerr_{12}=0$, this simplifies in the same way as for LRR, and has the positive solution
\begin{equation}
\label{eq: cho suf LOR pre}
\tarvar= \frac 1 {(3\suf_1/2+\cerr_1+\susrou)(\suf_1-1)}.
\end{equation}
The LOR estimator, like previous ones, uses $\susrou=1$ and $\suf_1$ given by \eqref{eq: cho suf}, which in this case, taking into account \eqref{eq: cho suf LOR pre}, is written as
\begin{equation}
\label{eq: cho suf LOR}
\suf_1 =
\displaystyle
\max\left\{ 3, \left\lceil
-1/4 + \sqrt{(5/4)^2 + 2/(3\tarvar)}
\right\rceil \right\}.
\end{equation}
As with previous estimators, this choice of $\suf_1$ yields average numbers of samples close to their minimum values with respect to this parameter (results not shown). The curve $\fcurv(\tarvar,\suf_1,1)=0$, plotted in Figure~\ref{fig: curv_zero_1}, is almost indistinguishable from that for OR; and \eqref{eq: cho suf LOR} corresponds to rounding up the ordinate values of this curve, with a minimum of $3$.

The estimation procedure for LOR is summarized in Algorithm~\ref{algo: OR LOR}  (see Appendix~\ref{part: app algo}), which also lists the properties of the estimator.

\begin{figure}%
\begin{subfigure}{\tamdosfig}%
\includegraphics[width=\textwidth]{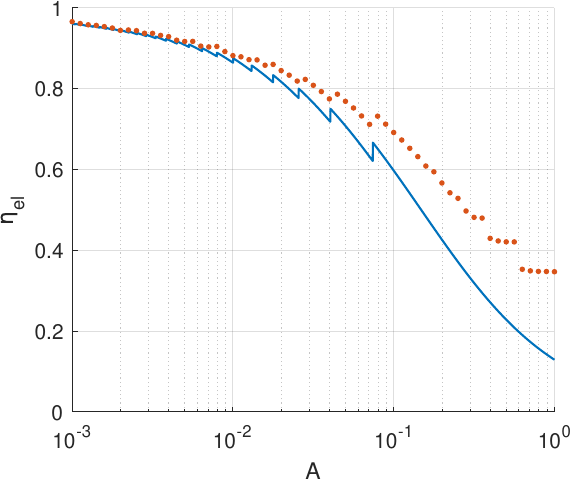}%
\captionsetup{justification=centering}%
\caption{$\tarsara=1$, $\rapr=16$, $\roprpr=0.1$\\ ($p_1 = 0.4$, $p_2 = 0.025$)}%
\label{fig: efs_LOR_1_11_16_0p1_cho}%
\end{subfigure}%
\hfill
\begin{subfigure}{\tamdosfig}%
\includegraphics[width=\textwidth]{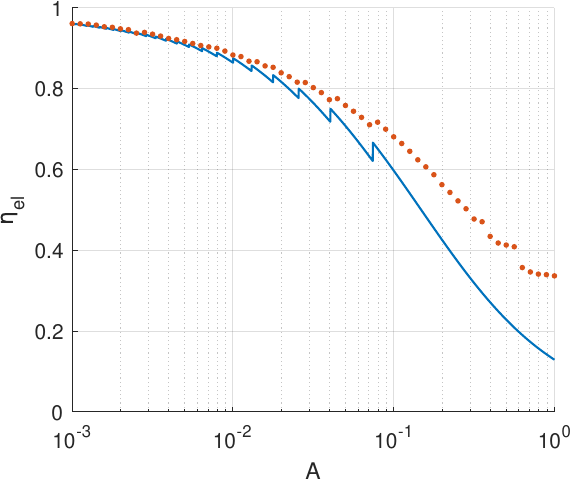}%
\captionsetup{justification=centering}%
\caption{$\tarsara=3$, $\rapr=1/16$, $\roprpr=0.01$\\ ($p_1 = 0.0025$, $p_2 = 0.04$)}%
\label{fig: efs_LOR_1_31_1d16_0p01_cho}%
\end{subfigure}%
\\[\splegbef]%
\centering%
\begin{small}%
Line: bound. Dots: simulation.
\end{small}%
\\[\splegaft]%
\caption{Efficiency for LOR with element sampling}
\label{fig: efs_LOR_1_cho}%
\end{figure}

Particularizing \eqref{eq: effic sep} for $\parg = \LOR$, with $\partial\LOR/\partial p_1 = 1/\pco p_1$, $\partial\LOR/\partial p_2 = -1/\pco p_2$, and then using \eqref{eq: E vasa 1 norm res OR LOR <}, \eqref{eq: E vasa 2 norm res OR LOR <} and the fact that $\Var[\vahLOR] < \tarvar$, it follows that the efficiency with element sampling is bounded by \eqref{eq: effic OR LOR sep bound} (with the value of $\cerr_1$ corresponding to LOR and with $\suf_1$ computed accordingly). Figure~\ref{fig: efs_LOR_1_cho} shows the results, which are very similar to those for OR (Figure~\ref{fig: efs_OR_0_cho}).

\subsection{Group sampling}

The group sampling procedure is analogous to that described for the other estimators. The average number of groups $\E[\vagr]$ is computed from \eqref{eq: gr approx RR bis}, where  $\E[|\diffgr|]$ satisfies \eqref{eq: pre diffgr OR}, as shown in Appendix~\ref{part: app OR LOR gr var}. Thus $\E[\vagr]$ is approximately given by \eqref{eq: gr approx OR LOR ter} and \eqref{eq: varafnbor OR LOR bis}, as for OR (but with the value of $\cerr_1$ corresponding to LOR). Simulation results are omitted.

\begin{figure}%
\begin{subfigure}{\tamdosfig}%
\includegraphics[width=\textwidth]{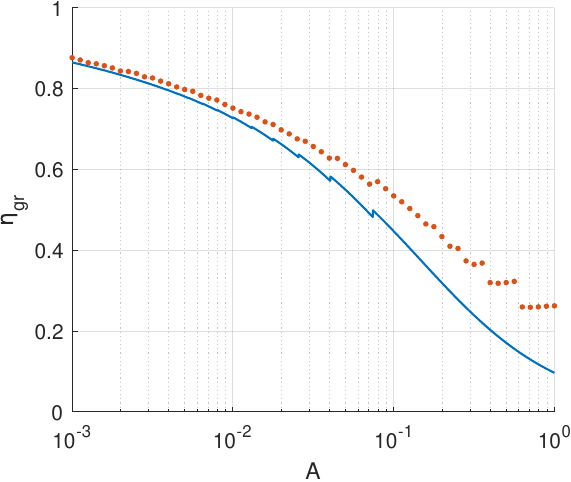}%
\captionsetup{justification=centering}%
\caption{$\tarsagr_1=1$, $\tarsagr_2=1$, $\rapr=16$, $\roprpr=0.1$\\ ($p_1 = 0.4$, $p_2 = 0.025$)}%
\label{fig: efg_LOR_1_11_16_0p1_cho}%
\end{subfigure}%
\hfill
\begin{subfigure}{\tamdosfig}%
\includegraphics[width=\textwidth]{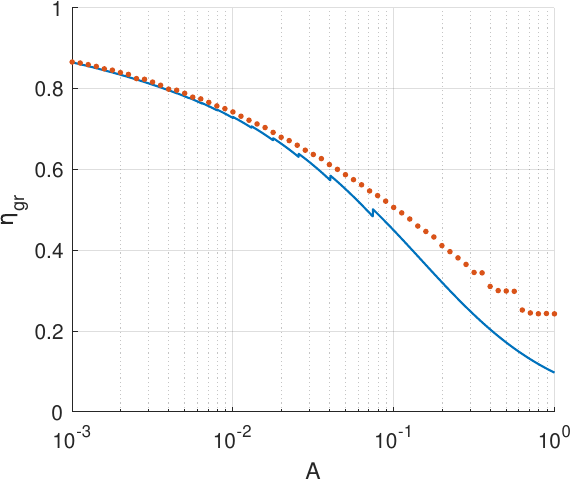}%
\captionsetup{justification=centering}%
\caption{$\tarsagr_1=3$, $\tarsagr_2=1$, $\rapr=1/16$, $\roprpr=0.01$\\ ($p_1 = 0.0025$, $p_2 = 0.04$)}%
\label{fig: efg_LOR_1_31_1d16_0p01_cho}%
\end{subfigure}%
\\[\splegbef]%
\centering%
\begin{small}%
Line: approximation. Dots: simulation.
\end{small}%
\\[\splegaft]%
\caption{Efficiency for LOR with group sampling}
\label{fig: efg_LOR_1_cho}%
\end{figure}

The efficiency with group sampling is expressed by \eqref{eq: effic OR LOR gr}, as for OR. The results, plotted in Figure~\ref{fig: efg_LOR_1_cho}, are not the same as in that case (Figure~\ref{fig: efg_OR_1_cho}) because of the differences in $\cerr_1$ and $\suf_1$, which affect $\E[\vagr]$ and thus the efficiency; but they are seen to be very similar.

\section{Conclusions}
\label{part: concl}

Two-stage sequential methods have been presented to estimate the RR $p_1/p_2$, the OR $p_1(1-p_2)/(p_2(1-p_1))$ or their logarithmic versions using independent binary observations
from two populations with parameters $p_1$ and $p_2$. The estimators are unbiased; guarantee that the relative mean square error, or the mean square error for the logarithmic versions, is less than a target value $\tarvar$ irrespective of $p_1 $ and $p_2$; and approximately achieve a prescribed ratio of average sample sizes when samples are taken from each population individually (element sampling). The estimators can also be used with group sampling. In this case,
samples are taken simultaneously from the two populations in fixed-size groups, and individual samples are extracted from those groups as needed, with a number of samples possibly discarded at the end of the process. The properties of unbiasedness and guaranteed accuracy are maintained with group sampling (and an exact sample size ratio is imposed by the sampling). Bounds and approximate expressions have been derived for the average sample sizes and the average number of sample groups, respectively. The estimation efficiency, defined in terms of the \CR{} bound, has been characterized in the same way. The efficiency is generally good, both with element sampling and with group sampling; and is close to $1$ for small $\tarvar$. Algorithms~\ref{algo: RR LRR} and \ref{algo: OR LOR}  (Appendix~\ref{part: app algo}) specify the estimation procedure and summarize the properties of the estimators.

The described method can be extended to estimate other functions of $p_1$ and $p_2$, provided that an error function can be defined as in \eqref{eq: ferr} and \eqref{eq: cond on cerr}, and that a Bernoulli factory can be found, if needed, to generate samples with parameters equal to the probabilities to which $\E[\vasas_1]$ and $\E[\vasas_2]$ are approximately inversely proportional (as was done for OR and LOR), where $\vasas_1$ and $\vasas_2$ are the numbers of observations required in the second stage. In addition, the Bernoulli factory needs to have an average number of inputs per output equal to a constant $\bfio$.

As an example, consider the estimation of $p_1 p_2$. A straightforward approach would be to generate samples with parameter $p_1 p_2$ from samples with parameters $p_1$ and $p_2$, using a procedure analogous to the Bernoulli factory described in Section~\ref{part: OR}, and then apply IBS to the generated samples.
However, this does not give any control on the proportion of samples from the two populations, and the average number of samples required from each population scales with the inverse of $p_1 p_2$. Instead, the method presented for RR and LRR can be used: two IBS processes with parameters $\sus_1$ and $\sus_2$ are respectively applied to the two populations (second stage), which requires $\vasas_1$ and $\vasas_2$ samples, where $\sus_1$ and $\sus_2$ are obtained from a previous pair of IBS processes with fixed parameters $\suf_1$ and $\suf_2$ (first stage). Then,
\[
\frac{(\sus_1-1)(\sus_2-1)}{(\vasas_1-1)(\vasas_2-1)}
\]
is an unbiased estimator of $p_1 p_2$, and by means of \eqref{eq: neg bin: Var inv minus 1} a target relative error can be guaranteed. The same expressions as in RR and LRR apply for the average sample sizes, average number of groups and efficiency. In particular, the average sample sizes with element sampling are approximately in the desired ratio, and, for $p_1/p_2$ fixed, they scale with the inverse of $\sqrt{p_1 p_2}$.

\appendix

\section{Estimation procedure and properties of the estimators}
\label{part: app algo}

The estimation procedure for RR and LRR, and that for OR and LOR, are respectively described in Algorithms~\ref{algo: RR LRR} and \ref{algo: OR LOR}. The properties of the estimators are also summarized.


\begin{algorithm}
\caption{Estimator of RR or LRR}
\label{algo: RR LRR}
\algfontsize

\vspace{\algvertspace}%
\textbf{Input parameters}:

\begin{enumerate}
\algfontsize
\item[]
\vspace{-\topsep}%
Target $\tarvar$, interpreted as relative MSE for RR, or MSE for LRR.
\item[]
Desired ratio of average sample sizes $\tarsara$, or group sizes $\tarsagr_1$, $\tarsagr_2$.
\vspace{-\topsep}%
\end{enumerate}

\vspace{\algvertspace}%
\textbf{Estimation procedure}:

\begin{enumerate}
\algfontsize
\item
\vspace{-\topsep}%
Define $\cerr_1, \cerr_2, \cerr_{12}$ as in \eqref{eq: cerr RR} for RR, or as in \eqref{eq: cerr LRR} for LRR.

Define $\cdesm_1 = \cdesm_2 = 1/2$ and $\susrou=1$.

Define $\tarsara = \tarsagr_1/\tarsagr_2$ if group sampling is applied.

\item
Compute $\suf_1$ from \eqref{eq: cho suf} with $\fcurv(\tarvar, \suf_1, \susrou)$ given by \eqref{eq: fcurv RR LRR} (for LRR the explicit expression \eqref{eq: cho suf LRR} can equivalently be used), or obtain it by rounding up the value from Figure~\ref{fig: curv_zero_1}, with a minimum of $3$. Compute $\suf_2$ as $\suf_1+\cerr_1-\cerr_2$.
\item
Compute $\cdemul$ and $\cdeadd_1$ from \eqref{eq: cdemul} and \eqref{eq: cdeadd RR LRR}.
\item
(\emph{First sampling stage}): For each $i=1,2$, repeatedly sample from population~$i$ until $\suf_i$ successes are obtained. Let $\vasaf_i$  be the number of samples.
\item
Compute $\varaf$ from \eqref{eq: varaf RR LRR}. Compute $\sus_1$, $\sus_2$ from \eqref{eq: nsus 1 bis}--\eqref{eq: disc bis}, 
and then round one of them up and the other down, or both up, to fulfill \eqref{eq: ferr leq tarvar}.
\item 
(\emph{Second sampling stage}): For each $i=1,2$, repeatedly sample from population~$i$ until $\sus_i$ successes are obtained. Let $\vasas_i$  be the number of samples.
\item
Compute $\vahRR$ from \eqref{eq: vahRR}, or $\vahLRR$ from \eqref{eq: vahLRR}.
\vspace{-\topsep}%
\end{enumerate}

\vspace{\algvertspace}%
\textbf{Output}:

\begin{enumerate}
\algfontsize
\vspace{-\topsep}%
\item[]
Estimation $\vahRR$ or $\vahLRR$.
\vspace{-\topsep}%
\end{enumerate}

\vspace{\algvertspace}%
\textbf{Properties}:

\begin{enumerate}
\algfontsize
\vspace{-\topsep}%
\item[]
Unbiased, with relative MSE (RR) or MSE (LRR) less than $\tarvar$.
\item[]
Average sample sizes: \eqref{eq: E vasa 1 norm res RR LRR <}, \eqref{eq: E vasa 2 norm res RR LRR <}; approximate ratio $\tarsara$.
\item[]
Efficiency with element sampling: \eqref{eq: effic RR LRR sep bound}; approaches $1$ for $\tarvar$, $\roprpr$ small.
\item[]
Average number of groups: \eqref{eq: varafnbor RR LRR bis}, \eqref{eq: gr approx RR LRR ter}.
\item[]
Efficiency with group sampling: \eqref{eq: effic RR LRR gr}.
\vspace{-\topsep}%
\end{enumerate}

\vspace{\algvertspace}%

\end{algorithm}

\begin{algorithm}
\caption{Estimator of OR or LOR}
\label{algo: OR LOR}
\algfontsize

\vspace{\algvertspace}%
\textbf{Input parameters}:

\begin{enumerate}
\algfontsize
\item[]
\vspace{-\topsep}%
Target $\tarvar$, interpreted as relative MSE for OR, or MSE for LOR.
\item[]
Desired ratio of average sample sizes $\tarsara$, or group sizes $\tarsagr_1$, $\tarsagr_2$.
\vspace{-\topsep}%
\end{enumerate}

\vspace{\algvertspace}%
\textbf{Estimation procedure}:

\begin{enumerate}
\algfontsize
\item
\vspace{-\topsep}%
Define $\cerr_1, \cerr_2, \cerr_{12}$ as in \eqref{eq: cerr OR} for OR, or as in \eqref{eq: cerr LOR} for LOR.

Define $\cdesm_1 = \cdesm_2 = 1/2$ and $\susrou=1$.

Define $\adjsus=2$ for OR, or $\adjsus=0$ for LOR.

Define $\tarsara = \tarsagr_1/\tarsagr_2$ if group sampling is applied.

\item
Compute $\suf_1$ from \eqref{eq: cho suf} with $\fcurv(\tarvar, \suf_1, \susrou)$ given by \eqref{eq: fcurv OR LOR} (for LOR the explicit expression \eqref{eq: cho suf LOR} can equivalently be used), or obtain it by rounding up the value from Figure~\ref{fig: curv_zero_1}, with a minimum of $3$. Set $\suf_2 = \suf_1$.
\item
Set $\cdemul = \tarsara$ and compute $\cdeadd_1$ from \eqref{eq: cdeadd OR LOR}.
\item
(\emph{First sampling stage}): For each $i=1,2$, using the method below, generate samples with parameter $\pco p_i$ until $\suf_i$ successes are obtained. Let $\pco \vasaf_i$ be the number of generated samples.

To generate a sample with parameter $\pco p_i$,
choose one of these options equally likely:%
\begin{itemize}
\algfontsize
\item[\bfa]
\vspace{-\topsep}%
Take a sample from population $i$. If it is a failure, output failure. Else, take another sample from population $i$ and output the opposite of its value.

\item[\bfb]
Take a sample from population $i$. If it is a success, output failure. Else, take another sample from population $i$ and output its value.
\vspace{-\topsep}%
\end{itemize}

Let $\vasaf_i$ be the total number of samples used from population $i$.
\item
Compute $\varaf$ from \eqref{eq: varaf OR LOR}. Compute $\sus_1$, $\sus_2$ from \eqref{eq: nsus 1 bis}--\eqref{eq: disc bis}, 
and then round one of them up and the other down, or both up, to fulfill \eqref{eq: ferr leq tarvar}.
\item 
(\emph{Second sampling stage}):
From population~$1$, take as many samples as needed, $\vasas'_1$, to obtain $\sus_1$ successes; then as many as needed, $\vasas''_1$, to obtain ${\sus_1-\adjsus}$ failures. From population~$2$, take as many samples as needed, ${\vasas'_2}$, to obtain ${\sus_2-\adjsus}$ successes; then as many as needed, $\vasas''_2$, to obtain $\sus_2$ failures.
Let $\vasas_1 = \vasas_1'+\vasas_1''$ and $\vasas_2 = \vasas_2'+\vasas_2''$.
\item
Compute $\vahOR$ from \eqref{eq: vahOR}, or $\vahLOR$ from \eqref{eq: vahLOR}.
\vspace{-\topsep}%
\end{enumerate}

\vspace{\algvertspace}%
\textbf{Output}:

\begin{enumerate}
\algfontsize
\vspace{-\topsep}%
\item[]
Estimation $\vahOR$ or $\vahLOR$.
\vspace{-\topsep}%
\end{enumerate}

\vspace{\algvertspace}%
\textbf{Properties}:

\begin{enumerate}
\algfontsize
\vspace{-\topsep}%
\item[]
Unbiased, with relative MSE (OR) or MSE (LOR) less than $\tarvar$.
\item[]
Average sample sizes: \eqref{eq: E vasa 1 norm res OR LOR <}, \eqref{eq: E vasa 2 norm res OR LOR <}; approximate ratio $\tarsara$.
\item[]
Efficiency with element sampling: \eqref{eq: effic OR LOR sep bound}; approaches $1$ for $\tarvar$ small.
\item[]
Average number of groups: \eqref{eq: gr approx OR LOR ter}, \eqref{eq: varafnbor OR LOR bis}.
\item[]
Efficiency with group sampling: \eqref{eq: effic OR LOR gr}.
\vspace{-\topsep}%
\end{enumerate}

\vspace{\algvertspace}%

\end{algorithm}

\section{Approximation of $\E[|\diffgr|]$}

\subsection{For relative risk and log relative risk}
\label{part: app RR LRR gr var}

The random variable $\vasaf_1$ in RR and LRR has a negative binomial distribution with parameters $\suf_1$ and $p_1 = \roprpr \sqrt{\rapr}$. For $\roprpr \rightarrow 0$, the distribution of $\vasaf_1\roprpr$ tends to a gamma distribution with location parameter $\suf_1$ and scale parameter $1/\sqrt{\rapr}$, because 
\begin{equation}
\begin{split}
\lim_{\roprpr \rightarrow 0} \Pr[\vasaf_1\roprpr \leq \nug] &=
\lim_{\roprpr \rightarrow 0} \sum_{k=\suf_1}^{\lfloor \nug/\roprpr \rfloor} \binom{k-1}{\suf_1-1} \left(\roprpr\sqrt{\rapr}\right)^{\suf_1} \left(1-\roprpr\sqrt{\rapr}\right)^{k-\suf_1} \\
&= \frac {\rapr^{\suf_1/2}} {(\suf_1-1)!} \int_0^\nug \vg^{\suf_1-1} \exp(-\vg\sqrt{\rapr}) \, \diff\mkern .8mu\vg.
\end{split}
\end{equation}
Likewise, the distribution of $\vasaf_2 \roprpr$ tends to that of a gamma random variable with location parameter $\suf_2$ and scale parameter $\sqrt{\rapr}$. By the continuous mapping theorem \citep[theorem 2.3]{vanderVaart98}, the variable $\varaf$ defined in \eqref{eq: varaf RR LRR} converges in distribution to the ratio of these two gamma random variables,
and therefore $\varafn = \varaf/\rapr$ and $1/\varafn$ converge in distribution to beta prime random variables with parameters $\suf_2$, $\suf_1$ and $\suf_1$, $\suf_2$ respectively \citep[section~4.4]{Chattamvelli21a}. It is easy to see that $\varafn^2$ and $1/\varafn^2$ are uniformly integrable as $\roprpr \rightarrow 0$,
which implies \citep[theorem~25.12]{Billingsley95} that the variances of $\varafn$ and $1/\varafn$ converge to those of the referred beta prime distributions. Thus, approximating $\Var[\varaf]$ and $\Var[1/\varaf]$ by their values for $\roprpr \rightarrow 0$, the following expressions are obtained:
\begin{align}
\label{eq: Var varaf approx RR LRR}
\Var[\varaf] &\approx \frac{\suf_2(\suf_1+\suf_2-1) \rapr^2 }{(\suf_1-2)(\suf_1-1)^2}, \\
\label{eq: Var 1/varaf approx RR LRR}
\Var\left[\frac 1 {\varaf} \right] &\approx \frac{\suf_1(\suf_1+\suf_2-1)}{(\suf_2-2)(\suf_2-1)^2 \rapr^2}.
\end{align}

The value of $\E[|\diffgr|]$ obviously depends both on the mean of $\diffgr$ and on the variability of $\diffgr$ with respect to its mean. In view of \eqref{eq: diffgr}, this variability can be understood as arising from three sources: \dgvi{} the variability of $\vasaf_1$ and $\vasaf_2$ as they \emph{directly} appear in that expression; \dgvii{} the variability of $\vasas_1$ and $\vasas_2$ conditioned on $\sus_1$ and $\sus_2$; and \dgviii{} the variability of $\vasas_1$ and $\vasas_2$ caused by variations of $\sus_1$ and $\sus_2$ (since the values of $\sus_1$ and $\sus_2$ are obtained from $\varaf$ this is another, \emph{indirect} effect of $\vasaf_1$ and $\vasaf_2$). As will be seen, the contributions of the first two sources of variability are small compared to that of the third, and can be neglected with little error. This statement can be made more precise by expressing it in terms of variance. Applying the law of total variance \cite[theorem~12.2.6]{Athreya06},
\begin{equation}
\label{eq: maxg variability RR}
\textstyle
\Var\left[\diffgr \right]
=
\E\left[\Var\left[\diffgr \cond \vasaf_1, \vasaf_2 \right]\right] + \Var\left[\E\left[\diffgr \cond \vasaf_1, \vasaf_2 \right]\right].
\end{equation}
Substituting \eqref{eq: diffgr} into \eqref{eq: maxg variability RR}, and noting that $\vasas_1$, $\vasas_2$ depend on $\vasaf_1$, $\vasaf_2$ only through $\sus_1$ and $\sus_2$,
\begin{equation}
\label{eq: maxg variability RR bis}
\begin{split}
\displaystyle
\Var\left[\diffgr \right] &=
\Var\left[\frac{\vasaf_1}{\tarsagr_1} - \frac{\vasaf_2}{\tarsagr_2}\right] +
\E\left[\Var\left[\frac{\vasas_1}{\tarsagr_1} - \frac{\vasas_2}{\tarsagr_2} \cond* \sus_1, \sus_2 \right]\right]
+ \Var\left[\E\left[\frac{\vasas_1}{\tarsagr_1} - \frac{\vasas_2}{\tarsagr_2} \cond* \sus_1, \sus_2 \right]\right] \\
& \quad + 2\Cov\left[ \frac{\vasaf_1}{\tarsagr_1} - \frac{\vasaf_2}{\tarsagr_2},\, \E\left[\frac{\vasas_1}{\tarsagr_1} - \frac{\vasas_2}{\tarsagr_2} \cond* \sus_1, \sus_2 \right] \right].
\end{split}
\end{equation}
The first three summands on the right-hand side of \eqref{eq: maxg variability RR bis} correspond to \dgvi{}, \dgvii{} and \dgviii{} respectively, and the fourth represents the statistical relationship between \dgvi{} and \dgviii{} (both of which stem from the variability of $\vasaf_1$ and $\vasaf_2$).

The first summand in \eqref{eq: maxg variability RR bis} can be written, taking into account that $\vasaf_1$ and $\vasaf_2$ are independent, as $\Var[\vasaf_1/\tarsagr_1] + \Var[\vasaf_2/\tarsagr_2]$, and then applying \eqref{eq: neg bin: Var} gives
\begin{equation}
\label{eq: Var vasaf RR}
\Var\left[\frac{\vasaf_1}{\tarsagr_1} - \frac{\vasaf_2}{\tarsagr_2}\right] \roprpr^2\tarsapr = \frac{\suf_1(1-p_1)}{\tarsara \rapr}  + \suf_2(1-p_2) \tarsara\rapr \approx \frac{\suf_1}{\tarsara \rapr} + (\suf_1+\cerr_1-\cerr_2) \tarsara\rapr.
\end{equation}

The second summand is computed as follows. Noting that $\vasas_1$ and $\vasas_2$ are conditionally independent given $\sus_1$ and $\sus_2$, the conditional variance $\Var[\vasas_1/\tarsagr_1 - \vasas_2/\tarsagr_2 \cond \sus_1, \sus_2]$ can be expressed as $\Var[\vasas_1/\tarsagr_1 \cond \sus_1] + \Var[\vasas_2/\tarsagr_2 \cond \sus_2]$, and using \eqref{eq: neg bin: Var} again yields
\begin{equation}
\label{eq: E Var vasas cond RR}
\begin{split}
\E\left[\Var\left[\frac{\vasas_1}{\tarsagr_1} - \frac{\vasas_2}{\tarsagr_2} \cond* \sus_1, \sus_2 \right]\right] \roprpr^2\tarsapr
&= \E\left[ \frac{\sus_1 (1-p_1)}{\tarsara\rapr} \right] + \E\left[ \sus_2 (1-p_2)\tarsara\rapr \right] \\
&\approx \frac{\E[\sus_1]}{\tarsara\rapr} + \E[\sus_2] \tarsara\rapr.
\end{split}
\end{equation}
Then, from \eqref{eq: nsus 1 approx}--\eqref{eq: csufva 2}, 
\eqref{eq: E varaf approx RR LRR}, \eqref{eq: E 1/varaf approx RR LRR} and \eqref{eq: suf 1 suf 2, cdeadd 1 cdeadd 2}--\eqref{eq: cdeadd RR LRR}, 
and including the rounding term $\susrou$ when computing $\E[\sus_1]$ and $\E[\sus_2]$,
\begin{equation}
\label{eq: E Var vasas cond RR bis}
\begin{split}
\E\left[\Var\left[\frac{\vasas_1}{\tarsagr_1} - \frac{\vasas_2}{\tarsagr_2} \cond* \sus_1, \sus_2 \right]\right] \roprpr^2\tarsapr &\approx
\frac{1}{\tarsara \rapr} \left(\frac 1 {\tarvar} + \cerr_1 + \susrou\right)
+ \tarsara \rapr \left(\frac 1 {\tarvar} + \cerr_2 + \susrou\right) \\
& \quad + 2\left(\frac 1 {\tarvar} + \suf_1 + \cerr_1 + \susrou \right).
\end{split}
\end{equation}

As for the third summand, from \eqref{eq: neg bin: E}, \eqref{eq: nsus 1 approx} and \eqref{eq: nsus 2 approx} it can be written as
\begin{multline}
\label{eq: Var E vasas cond RR}
\Var\left[\E\left[\frac{\vasas_1}{\tarsagr_1} - \frac{\vasas_2}{\tarsagr_2} \cond* \sus_1, \sus_2 \right]\right]  \roprpr^2\tarsapr =
\Var\left[\frac{\sus_1}{\tarsagr_1 p_1} - \frac{\sus_2}{\tarsagr_2 p_2}\right] \roprpr^2\tarsapr \\
\approx \frac {\csusva_1^2 \Var[\varaf]} {\tarsara \rapr} + \tarsara \rapr \csusva_2^2 \Var\left[\frac 1 {\varaf} \right] + 2\csusva_1 \csusva_2\Cov\left[\varaf, -\frac{1}{\varaf}\right].
\end{multline}
The term $\Cov[\varaf, -1/\varaf]$ is easily obtained using \eqref{eq: E varaf approx RR LRR}, \eqref{eq: E 1/varaf approx RR LRR} and \eqref{eq: suf 1 suf 2, cdeadd 1 cdeadd 2}:
\begin{equation}
\label{eq: Cov varaf -1/varaf RR}
\Cov\left[\varaf, -\frac{1}{\varaf}\right] = -1 + \E\left[\varaf\right] \E\left[\frac 1 {\varaf}\right] \approx
\frac{\suf_1 + \suf_2- 1}{(\suf_1-1)(\suf_2-1)}.
\end{equation}
Substituting \eqref{eq: csufva 1}, \eqref{eq: csufva 2}, \eqref{eq: Var varaf approx RR LRR}, \eqref{eq: Var 1/varaf approx RR LRR} and \eqref{eq: Cov varaf -1/varaf RR} into \eqref{eq: Var E vasas cond RR} and using \eqref{eq: suf 1 suf 2, cdeadd 1 cdeadd 2}--\eqref{eq: cdeadd RR LRR},
\begin{multline}
\label{eq: Var E vasas cond RR bis}
\Var\left[\E\left[\frac{\vasas_1}{\tarsagr_1} - \frac{\vasas_2}{\tarsagr_2} \cond* \sus_1, \sus_2 \right]\right]  \roprpr^2\tarsapr \approx
\left(\frac 1 {\tarvar} + \suf_1 + \cerr_1 + \susrou \right)^{\!2} (2\suf_1+\cerr_1-\cerr_2-1) \\
\cdot \left( \frac{\tarsara\rapr}{(\suf_1-2)(\suf_1+\cerr_2-\cerr_1)} + \frac{1}{\tarsara\rapr\suf_1(\suf_1+\cerr_2-\cerr_1-2)} + \frac 2 {\suf_1 (\suf_1+\cerr_2-\cerr_1)} \right).
\end{multline}

Lastly, the fourth summand in \eqref{eq: maxg variability RR bis} can be bounded using the Cauchy-Schwarz inequality~\cite[proposition 6.2.8]{Athreya06}:
\begin{multline}
\label{eq: Cov vasaf vasas RR}
\left| 2 \Cov\left[ \frac{\vasaf_1}{\tarsagr_1} - \frac{\vasaf_2}{\tarsagr_2},\, \E\left[\frac{\vasas_1}{\tarsagr_1} - \frac{\vasas_2}{\tarsagr_2} \cond* \sus_1, \sus_2 \right] \right] \right| \\
\leq 2 \sqrt{ \Var\left[\frac{\vasaf_1}{\tarsagr_1} - \frac{\vasaf_2}{\tarsagr_2}\right]
\Var\left[\E\left[\frac{\vasas_1}{\tarsagr_1} - \frac{\vasas_2}{\tarsagr_2} \cond* \sus_1, \sus_2 \right]\right] }.
\end{multline}

It will be established next that \eqref{eq: Var vasaf RR}, \eqref{eq: E Var vasas cond RR bis} and the right-hand side of \eqref{eq: Cov vasaf vasas RR} are much smaller than \eqref{eq: Var E vasas cond RR bis}. To this end, the following observations will be useful. From \eqref{eq: suf 1 suf 2, cdeadd 1 cdeadd 2} and \eqref{eq: cond on cerr} it stems that $\suf_1 \approx \suf_2 = \suf_1+\cerr_1-\cerr_2$ with little approximation error. In practice, the target $\tarvar$ will typically be smaller, or much smaller, than $1$. For example, a relative RMSE of $10\%$ corresponds to $\tarvar = 0.01$. On the other hand, $\susrou = 1$ is small compared to $1/\tarvar$, and so are the values of $\suf_1$ obtained from \eqref{eq: cho suf}; namely $\suf_1 \approx \sqrt{1/A}$, as can be seen from \eqref{eq: curv zero bis RR LRR} or in Figure~\ref{fig: curv_zero_1}. In the example, $\tarvar = 0.01$ gives $\suf_1 = 9$ for RR and $10$ for LRR.

To show that \eqref{eq: Var vasaf RR} is much smaller than \eqref{eq: Var E vasas cond RR bis}, it is convenient to study the cases $\tarsara \rapr \approx 1$, $\tarsara \rapr \gg 1$ and $\tarsara \rapr \ll 1$ separately. For $\tarsara \rapr \approx 1$, the approximations in the above paragraph imply that \eqref{eq: Var vasaf RR} and \eqref{eq: Var E vasas cond RR bis} reduce to $2\suf_1$ and $8(1/\tarvar + \suf_1)^2/\suf_1$ respectively, and their ratio,
\[
\frac{\suf_1^2}{4(1/\tarvar + \suf_1)^2},
\]
is much smaller than $1$. For $\tarsara \rapr \gg 1$, \eqref{eq: Var vasaf RR} and \eqref{eq: Var E vasas cond RR bis} are approximated as $\suf_1 \tarsara \rapr$ and $2\tarsara \rapr(1/\tarvar + \suf_1)^2/\suf_1$ respectively, and their ratio is
\[
\frac{\suf_1^2}{2(1/\tarvar + \suf_1)^2},
\]
which is again small compared with $1$. The case $\tarsara \rapr \ll 1$ gives the same result. Thus, for any value of $\tarsara \rapr$, the first summand in \eqref{eq: maxg variability RR bis} can be approximately neglected in comparison with the third.

Proceeding analogously to compare \eqref{eq: E Var vasas cond RR bis} with \eqref{eq: Var E vasas cond RR bis}, for $\tarsara \rapr \approx 1$ their approximate values are $4/\tarvar + 2\suf_1$ and $8(1/\tarvar + \suf_1)^2/\suf_1$ respectively, with a ratio
\[
\frac{ (4/\tarvar + 2\suf_1)\suf_1 } { 8(1/\tarvar + \suf_1)^2 } = \frac{ (1/\tarvar + \suf_1/2)\suf_1 } { 2(1/\tarvar + \suf_1)^2 } < \frac{ \suf_1 } { 2(1/\tarvar + \suf_1) },
\]
which is significantly smaller than $1$. For $\tarsara \rapr \gg 1$, \eqref{eq: E Var vasas cond RR bis} and \eqref{eq: Var E vasas cond RR bis} are approximated as $\tarsara \rapr / \tarvar$ and $2\tarsara \rapr(1/\tarvar + \suf_1)^2/\suf_1$ respectively, and this gives a ratio
\[
\frac{ (1 / \tarvar)\suf_1 } { 2(1/\tarvar + \suf_1)^2 } < \frac{ \suf_1 } { 2(1/\tarvar + \suf_1) },
\]
which is again small compared with $1$. For $\tarsara \rapr \ll 1$ the result is the same. Thus the second summand in \eqref{eq: maxg variability RR bis} can also, to a good approximation, be neglected in comparison with the third.

As for \eqref{eq: Cov vasaf vasas RR}, dividing its right-hand side by the left-hand side of \eqref{eq: Var E vasas cond RR bis} gives twice the square root of the ratio between \eqref{eq: Var vasaf RR} and \eqref{eq: Var E vasas cond RR bis}.
Therefore the fourth summand in \eqref{eq: maxg variability RR bis} is also significantly smaller than the third.

The conclusion of the preceding analysis is that \eqref{eq: maxg variability RR bis} can be approximated by keeping only the third summand in the right-hand side, as it is significantly larger than the others. This means that the variability in $\diffgr$ is mostly due to the variability of $\vasas_1$ and $\vasas_2$ caused by variations of $\sus_1$ and $\sus_2$, i.e.~\dgviii{} as defined above. The variability of $\vasaf_1$ and $\vasaf_2$, \dgvi{}, and that of $\vasas_1$ and $\vasas_2$ conditioned on $\sus_1$ and $\sus_2$, \dgvii{}, are comparatively smaller. Therefore, to compute $\E[|\diffgr|]$, the variables $\vasaf_1$, $\vasaf_2$ in \eqref{eq: diffgr} can be replaced by their means, and $\vasas_1$, $\vasas_2$ can be replaced by their conditional means given $\sus_1$, $\sus_2$, which yields \eqref{eq: pre diffgr RR}.

\subsection{For odds ratio and log odds ratio}
\label{part: app OR LOR gr var}

The law of total variance \cite[theorem~12.2.6]{Athreya06} conditioning on $\pco\vasaf_1$ and $\pco\vasaf_2$ gives, for OR and LOR,
\begin{equation}
\label{eq: maxg variability OR}
\textstyle
\Var\left[\diffgr \right]
=
\E\left[\Var\left[\diffgr \cond* \pco\vasaf_1, \pco\vasaf_2 \right]\right] + \Var\left[\E\left[\diffgr \cond* \pco\vasaf_1, \pco\vasaf_2 \right]\right].
\end{equation}
The variables $\vasas_1'$, $\vasas_1''$, $\vasas_2'$, $\vasas_2''$, and therefore $\vasas_1$, $\vasas_2$, depend on $\vasaf_1$, $\vasaf_2$ only through $\pco \vasaf_1$, $\pco \vasaf_2$. This implies that $\vasaf_1$, $\vasaf_2$, $\vasas_1$, $\vasas_2$, are conditionally independent given $\pco\vasaf_1$, $\pco\vasaf_2$. Thus, using \eqref{eq: diffgr}, the first term in the right-hand side of \eqref{eq: maxg variability OR} is written as
\begin{equation}
\label{eq: E Var diffgr cond OR}
\displaystyle
\E\left[\Var\left[ \diffgr \cond* \pco\vasaf_1, \pco\vasaf_2 \right]\right] =
\E\left[\Var\left[ \frac{\vasaf_1}{\tarsagr_1}-\frac{\vasaf_2}{\tarsagr_2} \cond* \pco\vasaf_1, \pco\vasaf_2 \right]\right] +
\E\left[\Var\left[ \frac{\vasas_1}{\tarsagr_1}-\frac{\vasas_2}{\tarsagr_2} \cond* \pco\vasaf_1, \pco\vasaf_2 \right]\right],
\end{equation}
whereas the second term is
\begin{equation}
\label{eq: Var E diffgr cond OR}
\begin{split}
\displaystyle
\Var\left[\E\left[ \diffgr \cond* \pco\vasaf_1, \pco\vasaf_2 \right]\right] &=
\Var\left[\E\left[ \frac{\vasaf_1}{\tarsagr_1} - \frac{\vasaf_2}{\tarsagr_2} \cond* \pco\vasaf_1, \pco\vasaf_2 \right]\right]
+ \Var\left[\E\left[ \frac{\vasas_1}{\tarsagr_1} - \frac{\vasas_2}{\tarsagr_2} \cond* \pco\vasaf_1, \pco\vasaf_2 \right]\right] \\
&\quad + 2\Cov\left[\E\left[ \frac{\vasaf_1}{\tarsagr_1} - \frac{\vasaf_2}{\tarsagr_2} \cond* \pco\vasaf_1, \pco\vasaf_2 \right]\!\!,\, \E\left[ \frac{\vasas_1}{\tarsagr_1} - \frac{\vasas_2}{\tarsagr_2} \cond* \pco\vasaf_1, \pco\vasaf_2 \right] \right].
\end{split}
\end{equation}
On the other hand, making use of the law of total variance again,
\begin{equation}
\label{eq: E Var vasaf cond OR}
\E\left[\Var\left[ \frac{\vasaf_1}{\tarsagr_1}-\frac{\vasaf_2}{\tarsagr_2} \cond* \pco\vasaf_1, \pco\vasaf_2 \right]\right] + \Var\left[\E\left[ \frac{\vasaf_1}{\tarsagr_1}-\frac{\vasaf_2}{\tarsagr_2} \cond* \pco\vasaf_1, \pco\vasaf_2 \right]\right]
= \Var\left[ \frac{\vasaf_1}{\tarsagr_1}-\frac{\vasaf_2}{\tarsagr_2} \right].
\end{equation}
Combining \eqref{eq: E Var diffgr cond OR}--\eqref{eq: E Var vasaf cond OR} with \eqref{eq: maxg variability OR}, and noting that conditioning $\vasas_1$ or $\vasas_2$ on $\pco \vasaf_1$, $\pco \vasaf_2$ is equivalent to conditioning on $\sus_1$, $\sus_2$, the following expression is obtained for $\Var[\diffgr]$:
\begin{equation}
\label{eq: maxg variability OR bis}
\begin{split}
\displaystyle
\Var\left[\diffgr \right] &=
\Var\left[ \frac{\vasaf_1}{\tarsagr_1}-\frac{\vasaf_2}{\tarsagr_2} \right]
+ \E\left[\Var\left[ \frac{\vasas_1}{\tarsagr_1}-\frac{\vasas_2}{\tarsagr_2} \cond* \sus_1, \sus_2 \right]\right]
+ \Var\left[\E\left[ \frac{\vasas_1}{\tarsagr_1} - \frac{\vasas_2}{\tarsagr_2} \cond* \sus_1, \sus_2 \right]\right] \\
&\quad + 2\Cov\left[\E\left[ \frac{\vasaf_1}{\tarsagr_1} - \frac{\vasaf_2}{\tarsagr_2} \cond* \pco\vasaf_1, \pco\vasaf_2 \right]\!\!,\, \E\left[ \frac{\vasas_1}{\tarsagr_1} - \frac{\vasas_2}{\tarsagr_2} \cond* \sus_1, \sus_2 \right] \right].
\end{split}
\end{equation}
This is similar to the decomposition \eqref{eq: maxg variability RR bis} in RR and LRR, where only the covariance term is different. Although the approximate expressions of the summands for small $\roprpr$, to be computed next, are different from that case, it will be shown that for OR and LOR the third summand again dominates the other three.

The first summand in \eqref{eq: maxg variability OR bis} is written, thanks to the independence of $\vasaf_1$ and $\vasaf_2$, as 
\begin{equation}
\label{eq: Var vasaf OR}
\Var\left[\frac{\vasaf_1}{\tarsagr_1} - \frac{\vasaf_2}{\tarsagr_2}\right] =
\frac{\Var[\vasaf_1]}{\tarsagr_1^2} + \frac{\Var[\vasaf_2]}{\tarsagr_2^2}.
\end{equation}
From the law of total variance, $\Var[\vasaf_i]$, $i=1,2$ is expressed as
\begin{equation}
\label{eq: vasaf LTV pco}
\Var\left[ \vasaf_i \right] = \E\left[ \Var\left[\vasaf_i \cond \pco\vasaf_i\right] \right]
+ \Var\left[ \E\left[\vasaf_i \cond \pco\vasaf_i\right] \right].
\end{equation}
The first stage applies IBS to samples with parameter $\pco p_i$, using a number $\pco \vasaf_i$ of those samples, of which $\suf_i$ are successes. Each sample with parameter $\pco p_i$ is generated by the Bernoulli factory described in Section~\ref{part: OR sep}, taking samples with parameter $p_i$ as inputs. With this factory, the average number of inputs needed to produce an output is $3/2$; a success output always uses $2$ inputs, and a failure output uses either $1$ or $2$ inputs. Let $\pbfe_i$ denote the probability that $2$ inputs are used, conditioned on the output being a failure. Then $3 / 2 = 2 \pco p_i + (1+\pbfe_i)(1-\pco p_i)$, from which
\begin{equation}
\label{eq: pbfe}
\pbfe_i = 1 - \frac{1}{2(1-\pco p_i)} = \frac{1 - 2\pco p_i}{2(1-\pco p_i)}.
\end{equation}
Thus, conditioned on $\pco\vasaf_i$, the average number of required inputs is
\begin{equation}
\label{eq: E vasaf cond pco vasaf}
\E[\vasaf_i \cond \pco\vasaf_i] = 2 \suf_i + (1+\pbfe_i)(\pco \vasaf_i-\suf_i) =
\frac{(3-4\pco p_i) \pco\vasaf_i + \suf_i}{2(1-\pco p_i)}.
\end{equation}
The term $\Var[\vasaf_i \cond \pco\vasaf_i]$ in \eqref{eq: vasaf LTV pco} equals the variance of a binomial random variable with parameters $\pco\vasaf_i-\suf_i$ and $\pbfe_i$. Therefore, computing $\E[\pco \vasaf_i]$ from \eqref{eq: neg bin: E} and substituting \eqref{eq: pbfe},
\begin{equation}
\label{eq: E Var vasaf con pco}
\E\left[ \Var\left[\vasaf_i \cond* \pco\vasaf_i\right] \right] =
\pbfe_i(1-\pbfe_i) \E[ \pco\vasaf_i-\suf_i] = 
\frac {(1-2\pco p_i) \suf_i} {4\pco p_i(1-\pco p_i)}.
\end{equation}
On the other hand, using \eqref{eq: E vasaf cond pco vasaf} and computing $\Var[\pco \vasaf_i]$ from \eqref{eq: neg bin: Var},
\begin{equation}
\label{eq: Var E vasaf con pco}
\Var\left[ \E\left[\vasaf_i \cond* \pco\vasaf_i\right] \right]
= \frac{(3-4\pco p_i)^2}{4(1-\pco p_i)^2} \Var\left[ \pco\vasaf_i \right]
= \frac { (3-4\pco p_i)^2 \suf_i } { 4\pco p_i^2 (1-\pco p_i) }.
\end{equation}
From \eqref{eq: vasaf LTV pco}, \eqref{eq: E Var vasaf con pco} and \eqref{eq: Var E vasaf con pco},
\begin{equation}
\label{eq: vasaf LTV pco bis}
\Var\left[ \vasaf_i \right]
= \frac { (14\pco p_i^2 - 23\pco p_i + 9) \suf_i } { 4\pco p_i^2 (1-\pco p_i) }
= \frac { (9 - 14\pco p_i) \suf_i } { 4\pco p_i^2 }.
\end{equation}
Thus, for $\roprpr$ small, which implies $\pco\roprpr$ small, substituting \eqref{eq: vasaf LTV pco bis} into \eqref{eq: Var vasaf OR} and using \eqref{eq: suf 1 suf 2, cdeadd 1 cdeadd 2, OR LOR},
\begin{equation}
\label{eq: Var vasaf OR bis}
\Var\left[ \frac{\vasaf_1}{\tarsagr_1} - \frac{\vasaf_2}{\tarsagr_2} \right] \pco\roprpr^2 \tarsapr
= \frac { 9 \suf_1 } 4 \left( \frac {1-14\pco p_1/9} {\tarsara \pco\rapr} + (1-14\pco p_2/9) \tarsara \pco\rapr \right)
\approx \frac { 9 \suf_1 } 4 \left( \frac 1 {\tarsara \pco\rapr} + \tarsara \pco\rapr \right).
\end{equation}

The second summand in \eqref{eq: maxg variability OR bis} is computed as follows. Using the fact that $\vasas_1'$, $\vasas_1''$, $\vasas_2'$, $\vasas_2''$ are conditionally independent given $\sus_1$, $\sus_2$, taking into account \eqref{eq: neg bin: Var} and approximating for $\roprpr$ small,
\begin{equation}
\label{eq: E Var vasas cond OR}
\begin{split}
& \E\left[\Var\left[\frac{\vasas_1}{\tarsagr_1} - \frac{\vasas_2}{\tarsagr_2}\cond* \sus_1, \sus_2 \right]\right] \pco\roprpr^2 \tarsapr \\
&\ = \frac{ \E[\sus_1](1-p_1) \pco\roprpr^2}{\tarsara p_1^2} + \frac{\E[\sus_1-\adjsus] p_1 \pco\roprpr^2}{\tarsara (1-p_1)^2} + \frac{ \tarsara \E[\sus_2-\adjsus](1-p_2) \pco\roprpr^2}{p_2^2} + \frac{\tarsara \E[\sus_2] p_2 \pco\roprpr^2}{(1-p_2)^2} \\
&\ = \left( \frac{\E[\sus_1] ((1-p_1)^3 + p_1^3) - \adjsus p_1^3} {\tarsara \pco p_1^2}
+ \frac{\tarsara\left(\E[\sus_2] ((1-p_2)^3 + p_2^3) - \adjsus (1-p_2)^3\right)} {\pco p_2^2} \right) \pco\roprpr^2 \\
&\ = \frac {\E[\sus_1] (1-3\pco p_1) - \adjsus p_1^3} {\tarsara \pco\rapr} + \tarsara \pco\rapr  \left(\E[\sus_2] (1-3\pco p_2) - \adjsus (1-p_2)^3 \right) \\
&\ \approx \frac {\E[\sus_1]} {\tarsara \pco\rapr} + \tarsara \pco\rapr (\E[\sus_2]- \adjsus ),
\end{split}
\end{equation}
with $\adjsus$ equal to $2$ for OR and $0$ for LOR. Using \eqref{eq: nsus 1 approx} and \eqref{eq: nsus 2 approx}, including the rounding term $\susrou$, and then substituting \eqref{eq: csufco 1}--\eqref{eq: csufva 2}, \eqref{eq: E varaf approx OR LOR}, \eqref{eq: E 1/varaf approx OR LOR} and \eqref{eq: cerr 1 eq cerr 2}--\eqref{eq: cdeadd OR LOR}, \eqref{eq: E Var vasas cond OR} becomes 
\begin{equation}
\begin{split}
\label{eq: E Var vasas cond OR bis}
\E\left[\Var\left[\frac{\vasas_1}{\tarsagr_1} - \frac{\vasas_2}{\tarsagr_2}\cond* \sus_1, \sus_2 \right]\right] \pco\roprpr^2 \tarsapr
&\approx \frac 1 {\tarsara \pco\rapr} \left( \frac 1 {\tarvar} + \cerr_1 + \susrou \right) + \tarsara \pco\rapr \left( \frac 1 {\tarvar} + \cerr_1 + \susrou - \adjsus \right) \\
&\quad + 2 \left(\frac 1 {\tarvar} + \frac{3\suf_1} 2 + \cerr_1 + \susrou\right).
\end{split}
\end{equation}

As for the third summand, using \eqref{eq: E vasa 1 cond OR}--\eqref{eq: pco p} 
and then taking into account \eqref{eq: nsus 1 approx} and \eqref{eq: nsus 2 approx},
\begin{multline}
\label{eq: Var E vasas cond OR}
\Var\left[\E\left[\frac{\vasas_1}{\tarsagr_1} - \frac{\vasas_2}{\tarsagr_2} \cond* \sus_1, \sus_2 \right]\right]  \pco\roprpr^2\tarsapr \\
\approx \frac {\csusva_1^2 \Var[\varaf]} {\tarsara \pco\rapr} + \tarsara \pco\rapr \csusva_2^2 \Var\left[\frac 1 {\varaf} \right] + 2\csusva_1 \csusva_2\Cov\left[\varaf, -\frac{1}{\varaf}\right].
\end{multline}
For $\roprpr$ small, $\Var[\varaf]$, $\Var[1/\varaf]$ and $\Cov[\varaf, -1/\varaf]$ are approximately given by the same expressions \eqref{eq: Var varaf approx RR LRR}, \eqref{eq: Var 1/varaf approx RR LRR} and \eqref{eq: Cov varaf -1/varaf RR} as in RR and LRR except with $\rapr$ replaced by $\pco \rapr$ and with $\suf_1=\suf_2$. According to this, and making use of \eqref{eq: cerr 1 eq cerr 2}--\eqref{eq: cdeadd OR LOR}, 
\eqref{eq: csufva 1} and \eqref{eq: csufva 2}, \eqref{eq: Var E vasas cond OR} yields
\begin{multline}
\label{eq: Var E vasaf cond OR bis}
\Var\left[\E\left[\frac{\vasas_1}{\tarsagr_1} - \frac{\vasas_2}{\tarsagr_2} \cond* \sus_1, \sus_2 \right]\right]  \pco \roprpr^2\tarsapr \\
\approx
\left(\frac 1 {\tarvar} + \frac{3 \suf_1} 2 + \cerr_1 + \susrou \right)^{\!2} (2\suf_1-1)
\left( \frac 1 {\suf_1(\suf_1-2)} \left(\tarsara \pco\rapr + \frac{1}{\tarsara \pco\rapr} \right) + \frac 2 {\suf_1^2} \right).
\end{multline}

The fourth summand in \eqref{eq: maxg variability OR bis} is bounded by means of the Cauchy-Schwarz inequality as in Appendix~\ref{part: app RR LRR gr var}; and then, using \eqref{eq: E vasaf cond pco vasaf} and the fact that $(3-4\pco p_i)/(2(1-\pco p_i))$ is less than $3/2$ (and approaches that value for $\roprpr$ small),
\begin{equation}
\begin{split}
\label{eq: Cov vasaf vasas OR}
& \left| 2 \Cov\left[ \E\left[ \frac{\vasaf_1}{\tarsagr_1} - \frac{\vasaf_2}{\tarsagr_2} \cond* \pco\vasaf_1, \pco\vasaf_2 \right]\!\!,\, \E\left[\frac{\vasas_1}{\tarsagr_1} - \frac{\vasas_2}{\tarsagr_2} \cond* \sus_1, \sus_2 \right] \right] \right| \\
&\quad \leq 2 \sqrt{ \Var\left[ \E\left[ \frac{\vasaf_1}{\tarsagr_1} - \frac{\vasaf_2}{\tarsagr_2} \cond* \pco\vasaf_1, \pco\vasaf_2 \right] \right]
\Var\left[ \E\left[\frac{\vasas_1}{\tarsagr_1} - \frac{\vasas_2}{\tarsagr_2} \cond* \sus_1, \sus_2 \right] \right] } \\
&\quad < 3 \sqrt{ \Var\left[ \frac{\pco \vasaf_1}{\tarsagr_1} - \frac{\pco \vasaf_2}{\tarsagr_2} \right]
\Var\left[ \E\left[\frac{\vasas_1}{\tarsagr_1} - \frac{\vasas_2}{\tarsagr_2} \cond* \sus_1, \sus_2 \right] \right] }.
\end{split}
\end{equation}
The variables $\pco \vasaf_1$ and $\pco \vasaf_2$ are independent. Combined with \eqref{eq: neg bin: Var} and \eqref{eq: suf 1 suf 2, cdeadd 1 cdeadd 2, OR LOR}, this yields
\begin{equation}
\label{eq: Var pco vasaf OR}
\Var\left[ \frac{\pco \vasaf_1}{\tarsagr_1} - \frac{\pco \vasaf_2}{\tarsagr_2} \right] \pco \roprpr^2\tarsapr = \frac{\suf_1(1-\pco p_1)}{\tarsara \pco\rapr} + \suf_2(1-\pco p_2) \tarsara \pco\rapr \approx \suf_1 \left( \frac 1 {\tarsara \pco\rapr} + \tarsara \pco\rapr \right).
\end{equation}
From \eqref{eq: Var vasaf OR bis}, \eqref{eq: Cov vasaf vasas OR} and \eqref{eq: Var pco vasaf OR} it follows that the fourth summand is asymptotically bounded for $\roprpr$ small as
\begin{multline}
\label{eq: Cov vasaf vasas OR bis}
\left| 2 \Cov\left[ \E\left[ \frac{\vasaf_1}{\tarsagr_1} - \frac{\vasaf_2}{\tarsagr_2} \cond* \pco\vasaf_1, \pco\vasaf_2 \right]\!\!,\, \E\left[\frac{\vasas_1}{\tarsagr_1} - \frac{\vasas_2}{\tarsagr_2} \cond* \sus_1, \sus_2 \right] \right] \right| \\
\quad < 2 \sqrt{ \Var\left[ \frac{\vasaf_1}{\tarsagr_1} - \frac{\vasaf_2}{\tarsagr_2} \right]
\Var\left[ \E\left[\frac{\vasas_1}{\tarsagr_1} - \frac{\vasas_2}{\tarsagr_2} \cond* \sus_1, \sus_2 \right] \right] }.
\end{multline}

Using \eqref{eq: Var vasaf OR bis}, \eqref{eq: E Var vasas cond OR bis}, \eqref{eq: Var E vasaf cond OR bis} and \eqref{eq: Cov vasaf vasas OR bis}, a similar assessment as in Appendix~\ref{part: app RR LRR gr var} can be made: for $\roprpr$ small; typical values of $\tarvar$, for which $1/\tarvar$ is large and $\suf_1$ is considerably smaller, namely $\suf_1 \approx \sqrt{2/(3\tarvar)}$; both for OR ($\cerr_1=\cerr_2=2$, $\adjsus=2$) and LOR ($\cerr_1=\cerr_2=5/4$, $\adjsus=0$); and for any $\tarsara$ and $\pco\rapr$, the third summand in the right-hand side of \eqref{eq: maxg variability OR bis} dominates the result for $\Var[\diffgr]$. Specifically, the first summand divided by the third takes the following approximate values, where the first corresponds to the case $\tarsara \pco\rapr \approx 1$ and the second to $\tarsara \pco\rapr \gg 1$ and to $\tarsara \pco\rapr \ll 1$: 
\[
\frac{(3\suf_1/2)^2}{4(1/\tarvar+3\suf_1/2)^2}, \quad \frac{(3\suf_1/2)^2}{2(1/\tarvar+3\suf_1/2)^2}.
\]
The ratio of the second and third summands has, for the three cases, the upper bound
\[
\frac{\suf_1}{2(1/\tarvar+3\suf_1/2)}.
\]
Lastly, the absolute value of the fourth summand in \eqref{eq: maxg variability OR bis} divided by the third is, in view of \eqref{eq: Cov vasaf vasas OR bis}, less than twice the square root of the ratio of first to third summands. Thus, all these ratios are seen to be significantly less than $1$.

It is concluded from the above analysis that the variability of $\diffgr$ in OR and LOR, as in RR and LRR, is mostly caused by that of $\vasas_1$ and $\vasas_2$ induced by the variations of $\sus_1$ and $\sus_2$. This implies that to compute $\E[|\diffgr|]$ each variable $\vasaf_i$, $i=1,2$ in \eqref{eq: diffgr} can be replaced by its mean, given by \eqref{eq: vasaf pco vasaf}, and $\vasas_i$ can be replaced by its conditional mean given $\sus_i$, which is approximately $\sus_i/\pco p_i$ according to \eqref{eq: E vasa 1 cond OR} and \eqref{eq: E vasa 2 cond OR}. This establishes \eqref{eq: pre diffgr OR}.



\section*{Acknowledgments}

The author would like to thank Dr.~Ben O'Neill for advice on terminology.

%

%


\end{document}